\documentclass[preprint2]{aastex631}
\usepackage{graphicx}
\usepackage{natbib}

\begin{document}

\title[WIYN Open Cluster Survey XC: NGC 2506]{WIYN Open Cluster Study. XC. Radial-velocity Measurements and Spectroscopic Binary Orbits in the Open Cluster NGC 2506}

\author{Evan Linck}
\affiliation{University of Wisconsin-Madison, 2535 Sterling Hall
475 N. Charter Street
Madison, WI 53706; elinck@wisc.edu}
\author{Robert D. Mathieu}
\affiliation{University of Wisconsin-Madison, 2535 Sterling Hall
475 N. Charter Street
Madison, WI 53706; elinck@wisc.edu}
\author{David W. Latham}
\affiliation{Center for Astrophysics ${\rm \mid}$  Harvard \& Smithsonian, 60 Garden Street, Cambridge, MA 02138}
\vspace{10pt}

\begin{abstract}

NGC 2506 is a rich, intermediate-age (2.0 Gyr), metal-poor ([Fe/H] $\sim$ -0.2) open cluster. This work presents the results of 12,157 spectroscopic radial-velocity measurements of 2,442 stars in the NGC 2506 field over 41 years, made as part of the WIYN Open Cluster Study. Radial-velocity measurements are complete for the population of proper-motion member stars brighter than Gaia G magnitude of 15.5, in which 320 proper-motion and radial-velocity cluster members were identified. Within the observation limit of G $<$ 16.5, 469 proper-motion and radial-velocity members were identified. This work reports on the characteristics of NGC 2506, including projected spatial distribution, radial-velocity dispersion, and virial mass. This work also presents orbital solutions for 49 binary members with periods between 1 and 7,580 days. NGC 2506 has an incompleteness-corrected binary frequency for binaries with periods less than $10^4$ days of $35 \pm 5\%$. This work also discusses in detail the 14 blue stragglers stars of NGC 2506---finding at least 64 $\pm$ 21\% to be in binaries, 5 yellow straggler stars, and several other stars of note.

\end{abstract}

\section{Introduction}\label{introduction}

NGC 2506 is a rich, intermediate-age, metal-poor open cluster ($\mathrm{\alpha=08^h\;00^m\; 02.10^s,\;\delta=}$ $\mathrm{-10^{\circ}\; 46'\; 45.15"}$, J2000). As noted by Anthony-Twarog et al.  \citeyearpar{anthony-twarogWIYNOPENCLUSTER2016, anthony-twarogWIYNOpenCluster2018}, at 2.0 Gyr, NGC 2506 is an important cluster to study as it occupies a unique age between other well-studied open clusters---including the younger clusters NGC 752 (1.4 Gyr), NGC 7789 (1.4 Gyr) and the older open cluster NGC 6819 (2.5 Gyr). These authors note that NGC 2506 is a prominent example of metal-deficient open clusters outside of the solar circle; and, due to its richness, has stars that clearly trace many stages of single-star stellar evolution. 

The first major investigations of NGC 2506 were complementary photometric and astrometric studies that found the cluster to have an age of 3.4 Gyr, estimated that roughly half of the stars are binaries, and identified 340 cluster members and 4 blue stragglers \citep{mcclureOldOpenCluster1981,chiuMembershipOldOpen1981}. Since then, determinations of NGC 2506's age have decreased to between 1.8 and 2.1 Gyr \citep{knudstrupExtremelyPreciseAge2020,anthony-twarogWIYNOPENCLUSTER2016,rangwalAstrometricPhotometricStudy2019,netopilMetallicityOpenClusters2016}. Recent studies of NGC 2506 have found it to be metal poor \citep[{[Fe/H]} ranging from -0.20 to -0.36,][]{anthony-twarogWIYNOpenCluster2018,misheninaNewInsightsBa2015,mikolaitisAbundancesCarbonIsotope2012,reddyComprehensiveAbundanceAnalysis2012,knudstrupExtremelyPreciseAge2020}, have low reddening (E(b-y) = 0.042$\pm 0.001$ , \cite{anthony-twarogWIYNOPENCLUSTER2016}; E(b-y) = 0.057$\pm 0.004$ , \cite{knudstrupExtremelyPreciseAge2020}; E(B-V) = 0.04, \cite{netopilMetallicityOpenClusters2016}), and be at a distance from 3.1 to 3.9 kpc  \citep{knudstrupExtremelyPreciseAge2020,gaoDiscoveryTidalTails2020,cantat-gaudinGaiaDR2View2018, anthony-twarogWIYNOPENCLUSTER2016,netopilMetallicityOpenClusters2016,rangwalAstrometricPhotometricStudy2019}. Here, we adopt the cluster distance of 3,101$\pm 17$ pc from \cite{knudstrupExtremelyPreciseAge2020}.

NGC 2506 also hosts a number of pulsating stars, including 6 $\delta$ Scuti stars and 15 $\gamma$ Doradus stars \citep{kimThreeDeltaScuti2000, arentoftOscillatingBlueStragglers2007}. NGC 2506 is dynamically relaxed and has undergone at least some mass segregation, the extent of which is debated \citep{leeDeepWidePhotometry2013, rangwalAstrometricPhotometricStudy2019}. Using Gaia DR2 data \citep{gaiacollaborationGaiaDataRelease2018}, \cite{gaoDiscoveryTidalTails2020} has found 147 stars that have parallaxes and proper motions consistent with those of the cluster but that are beyond the tidal radius, and suggests that these stars are part of two extra-tidal tails. 
 
This paper presents the results from an extensive radial-velocity (RV) survey of NGC 2506 undertaken with the WIYN 3.5m telescope\footnote{The WIYN 3.5m Observatory is a joint facility of the University of Wisconsin–Madison, Indiana University, NSF’s NOIRLab, the Pennsylvania State University, Princeton University, and Purdue University.} and the CfA Digital Speedometers on the MMT Observatory and 1.5m Tillinghast Reflector from 1982 November 28 through 2023 March 16. Our observations of cluster members are complete for stars brighter than 15.5 magnitude in Gaia G out to a radius of $0.6^{\circ}$ and reach to stars as faint as G = 16.5 magnitude. NGC 2506 is the most metal-poor cluster in the WIYN Open Cluster Survey (WOCS). Previous WOCS RV time-series survey papers include M35 \citep{gellerWIYNOPENCLUSTER2010, leinerWIYNOPENCLUSTER2015}, NGC 7789 \citep{nineWIYNOpenCluster2020}, NGC 6819 \citep{tabethaholeWIYNOPENCLUSTER2009, millimanWIYNOPENCLUSTER2014}, M67 \citep{gellerLARRADIALVELOCITIES2015, gellerStellarRadialVelocities2021}, NGC 188 \citep{gellerWIYNOPENCLUSTER2008, gellerWIYNOPENCLUSTER2009,gellerWIYNOPENCLUSTER2012}, and NGC 6791 \citep{tofflemireWIYNOPENCLUSTER2014, millimanWIYNOPENCLUSTER2016}.

Section \ref{sample} introduces our stellar sample and proper-motion membership determinations. Section \ref{sec_observations} discusses our spectroscopic observations and data reduction. Section \ref{sec_rv_observations} presents our RV measurements, RV membership calculations, and projected rotational velocity measurements. In Section \ref{Membership}, we categorize all RV and proper-motion member stars as single or binary stars and identify a complete sample of 320 member stars for G $<$ 15.5. Section \ref{results} focuses on the properties of NGC 2506, including the cluster isochrone, binary fraction, mean RV, velocity dispersion, virial mass, spatial distribution, and mass segregation. Section \ref{orbits} presents orbital solutions for 60 binary stars in this cluster and in the field. In Section \ref{starsofnote}, we discuss our findings on specific stars, including blue straggler stars (BSSs), yellow straggler stars (YSSs), and other stars of note.  In Section \ref{summary}, we summarize this work.

\section{Stellar Sample}\label{sample}

The original catalog of stars in the NGC 2506 field was developed by Imants Platais using astrometry from 2MASS \citep{2003_2MASS_pointsourcecat} and down-selected based on photometry from \cite{marconiOldOpenClusters1997} and \cite{momanyESOImagingSurvey2001} for stars brighter than V = 16.5 mag and within 30 arcmin of the center of the cluster to form our initial target sample. 

In 2020, the target sample was rebuilt using proper-motion membership and photometry from Gaia DR2 and again later from Gaia EDR3. The original catalog (23,491 stars) was matched with the roughly 58,000 Gaia stars within $0.6^{\circ}$ of the cluster center based on position and magnitude. Any match that was more than 1 arcsecond or 2 magnitudes different was checked by hand. We were able to match all the stars we had observed in the original catalog to Gaia stars. After conducting a proper-motion membership analysis (see Section \ref{PM_analysis}), we limited our target sample to stars with at least a 50\% proper-motion membership probability from HDBSCAN (see Section \ref{PM_analysis}) and that were brighter than G = 16.5, for a total sample size of 865 stars. Of these stars, only 22 stars were not within the original magnitude and radius limits of the catalog. 

In this paper, we use the term ``catalog" to refer to the entire list of stars in the NGC 2506 field for which we have RV observations; these stars are not necessarily proper-motion cluster members as many were observed prior to rebuilding our sample with Gaia proper motions. The term ``target sample" refers to the subset of stars that are proper-motion members within the magnitude and radius limits of our observations. 

Observation priorities were first set by higher membership proper-motion probabilities (e.g., $>$ 50\%), originally using the membership probabilities of \cite{chiuMembershipOldOpen1981} in the central field and subsequently throughout the cluster by Gaia membership probabilities in this paper. Priorities were then increased throughout the survey for RV variables or alternative stellar evolution products, such as BSSs. 

\subsection{Proper-motion Analysis}\label{PM_analysis}

Proper motions can be key to distinguishing cluster stars from those in the field. Clustering algorithms, such as Gaussian mixture models (GMM) or K-means, are frequently used to identify star cluster members based on astrometric information. Newer clustering algorithms, such as hierarchical density-based clustering \citep[HDBSCAN,][]{campelloDensityBasedClusteringBased2013}, were built to overcome some of the limitations of early clustering algorithms, such as requiring specific distributions of parameters or specifying the number of clusters in a dataset. Essentially, HDBSCAN identifies over-dense regions in a parameter space without specifying a type of probability distribution of those parameters.  

In an investigation of clustering algorithms for open cluster membership in Gaia DR2 data, \cite{huntImprovingOpenCluster2021} found that HDBSCAN is more sensitive than other clustering algorithms to discovering open clusters. Furthermore HDBSCAN is able to reproduce cluster membership lists of well-studied open clusters and is especially useful for detecting extra-tidal tails.  

In previous WOCS papers, we have used a simplified GMM to identify cluster members, typically identifying  stars with greater than 50\% membership probability as being cluster members. Here, we investigated both the Python scikit-learn implementation of GMM \citep{scikit-learn} and the Python implementation of HDBSCAN \citep{mcinnesHdbscanHierarchicalDensity2017} for identifying proper-motion members of NGC 2506. In this comparison, we limited our analyses to stars within $0.5^{\circ}$ of the cluster center, a parallax less than 1 mas, and G $<$ 15.5. We chose the 1 mas parallax cut-off as distance measurements in the literature place NGC 2506 between 3.1 and 3.9 kpc (see Section \ref{introduction}), which is equivalent to a parallax of 0.32 and 0.25 mas, respectively. To compare these algorithms, we ran HDBSCAN twice, once with only clustering stars based on the two dimensions of proper motion and again with clustering stars based on four-dimensional location on the sky and proper motion. 

\begin{figure}
    \centering
    \includegraphics[width=0.5\textwidth]{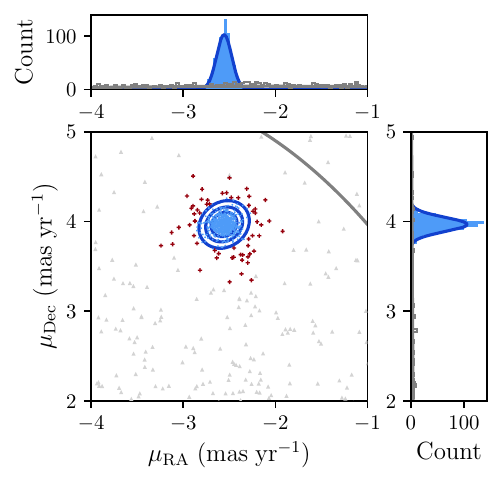}
    \caption{2-dimensional proper motions of stars in the NGC 2506 field. Proper-motion members of the cluster are well-defined compared to the field. Proper-motion cluster member stars from both the GMM and HDBSCAN analyses are plotted as small light blue circles, stars that are only HDBSCAN proper-motion members are plotted as red pluses, and field stars are plotted as grey triangles. Contours are plotted in the main panel for the 1-, 2-, and 3-$\sigma$ levels of the cluster Gaussian and for the 1-$\sigma$ level of the field Gaussian from the GMM analysis. Stars that are HDBSCAN proper-motion members but not GMM proper-motion members are in the over-dense regions just beyond the GMM proper-motion member distribution. The top and side panel show the cluster and field Gaussians projected into 1-dimension, plotted in blue and grey, respectively. Although binning was not used for the GMM analysis, we bin the data here for visual clarity.}
    \label{fig:GMM}
\end{figure}

We found that the GMM analysis identified 400 proper-motion members with membership probability greater than 50\%. Table \ref{pm_gmm_table} presents the parameters of our best fit Gaussian distributions for cluster and field proper motions.  The reported amplitude of the cluster and field are from scaling the Gaussian distributions from the GMM analysis to the binned data and are provided to show the relative scale of the field and cluster distributions. Errors were calculated by using a Monte Carlo simulation to fit a GMM to 1,000 realizations of the data shifted by a random sample of the normal distribution of each star's proper-motion errors. Figure \ref{fig:GMM} shows that NGC 2506 is well-separated from the field in proper-motion space. Even so, the field Gaussian distribution includes the cluster Gaussian; integrating over the two distributions suggests about 10\% of proper-motion members are falsely assigned to the cluster. The results of our RV membership study (Section \ref{Membership}) will help reduce false positives. 

Based on previous studies of NGC 2506 finding it to be a rich cluster of at least several hundred members, we set the minimum cluster size in HDBSCAN to 100 so as to avoid false positive over-dense groups in the field and left all other parameters as default \citep{mcinnesHdbscanHierarchicalDensity2017}. HDBSCAN determines a strength of membership for each star in the cluster grouping with which the star's astrometric characteristics are most associated. Because this strength of membership measures a different quantity than the probability of cluster membership from the GMM, we used a Monte Carlo simulation to assign membership probability with HDBSCAN, in a similar manner to \cite{cantat-gaudinGaiaDR2View2018}. We created 100 realizations of a star's proper motion from Gaia DR3 \citep{GaiaDataRelease2023} and shifted it by a random sample of the normal distribution of that star's proper-motion errors. We then ran HDBSCAN 100 times and assigned a membership probability based on the percentage of runs in which each star was grouped with the cluster. Probabilities of membership converged much quicker than the 100 trials. Stars with membership probabilities greater than 50\% are considered cluster members.

HDBSCAN identified 437 stars as cluster members based on proper motion alone and 436 stars as cluster members based on proper motion and location on the sky. Both HDBSCAN membership lists were fully inclusive of the GMM membership list of 400 stars. There are five stars that are different between the two HDBSCAN membership lists. These stars are all either at the extremes in radius or proper-motion difference from the cluster average. 

The distributions of member stars in the three algorithms are roughly the same in radius, but HDBSCAN includes stars that are further out in proper-motion space than the GMM does, as seen in Figure \ref{fig:GMM}. This suggests that HDBSCAN captures some over-dense, non-Gaussian structure in proper-motion space. As noted in Section \ref{Membership}, about 40\% of the stars that are proper-motion members according to the HDBSCAN analysis but not the GMM analysis are found to be RV members.

\begin{deluxetable}{lcr}
\tablehead{\colhead{Parameter} &\colhead{Cluster}
&\colhead{Field}}
\tabletypesize{\footnotesize}
\tablewidth{1.0\columnwidth}
\tablecaption{Cluster and Field Gaussian Distributions of Proper Motion\label{pm_gmm_table}}
\startdata
     $\mathrm{Ampl._{RA}\; (count)}$ & $104.78 \pm 1.38 $ & $6.50 \pm 0.02 $ \\
     $\mathrm{Ampl._{Dec}\; (count)}$ & $105.67 \pm 1.16 $ & $4.46 \pm 0.01 $ \\
     $\mathrm{\overline{\mu}_{RA}\;(mas\;yr^{-1})}$ & $-2.511 \pm  0.002$ & $-2.373\pm 0.002 $ \\
     $\mathrm{\overline{\mu}_{Dec}\;(mas\;yr^{-1})}$ & $3.966 \pm  0.001$ & $0.847 \pm 0.005 $ \\
     $\mathrm{\sigma_{RA} \;(mas\;yr^{-1})}$ & $0.0063\pm 0.0004 $ & $5.645 \pm 0.014 $ \\
     $\mathrm{\sigma_{Dec} \;(mas\;yr^{-1})}$ & $0.0062\pm 0.0003 $ & $ 9.513\pm  0.011 $ \\
\enddata
\end{deluxetable}

\subsection{Proper-motion Membership}\label{PMMembership}

For our final target sample of proper-motion members, we included all stars within a radius of $0.6^{\circ}$ of the cluster center (The field of the WIYN Hydra Multi-Object Spectrograph is $0.5^{\circ}$ in radius.). For our WIYN observation campaign, we choose to employ HDBSCAN membership probabilities using both location and proper motion to create our target sample in order to investigate the over-dense regions outside of the Gaussian distribution. 

Prior to doing our proper-motion membership analyses for all stars in the field, we removed foreground stars that had parallax measurements with uncertainties that do not overlap with the cluster parallax (generally less than 1,250 pc). As RV observations of 533 of these stars had been made prior to the availability of Gaia proper-motion memberships and are provided here, we assigned such stars a proper-motion membership probability of 0 and noted their parallax does not match the cluster's distance in Table \ref{memberships}. 22 stars for which we have made RV measurements do not have proper-motion or parallax measurements in Gaia DR3; these stars are listed in Table \ref{memberships} without a value for either proper-motion membership analysis and labelled ``No Gaia DR3 astrometry" in the comment. We treat these stars as non-members in analyses throughout this work. We did not use parallax otherwise as part of our selection criteria due to large uncertainties at the distance of the cluster. We also removed stars fainter than G=16.5. 

Our final target sample includes 865 stars within $0.6^{\circ}$ of the cluster, G $<$ 16.5 and HDBSCAN proper-motion membership $\geq$ 50\%. If we were to instead use our GMM proper-motion membership greater than 50\%, as we do in Sections \ref{Membership} and \ref{results}, our sample would be 797 stars. Membership probabilities for both analyses are given in Table \ref{memberships} for every star in our catalog (both proper-motion members and stars for which we had made RV measurements prior to the Gaia proper-motion analysis).

\subsection{Catalog of Stars in this Study}\label{star_summary}

Table \ref{memberships} provides the summarized information for each star in this study. The table provides the WOCS ID, the Gaia DR3 ID, position ($\mathrm{\alpha\;and\;\delta}$), Gaia photometry \citep{GaiaDataRelease2023}, number of RV measurements  (N, Table \ref{rv_observations}), mean RV ($\mathrm{\bar{RV}}$, Section \ref{sec_rv_observations}), standard deviation of RV (e, Section \ref{RV_measurements}), internal error (i, Section \ref{RV_measurements}), median $v \sin i $ measurement (Section \ref{vsini}), e/i, proper-motion membership probability from the GMM analysis ($\mathrm{P_{\mu,\;G}}$, Section \ref{PMMembership}), proper-motion membership probability from the HDBSCAN analysis ($\mathrm{P_{\mu,\;H}}$, Section \ref{PMMembership})), RV membership probability ($\mathrm{P_{RV}}$, Section \ref{rvmembership}), classification (Section \ref{Membership}), and comments. This table includes proper-motion members for which we do not have RVs for completeness of our sample. These stars are classified as U and do not have anything listed for values derived from RV measurements (e.g., $\mathrm{\bar{RV}}$,  $v \sin i $, $\mathrm{P_{RV}}$).

\begin{figure}
    \centering
    \includegraphics{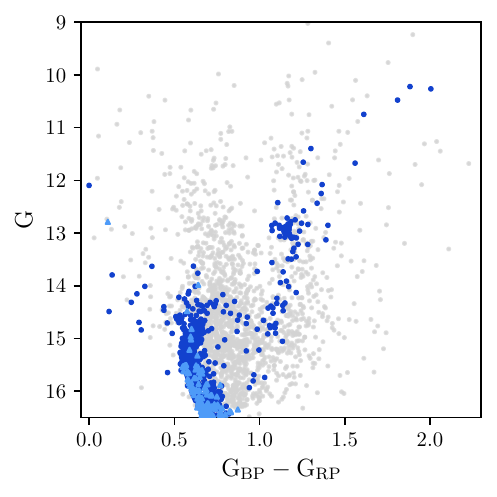}
    \caption{CMD of the catalog of stars in the NGC 2506 field for our RV survey. Blue markers are our final target sample of proper-motion members with dark blue circles representing stars for which we have at least one RV measurement and light blue triangles representing stars without an RV measurement. Grey circles are non-proper-motion members with RV measurements made prior to rebuilding the sample and constitute the remainder of our catalog; RVs of these stars are also reported in this paper.}
    \label{fig:pmm_sample_cmd}
\end{figure}

Figure \ref{fig:pmm_sample_cmd} shows a color-magnitude diagram (CMD) of proper-motion members and all stars with measured RVs (i.e., all stars in Table \ref{memberships}). Many RV measurements of field stars were obtained prior to rebuilding our sample with Gaia proper motions, in service at the time to an RV discovery survey of cluster membership.

\begin{deluxetable*}{l c c c c c c c c c c c c c c c p{1.3cm}}
\tabletypesize{\tiny}
\tablewidth{0pt}
\centering
\tablecaption{Star Summary Table\label{memberships}}
\tablehead{\colhead{WOCS} & \colhead{Gaia DR3} &\colhead{$\mathrm{\alpha}$} & \colhead{$\mathrm{\delta}$} & \colhead{$\mathrm{G}$} & \colhead{$\mathrm{G_{RP} - G_{BP}}$} & \colhead{$\mathrm{N}$} & \colhead{$\mathrm{\bar{RV}}$} & \colhead{$\mathrm{e}$} &\colhead{$\mathrm{i}$} & \colhead{$v\sin i$}& \colhead{$\mathrm{e/i}$} &  \colhead{$\mathrm{P_{\mu, G}}$}&\colhead{$\mathrm{P_{\mu, H}}$} &  \colhead{$\mathrm{P_{RV}}$}& \colhead{$\mathrm{Class}$}& \colhead{$\mathrm{Comm.}$}
} 
\startdata
1001  & 3038046289357637632 & 120.013718 & -10.762228 & 12.71 & 1.16 & 8 &  83.21  &  0.70 &  0.40 & <10   &   1.8 & 1.00 & 1.00  & 0.89 &  SM   &   Knudstrup: 1112. \\ 1002  & 3038044051673723008 & 120.023897 & -10.778706 & 11.72 & 0.74 & 3 &  40.73  &  0.19 &  0.40 & <10   &   0.5 & 0.00 & 0.00 & 0.00 & SNM    & Knudstrup: K1451. Parallax greater than 1 mas.\\ 
1003  & 3038043987255173888 & 120.024473 & -10.787090 & 11.40 & 1.30 & 3  & 81.76  &  0.50 &  0.40 & <10   &   1.2 & 0.00 & 1.00 & 0.53 & SNM   &  Knudstrup: K1320.\\ 
1004  & 3038044777529199744 & 119.985970 & -10.777444 & 10.89 & 0.38 & 7  &  6.83  &  0.53 &  1.58 & 99.7  &   0.3 & 0.00 & 0.00 & 0.00 & SNM   &  Parallax greater than 1 mas.\\
1017  & 3038059376119812096 & 120.021985 & -10.629734 & 10.22 & 1.88 & 28 &  84.27 &   3.22 &  0.40 & <10   &   8.1 &0.97 & 1.00 & 0.93 & BM    &\\
1025  & 3038084085069718656 & 120.010435 & -10.563982 & 10.99 & 0.15 & 0   &       &        &       & >120  &      & 0.00 &  0.00 &     & VRR, NM  &\\
3001  & 3038044055974650368 & 120.016223 & -10.774118 & 14.54 & 0.56 & 11 &  92.56 &  13.95 &  1.14 & 63.6  &  12.2 &1.00 & 1.00 & 0.00 & BU    &  SB2.\\
13035 & 3038017839494070784 & 120.300650 & -10.822861 & 13.81 & 0.61 & 20 &  52.36 &  14.31  & 0.40  & <10  &   35.8 & 0.00 &  0.00 & 0.00 & BNM   &  SB2.\\
16001 & 3038044055974648960 & 120.014792 & -10.769699 & 16.32 & 0.70 & 0   &       &         &        &     &      & 1.00 & 1.00   &    & U      & \\
16014 & 3038046186278294272 & 120.111373 & -10.713544 & 15.22 & 0.57 & 1  & 86.37   &       & 1.18  & 66.7   &     & 1.00 & 1.00 & 0.74 & U     & \\
19003 & 3038046323717361664 & 120.017399 & -10.750538 & 14.82 & 0.60 & 0  &         &       &       & >120  &      & 0.99 &  1.00 &      & VRR, M  & \\
\enddata
\tablecomments{This table is available in machine-readable format. $\mathrm{\alpha\;and\;\delta}$ are at epoch J2015.5. $\mathrm{\bar{RV},\;e,\;i,\;and\;}v\sin i$ are in $\mathrm{km\;s^{-1}}$. $\mathrm{P_{\mu, G}}$ is the probability of proper-motion membership from the GMM analysis; $\mathrm{P_{\mu, H}}$ is the probability of proper-motion membership from the HDBSCAN analysis.}
\end{deluxetable*}

\section{Observations and Data Reduction}\label{sec_observations}

\subsection{Telescope and Instrument}

Between 2005 December 14 through 2023 March 16, 12,157 RV measurements from spectra of 2,442 stars in the NGC 2506 field have been obtained at WIYN. Spectra were taken on the WIYN 3.5m telescope at Kitt Peak, Arizona using the Hydra Multi-Object Spectrograph (MOS). The WIYN telescope has a $1^{\circ}$ field of view and 0.8" median seeing, making it ideal for studying open clusters. The Hydra MOS has approximately 80 3.1" aperture fibers that can be precisely placed on targets. Our spectra span 250 $\text{\AA}$ centered on 5,125 $\text{\AA}$. This wavelength range encompasses the Mg b triplet and a large number of narrow metal absorption lines. The spectra were obtained using an echelle grating in the 11th order and achieved a resolution of approximately 20,000 and dispersion of 0.13 $\text{\AA}$ per pixel. Further details of our setup can be found in \cite{gellerWIYNOPENCLUSTER2008}.

An additional 80 RV measurements of 33 giants in the NGC 2506 field were obtained between 1982 November 28 and 1984 March 23 using the CfA Digital Speedometers. Further details on these CfA RV measurements can be found in \cite{gellerLARRADIALVELOCITIES2015}. 

\subsection{Observation Procedure and Data Reduction}

Complete observation and data reduction procedures for the WIYN spectra can be found in \cite{gellerWIYNOPENCLUSTER2008}, and are summarized here for convenience. For each observation, we assign approximately 70 fibers to stars and 10 fibers to sky for later sky subtraction. Each observation consists of a set of three 20- to 40-minute science exposures, depending on faintness of targets and sky conditions. A 200-second dome-flat and a 300-second thorium-argon emission lamp spectra is obtained before and after each set of science exposures. IRAF routines are used to subtract bias, extract spectra, derive wavelength solutions, correct for fiber-to-fiber throughput, and subtract sky. The L.A. Cosmic routine \citep{dokkumCosmicRayRejection2001} was added to reject cosmic rays starting in 2014. RVs are then calculated using the IRAF cross-correlation function (CCF) routine fxcor (discussed further in section \ref{RV_measurements}).

Target priorities are then updated for subsequent observation runs to prioritize velocity-variable stars without orbital solutions, alternative stellar evolution pathway stars (e.g., BSSs), and likely RV cluster members with fewer than 3 observations. 

\section{Radial-Velocity Measurements}\label{sec_rv_observations}

Table \ref{rv_observations} presents the RV measurements for each star. The first column lists the WOCS ID of a star if available, and otherwise gives the 19-digit Gaia DR3 ID. Further information on each star can be found in Table \ref{memberships}. We also give a truncated heliocentric Julian date of observation (T - 2,400,000), the radial velocities of the primary ($RV_1$) and secondary peaks ($RV_2$)---if a star is a double-lined spectroscopic binary (SB2), the CCF peak height of the primary (discussed further in Section \ref{RV_measurements}), and the residuals ($O-C$) and phase of RV measurements for stars with orbital solutions (discussed in Section \ref{orbits}).

\begin{deluxetable*}{l c r r r r r r r r r r c c l}
\tabletypesize{\footnotesize}
\tablewidth{0pt}
\centering
\tablecaption{Radial-Velocity Measurements\label{rv_observations}}
\tablehead{\colhead{ID} & \colhead{HJD - 2,400,000} & \colhead{$\mathrm{RV_1}$} & \colhead{Correlation} & \colhead{$\mathrm{O-C_1}$} & \colhead{$\mathrm{RV_2}$} & \colhead{$\mathrm{O-C_2}$} & \colhead{phase} \\ 
\colhead{} & \colhead{(days)} & \colhead{$\mathrm{(km\,s^{-1})}$} & \colhead{Height} & \colhead{$\mathrm{(km\,s^{-1})}$} & \colhead{$\mathrm{(km\,s^{-1})}$} & \colhead{$\mathrm{(km\,s^{-1})}$} & \colhead{} }
\startdata
9004 & 56708.681 & 68.759 & 0.69 & -2.17 & \nodata & \nodata & 0.46 \\
& 57031.849 & 89.163 & 0.81 & -1.41 & \nodata & \nodata & 0.88 \\
 & 57087.667 & 74.330 & 0.75 & 2.87 & \nodata & \nodata & 0.65 \\
 & 57405.757 & 130.319 & 0.74 & -0.63 & 39.619 & 0.37 & 0.99 \\
 & 57406.808 & 132.518 & 0.71 & 1.21 & 38.462 & -0.43 & 0.01 \\
 & 57443.656 & 64.799 & 0.73 & -5.73 & 99.161 & 0.27 & 0.51 \\
& 57444.650 & 61.200 & 0.77 & -9.29 & 89.447 & -9.48 & 0.53 \\
 & 57445.649 & 64.336 & 0.75 & -6.14 & 96.898 & -2.05 & 0.54 \\
 & 57446.632 & 65.823 & 0.75 & -4.67 & 103.866 & 4.94 & 0.55 \\
& 57447.694 & 67.100 & 0.75 & -3.45 & 101.219 & 2.35 & 0.57 \\
\enddata
\tablecomments{This table is available in machine-readable format. A portion is shown here for guidance regarding its form and content.}
\end{deluxetable*}

\subsection{Radial-velocity Measurements and Precision}\label{RV_measurements}

For WIYN spectra, instrumental RV measurements of single-lined stars were obtained by cross-correlating observed spectra with an observed solar template using the IRAF fxcor routine. Double-lined SB2s were identified by eye and the two-dimensional cross-correlation algorithm TODCOR \citep{zuckerStudySpectroscopicBinaries1994} was used to measure the RV of each component. RVs were corrected for slight offsets between Hydra fibers and then converted to a heliocentric velocity. 

RVs of single-lined stars are found automatically by fitting a Gaussian to the peak of the CCF and using the mean of the Gaussian as the RV measurement. To reduce erroneous or imprecise RV measurements, we require the peak height of the CCF be greater than 0.4, following \cite{gellerWIYNOPENCLUSTER2008}, unless checked manually. Measurements from the CfA Digital Speedometers do not have a CCF peak height. 

Stars that display rapid rotation are fit using fxcor interactively. For stars with a projected rotational velocity $v\sin i > 120\mathrm{\;km\;s^{-1}}$, this method of measuring RVs degrades due to very broad CCFs with higher frequency noise structure, and so does not provide reportable RV measurements. Measurement of $v \sin i$ is discussed in Section \ref{vsini}. The 17 stars observed in this study to have $v\sin i > 120\mathrm{\;km\;s^{-1}}$ are marked in Table \ref{memberships} as very rapid rotators (VRR). The 8 VRR proper-motion member stars for which we do not publish RVs due to high uncertainties had between 3 and 17 observations each. For 5 of the BSSs (WOCS 10006, 2006, 7005, 6010, and 49003, which are hotter and/or rapidly rotating), the solar template does not produce CCFs with peak heights greater than 0.4. For these stars, the correlations were performed with a library of spun-up Synspec spectral templates \citep{hubenySynspecGeneralSpectrum2011}. RVs are reported for the library template that produced the highest CCF peak, hereinafter referred to as the best-fit template. 

\begin{figure}
    \centering
    \includegraphics[width=0.45\textwidth]{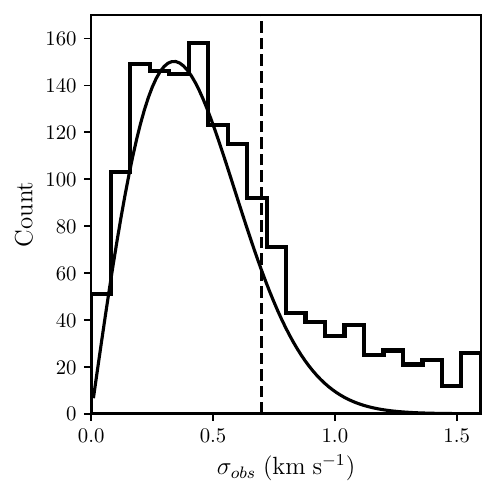}
    \caption{Histogram of the RV standard deviations of the narrow-line stars in our sample. The best-fit $\chi^2$ distribution function, with a $\sigma_i = \mathrm{0.5\; km\; s^{-1}}$, is shown as a solid curve. As $\sigma_{obs}$ will be higher for velocity variable stars, the fit is restricted to $\sigma_{obs} <\mathrm{0.7\; km\; s^{-1}}$ (dashed line). The excesses above the theoretical distribution beginning at $\mathrm{0.7\; km\; s^{-1}}$ are velocity-variable stars.}\label{RV_precision}
\end{figure}

We identify binarity in single-lined stars by examining RV variability over time. This requires quantifying the WIYN RV measurement precision, first for narrow-line stars ($v \sin i < 10 \mathrm{\;km\;s^{-1}}$, see Section \ref{vsini}). Following \cite{gellerWIYNOPENCLUSTER2008}, the distribution of RV standard deviations of stars with small RV measurement variation ($\mathrm{\sigma_{obs}< 1.6\; km\; s^{-1}}$) can be fit with a $\chi^2$ distribution. Any excess population of narrow-line stars above the $\chi^2$ distribution is presumed to be due to velocity variability. Figure \ref{RV_precision} shows our best-fit $\chi^2$ distribution with two degrees of freedom, fit only to the core of the observed distribution. We found our precision, $\sigma_i$, to be $0.5 \mathrm{\;km\; s^{-1}}$, which is similar to values from previous WOCS studies \citep[e.g., $\mathrm{0.4\; km\; s^{-1}}$,][]{gellerWIYNOPENCLUSTER2008}. For consistency with other papers, we adopt $\sigma_i = 0.4 \mathrm{\;km\; s^{-1}}$. 

Contributors to our precision measure include photon errors, observation-to-observation errors, fiber-to-fiber variations, and long-period binaries that are mistaken as single stars, as characterized in \cite{gellerWIYNOPENCLUSTER2008}. Following the convention of earlier WOCS papers, we identify as velocity variable those stars with RV standard deviations (external error, $\sigma_e$ or e) greater than 4 times the precision (the internal error, $\sigma_i$ or i). Specifically, for stars with low $v \sin i$, we count stars with RV standard deviation of $1.6 \mathrm{\;km\; s^{-1}}$ as velocity variable. For stars with higher $v \sin i$, \cite{gellerWIYNOPENCLUSTER2010a} found that the internal error scales as: 

\begin{equation}
    \sigma_i = 0.38 + 0.012*(v \sin i) \mathrm{\;km\;s^{-1}}
\end{equation}

\noindent Table \ref{memberships} provides e, i, e/i and $v \sin i$ for each star. 

\subsection{Radial-velocity Completeness}

\begin{figure*}
    \centering
    \includegraphics{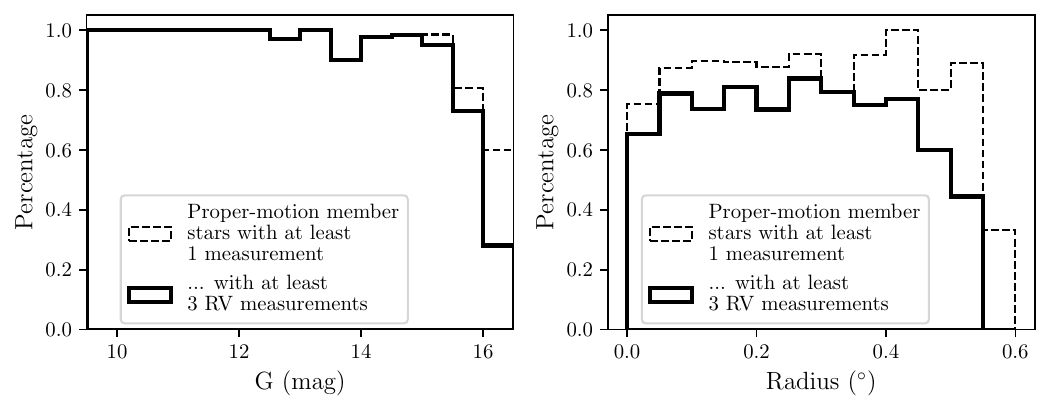}
    \caption{Left panel: percentage by G magnitude of proper-motion members that have at least one (dashed line) or three (solid line) RV measurements. Right panel: percentage by radius ($^\circ$) of proper-motion members that have at least one (dashed line) or three (solid line) RV measurements.}\label{vel_var_complete}
\end{figure*}

We have made at least one RV measurement of 727 of our 865 proper-motion members (G $<$ 16.5). 421 of the 436 proper-motion member stars brighter than G = 15.5 have 3 RV measurements. Of the 15 with fewer than 3 RV measurements, 13 have been observed at least 3 times but do not have 3 observations that meet our quality-control thresholds---including the 8 VRR stars---and 2 have fewer than 3 observations. As such, we can make RV membership and binarity determinations for 96.6\% of proper-motion member stars with G $<$ 15.5 (see Sections \ref{rvmembership} and \ref{Membership}). Figure \ref{vel_var_complete} shows the completeness for proper-motion members of G $<$ 16.5 with at least one and at least three RV measurements as functions of G magnitude (left panel) and radius (right panel). Among stars brighter than G = 15.5, most radial bins are complete for three measurements. The three innermost bins ($\mathrm{r}<0.15^{\circ}$) and the bin for $0.25^{\circ}<\mathrm{r} < 0.30^{\circ}$ are at least 95\% complete, and the outermost radial bin ($\mathrm{r}>0.55^{\circ}$), which contains one star in this magnitude range, does not have three observations.  

\subsection{Radial-Velocity Members}\label{rvmembership}

Radial velocities provide another dimension for cluster membership analyses and can eliminate field interlopers from the proper-motion membership list. We used both the HDBSCAN and GMM algorithms to group stars in either the field or cluster as a function of RV, following the procedures in Section \ref{PM_analysis}. To reduce the impact of artificially inflating the observed cluster velocity dispersion due to binaries, we used the mean of the RV measurements for each non-velocity-variable star to define the field and cluster distributions. 

\begin{figure}
    \centering
    \includegraphics{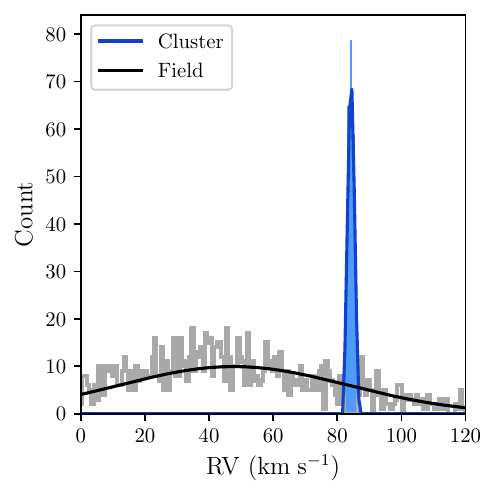}
    \caption{Distributions of measured RVs for non-velocity variable stars. RV measurements are binned for visual clarity on the plot, but these bins were not used in fitting. The Gaussian distributions representing cluster and field stars are plotted and their sums are scaled to the height of the highest count bin.}\label{rv_gmm}
\end{figure}

Figure \ref{rv_gmm} shows the Gaussian distribution fits of the GMM to the cluster and field. Table \ref{rv_gmm_table} gives the parameters of the best-fit Gaussian distributions for the field and cluster with errors estimated using a Monte Carlo simulation with the same procedure as the GMM proper-motion analysis in Section \ref{PM_analysis}. Like with our proper-motion analysis, HDBSCAN identified over-densities in the wings of the GMM cluster Gaussian distribution as cluster members. However, as the distribution of field stars with RV measurements is significantly more sparse than those with proper-motion measurements, HDBSCAN does not cleanly separate the boundary between the cluster and field. As such, the remainder of this work will use the results of the GMM RV membership analysis to determine cluster membership in order to be conservative in identifying the stars most associated with the kinematics of NGC 2506. 

After each grouping was defined, cluster membership probabilities were found for every star using the following values for RV. For non-velocity-variable stars and velocity-variable stars that do not have known orbital solutions, we used the mean RV measurement. For SB2s, we use the mean RV measurement of the primary star. For binary stars with orbital solutions, we use the systemic velocity of the orbit (see Section \ref{orbits}). Table \ref{memberships} lists the mean RV, RV standard deviation, and GMM RV membership probability of each star we have observed.

\cite{knudstrupExtremelyPreciseAge2020} measured RVs for 122 stars in the NGC 2506 field as part of their spectroscopic survey of the cluster. We matched every star in Appendix A of \cite{knudstrupExtremelyPreciseAge2020} against our catalog and list the ID given in \cite{knudstrupExtremelyPreciseAge2020} in the comment. Three stars (K2019 (V5), K682, and 3308) are not included in Table \ref{memberships}
as they were below our faint limit or not a proper-motion member. Our determination of RV membership largely agrees with the RV membership of the 123 stars in Appendix A of \cite{knudstrupExtremelyPreciseAge2020}. 
All stars with RV membership greater than 0.5 in this work are also Knudstrup RV members.
Of the 42 stars we determine to be non-RV members, \cite{knudstrupExtremelyPreciseAge2020} find 28 (66\%) to also be non-RV members. 11 of the 14 remaining stars have RVs from \cite{knudstrupExtremelyPreciseAge2020} that fall outside of the cluster RV distribution found in this work. The remaining 3 have RVs from \cite{knudstrupExtremelyPreciseAge2020} that are within our distribution, but our RV measurements place these stars slightly outside the cluster RV distribution.

We compared the mean RVs of the 70 stars that both this work and \cite{knudstrupExtremelyPreciseAge2020} consider non-velocity variable stars. The distribution of the RV differences show a core Gaussian distribution including 51 stars, with the velocity differences of the remaining 19 stars being outliers (and perhaps long-period binaries).
The mean of this distribution indicates that the RVs reported in \cite{knudstrupExtremelyPreciseAge2020} are 0.63 $\pm$ 0.09 km s$^{-1}$ higher than the RVs reported in this study.

\begin{deluxetable}{lcr}
\tablehead{\colhead{Parameter} &\colhead{\hspace{1.5cm}Cluster}
&\colhead{Field}}
\tablewidth{1.0\columnwidth}
\tabletypesize{\footnotesize}
\tablecaption{Cluster and Field Gaussian Distributions of RV\label{rv_gmm_table}}
\startdata
        $\mathrm{Ampl.\; (count)}$ & \hspace{1.5cm}$67.23 \pm 1.10$ & $10.02 \pm 0.27$ \\
         $\mathrm{\overline{RV}\;(km\;s^{-1})}$ & \hspace{1.5cm}$84.29 \pm  0.06$ & $47.68 \pm 0.08 $ \\
         $\mathrm{\sigma \;(km\;s^{-1})}$ & \hspace{1.5cm}$1.17\pm 0.16 $ & $ 35.55 \pm  1.37 $ \\
\enddata
\end{deluxetable}

\subsection{Projected Rotation Velocities}\label{vsini}

The full-width half-maximum (FWHM) of a CCF peak traces the projected rotational velocity, $v \sin i$. Following the procedure established in \cite{gellerWIYNOPENCLUSTER2010}, we calculate the $v \sin i$ of each star based on the median FWHM of the CCFs of all measurements of that star. Our spectral resolution limits our measurement of $v \sin i$ to a floor of 10 $\mathrm{km\;s^{-1}}$. For any star measured at this limit, we report $v \sin i < 10 \mathrm{\;km\;s^{-1}}$. 

For stars that are clearly or likely SB2s, we do not report a $v \sin i$ unless the CCFs of both binary members can be well separated, in which case we report the $v \sin i$ of the primary. Some stars may have a companion that is bright enough to contribute to a star's spectrum but too faint to easily de-convolve into separate CCFs. These may falsely broaden the CCF, which we would report as having a higher $v \sin i$. 

Table \ref{memberships} lists the $v \sin i$ for each star for which we have a measurement. 

\section{Three-dimensional Cluster Membership}\label{Membership}

\begin{figure}
    \centering
    \includegraphics{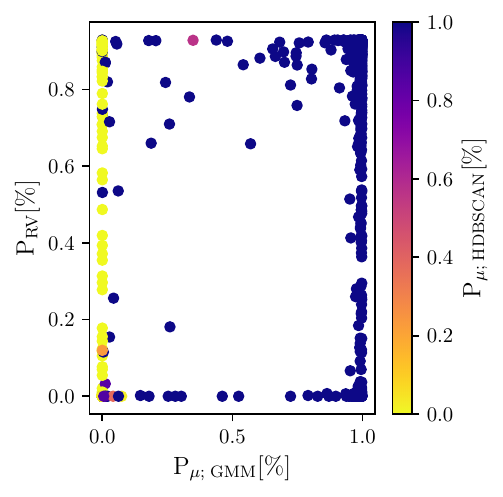}
    \caption{Comparison of proper-motion membership probabilities from GMM with RV membership probabilities from the GMM analysis. The color of each point represents the proper-motion membership probability from HDBSCAN. We find that including RV information is important to separating cluster stars from the field as it identifies the $\sim10\%$ of field interlopers with similar proper motions. \label{fig:rv_pm_comp}}
\end{figure}

Based on our Gaussian analysis in Section \ref{PM_analysis}, we expect that 10\% of our proper-motion members are false positives. To improve membership identification, we incorporate RV observations to further remove non-members. Figure \ref{fig:rv_pm_comp} compares RV membership probabilities with GMM and HDBSCAN proper-motion membership probabilities for all stars in our sample. Among stars that have RV measurements and $\mathrm{e/i}<4$, we find that 90.7\% of our GMM proper-motion members are also RV members, and 87.1\% of our HDBSCAN proper-motion members are also RV members. 

There are 68 stars that were found to not be proper-motion members in our GMM but are proper-motion members with HDBSCAN. Including RV information and limiting to stars with $\mathrm{e/i}<4$ to remove binaries with unknown RV membership, 15 of these stars are RV members and 20 are not. The rate of RV non-members among this population (57.1\%) is significantly higher than the rate of RV non-members among the proper-motion members from the GMM (9.3\%). We suggest the 15 stars that are RV members but are outside of the proper-motion Gaussian distribution are closely associated with the cluster and could be escaping stars. 

Because of this high false-positive rate of membership and to more accurately capture the kinematic properties of the stars of NGC 2506, for the remaining analyses in this paper we use the proper-motion and RV memberships from the GMM analyses to determine cluster membership.

Monte Carlo simulations show that three epochs of RV measurements over at least a year are required to give us a 95\% confidence that a star is either a binary or single star \citep{gellerWIYNOPENCLUSTER2008}. As such, we require at least three observations to identify a star as a binary. Combining the GMM proper-motion and RV analyses, we categorize each star in NGC 2506 as single stars or binaries and as members or non-members in Table \ref{memberships}. Following the conventions of previous WOCS papers, we classify each star as follows:

\begin{itemize}
    \item Single Member (SM): stars that have $\mathrm{e/i}<4$, $P_{RV} >50$\%, and $P_{\mu,\;GMM} >50$\%. 
    \item Single Non-member (SNM): stars that have $\mathrm{e/i}<4$ and either $P_{RV} <50$\% or  $P_{\mu,\;GMM} <50$\%. 
       \item Binary Member (BM): velocity-variable stars ($\mathrm{e/i\geq4}$) that have a completed orbital solution, $P_{RV} >50$\%, and $P_{\mu,\;GMM} >50$\%. 
     \item   Binary Non-member (BNM): velocity-variable stars that have a completed orbital solution and either $P_{RV} <50$\% or $P_{\mu,\;GMM} <50$\%.
     \item   Binary Likely Member (BLM): velocity-variable stars that do not have a completed orbital solution, $P_{RV} >50$\%, and $P_{\mu,\;GMM} >50$\%.
     \item   Binary Likely Non-member (BLN): velocity-variable stars that do not have a completed orbital solution. Either $P_{RV} <50$  or $P_{\mu,\;GMM} <50$ and the range of RVs does not include the cluster mean, making it unlikely that the orbital solution will place the star within the cluster distribution.
    \item  Binary Unknown (BU): velocity-variable stars that do not have a completed orbital solution. $P_{RV} <50$\%, but $P_{\mu,\;GMM} >50$\%, and the range of individual RVs includes the cluster mean, making it possible that the binary could be a member. 
    \item  Very Rapid Rotator Likely Member (VRR, M): stars that are too rapidly rotating ($v\sin i > 120\mathrm{\;km\;s^{-1}}$) for accurate RV measurements, $P_{\mu,\;GMM} >50$\%. We include these stars as cluster members for purposes of population analyses below.
    \item  Very Rapid Rotator Likely Non-member (VRR, NM): stars that are too rapidly rotating for accurate RV measurements, $P_{\mu,\;GMM} <50$\%.
    \item Unknown (U): Stars with $P_{\mu,\;GMM} >50$\% that have fewer than three RV measurements. Some of these stars are likely RV members and/or binaries, but we do not have enough measurements to conclusively make that determination.
    \item Unknown Non-member (UNM): $P_{\mu,\;GMM} < 50$\% stars that have fewer than three RV measurements. These stars are included as many had RV measurements taken prior to rebuilding our catalog with proper-motion membership.
\end{itemize}

 Table \ref{star_category_counts} tallies the number of stars from our sample in each membership category. We find 320 stars with G $<$ 15.5 that are members (SM, BM, BLM). As an upper bound, if we include proper-motion member binaries that have RV measurements that overlap the cluster's RV distribution (BU) and proper-motion member rapid rotators (VRR, M), we find 367 member stars with G $<$ 15.5. Including stars to our observation limit of G $<$ 16.5, we have identified 469 member stars (SM, BM, BLM). As an upper bound, if we also include proper-motion member BU and VRR,M stars, we find 553 member stars. 

\begin{deluxetable}{lr}
\tablehead{\colhead{Category} &\colhead{\hspace{3cm}Count}}
\tablewidth{1.0\columnwidth}

\tablecaption{Number of Stars within each Membership Category\label{star_category_counts}}
\startdata
      SM &    354 \\
     SNM &   1272 \\
      BM &     49 \\
     BNM &    11 \\
     BLM &     68 \\
     BLN &    243 \\
      BU &     78 \\
  VRR, M &      6 \\
 VRR, NM &     11 \\
       U &    205 \\
     UNM &    293 \\
\enddata
\end{deluxetable}

Additionally of note, we have found several stars that have appeared in the literature of NGC 2506 to not be member stars. Oscillating BSS V8 and eclipsing binaries V26 and V27 from \cite{arentoftOscillatingBlueStragglers2007} are not proper-motion members and, at least for V8 and V26 for which we have RV measurements, not RV members.

\section{Results}\label{results}

\subsection{Binary Frequency}

Among stars from the main-sequence turnoff (MSTO) (G = 14.5) through our completeness limit (G = 15.5), we identified 216 member stars (SM, BM, BLM), 56 of which are spectroscopic binaries (BM, BLM). This yields an observed main-sequence binary frequency for periods less than $10^4$ days of $25.9\pm3.5\%$. (We note that 6 VRR stars are not included in this analysis.) This frequency is at the high end of the range of uncorrected observed main-sequence binary frequencies of other open clusters studied by WOCS, including M35 \citep[15\%][]{leinerWIYNOPENCLUSTER2015}, NGC 7789 \citep[19\%][]{nineWIYNOpenCluster2020}, NGC 6819 \citep[16\%][]{millimanWIYNOPENCLUSTER2014}, M67 \citep[25\%][]{gellerStellarRadialVelocities2021}, and NGC 188 \citep[19\%][]{gellerWIYNOPENCLUSTER2012}.

Among non-YSS (or YSS candidates, see Section \ref{yss}) giant stars, we find $18.7 \pm 4.9 \%$ to be binary stars. Notably, of the 9 giant binaries with orbital solutions (see Section \ref{orbits}), the shortest period is 103 days, suggesting shorter period binaries did not survive evolution of the primary up the giant branch and instead produced BSS or other alternative evolution products. 

As our survey of NGC 2506 has a similar duration, number of observations, and sample size to our survey of M67 \citep{gellerStellarRadialVelocities2021}, we adopt the binary detection completeness of that paper (74\%) found through a Monte Carlo simulation of RVs from the field-binary distribution sampled at the cadence of our observation campaign. This yields an incompleteness-corrected binary fraction for binaries with periods less than $10^4$ days of $35 \pm 5\%$.

This binary fraction is within errors the same as the binary fractions of M67 \citep[34\% with e/i $\geq 3$; ][]{gellerStellarRadialVelocities2021} and NGC 188 \citep[32\%; ][]{gellerWIYNOPENCLUSTER2008, gellerStellarRadialVelocities2021} after corrections for incomplete detections of binaries. 

\subsection{Cluster Isochrone}\label{cluster_isochrone}

We fit MIST isochrones \citep{dotterMESAISOCHRONESLAR2016, choiMESAISOCHRONESSTELLAR2016, paxtonMODULESEXPERIMENTSSTELLAR2011, paxtonMODULESEXPERIMENTSSTELLAR2013,paxtonMODULESEXPERIMENTSSTELLAR2015} to the single member stars. Using the Python package isochrones \citep{mortonIsochronesStellarModel2015}, we implemented a grid of interpolated MIST isochrones with distance = 3,101 pc, 30 ages between 1.77 Gyr and 2.24 Gyr, and 16 metallicities between -0.5 and -0.1, and 15 A$_V$ values between 0.1 to 0.24. For each isochrone from every parameter combination, we calculated the distance in color and magnitude to every non-velocity variable star that was likely not an alternative evolution pathway star, and then chose the isochrone with the lowest residuals. 

\begin{figure*}
    \centering
    \includegraphics{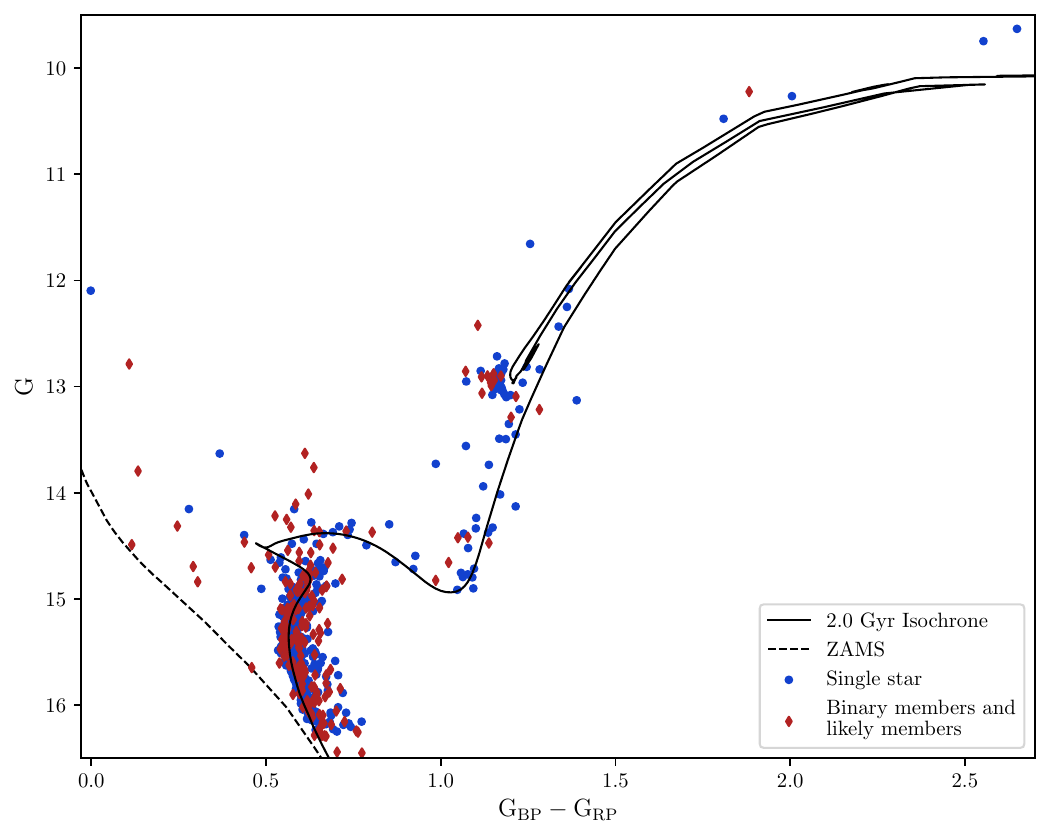}
    \caption{NGC 2506 three-dimensional member stars with the best-fit MIST isochrone (solid line). The isochrone was fit to non-BSS non-RV-variable stars, marked with blue circles. Red diamonds represent spectroscopic binary stars. We have also plotted a zero-age main-sequence MIST isochrone for reference (dashed line).}\label{isochrone}
\end{figure*}

Figure \ref{isochrone} shows the single and binary members of NGC 2506 with our best-fit isochrone reddened to an A$_V$ of 0.16. The best-fit isochrone had parameters of 2.03 Gyr, [Fe/H] = -0.15 , and A$_V$ = 0.16, in agreement with other recent determinations (Section \ref{introduction}). The agreement of isochrone and data is good, excepting the brightest giants and the color of the red clump, discrepancies that have been seen in some other clusters fit by MIST isochrones \citep[e.g., M67; ][]{choiStarClusterAges2018}.

\subsection{Cluster Radial Velocity and Velocity Dispersion}

The mean RV of NGC 2506 is $84.29 \pm 0.06 \mathrm{\;km\;s^{-1}}$, with an observed RV dispersion for single, narrow-lined RV and GMM proper-motion members of $0.89 \pm 0.07 \mathrm{\; km\; s^{-1}}$. 

This observed RV dispersion includes undetected long-period binaries ($P > 10^4$ days) and internal measurement errors, which will inflate the true velocity dispersion. \cite{gellerWIYNOPENCLUSTER2008} \citep[see also][]{mathieuStructureInternalKinematics1985} used Monte Carlo simulations to model the effects of undetected binaries in open clusters. They showed that measured velocity dispersions in massive open clusters are inflated by as much as $0.22\mathrm{\; km\; s^{-1}}$. We adopt this value to correct our observed dispersion for undetected binaries and further subtract in quadrature our internal error to find a true RV dispersion of $0.54 \pm 0.06 \mathrm{\; km\; s^{-1}}$. 

\cite{knudstrupExtremelyPreciseAge2020} found NGC 2506 to have a peak in the RV distribution of 83.8 km s$^{-1}$ with a FWHM of 4.7 km s$^{-1}$. Our measurement of the mean RV of NGC 2506 falls well within their distribution, even after accounting for a zero-point shift of 0.6 km s$^{-1}$  (see Section \ref{rvmembership}). In addition, \cite{anthony-twarogWIYNOpenCluster2018} find the cluster RV to be 83.2 $\pm$ 1.2 km s$^{-1}$ and \cite{carlbergROTATIONALRADIALVELOCITIES2014} find 83.4 $\pm$ 1.3 km s$^{-1}$, each using RV measurements from approximately two dozen giants in NGC 2506.

\subsection{Virial Mass, Radial Spatial Distribution, and Mass Segregation}

NGC 2506 has a projected half-mass radius of $R_{hm}$ = $4.95 \pm 0.75$ pc based on the upper-main-sequence SM stars. We compute the virial mass from \cite{spitzerDynamicalEvolutionGlobular1987}:

\begin{equation}
    M = \frac{10 <\sigma_c^2> R_{hm}}{G}
\end{equation}

\noindent where $\sigma_c$ is the RV dispersion. We find NGC 2506 to have a current virial mass of 3,400 $\pm\;900 M_\odot$. 

Using the MIST isochrone to assign masses to the upper main-sequence stars from the MSTO through our completeness limit and then extending to 0.1 $M_\odot$ with a Kroupa initial mass function (IMF; \citep{kroupaInitialMassFunction2002}, NGC 2506 would have between 5,200 and 6,000 $M_\odot$ of mass in stars less than the red clump mass of 1.6 $M_\odot$, depending on whether the slope of the IMF for stars of $M \geq 1 M_\odot$ is 2.3 or 2.7. This photometric mass is marginally higher than the virial mass, perhaps indicative of loss of low-mass stars due to dynamical evolution.

\begin{figure}
    \centering
    \includegraphics[width=0.45\textwidth]{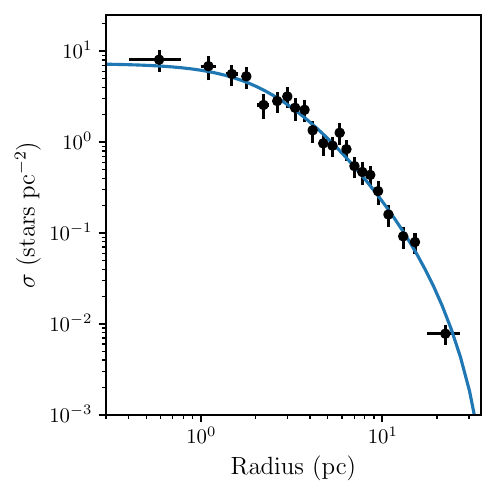}
    \caption{A King profile of core radius of $2.46 \pm 0.13$ pc and a tidal radius of $44.58 \pm 2.92$ pc fit to 405 upper-main-sequence stars of NGC 2506. }
    \label{fig:king_prof}
\end{figure}

In Figure \ref{fig:king_prof}, we show a single-mass King profile \citep{kingStructureStarClusters1966} as implemented by \cite{gielesFamilyLoweredIsothermal2015} in the Python software LIMEPY fit to the radial distribution of 405 upper-main-sequence stars (G $<$ 16.5). Stars were grouped into radial bins of equal counts of stars prior to finding the surface density of each bin. As bins contain equal numbers of stars, we used the equiprobable bin rule ($k = 2 n ^{\frac{2}{5}} = 22$ bins) to define the number of bins.

We used a Monte Carlo simulation with 1,000 realizations of a King profile fit to estimate errors of our fit. We find NGC 2506 to have a core radius of $2.46 \pm 0.13$ pc and a tidal radius of $44.58 \pm 2.92$ pc. The measured tidal radius is influenced by the number of bins we use, but the core radius is quite stable. The tidal radius is beyond the target sample radius ($0.6^\circ$ or 32.5 pc at a distance of 3,101 pc) of this work.

This result is in good agreement with \cite{vaidyaBlueStragglerPopulations2020}, who found a core radius of 2.7 arcminutes (2.4 pc at a distance of 3,101 pc) and tidal radius of 42.6 arcminutes  (38.4 pc at a distance of 3,101 pc), and is similar to the core radius found by \cite{rangwalAstrometricPhotometricStudy2019} of $3.23 \pm 0.15$ arcminutes ($2.91 \pm 0.14$ pc at a distance of 3,101 pc). 

For a cluster of 3,400 $\pm\;900 M_\odot$ and a distance from the galactic center of 10.84 kpc on a nearly circular galactic orbit \citep[e = 0.02;][]{rangwalAstrometricPhotometricStudy2019}, we expect a tidal radius \citep[using the Bovy 2015 Milky Way gravitational potential; ][]{bovyGalpyPythonLIBRARY2015} at the perigalactic point of the cluster's orbit \citep[equation 11 of ][]{kingStructureStarClusters1962} of $21.3 \pm 0.9$ pc, significantly smaller than the tidal radius we have found from fitting a King profile to cluster members. 

Using Gaia DR2 data, \cite{gaoDiscoveryTidalTails2020} has found 147 stars that are associated with the astrometric characteristics of the cluster, but are beyond 22.8 arcminutes (20.6 pc at a distance of 3,101 pc) from the cluster center (based on the tidal radii found by \cite{leeDeepWidePhotometry2013} of 19 arcminutes (17.1 pc at a distance of 3,101 pc) and \cite{rangwalAstrometricPhotometricStudy2019} of 12 arcminutes (10.8 pc at a distance of 3,101 pc)). Gao suggests that these stars are part of two tidal tails that extend in projection at least 54 pc (using a distance of 3,111 pc) from the cluster center in opposite directions. However, many of these stars are within the tidal radius reported in this work. 

As noted in Section \ref{Membership}, our membership analyses in both proper motion and RV showed slight over-densities of stars just outside of the wings of the fit cluster Gaussian distributions, suggesting NGC 2506 has closely associated stars that may be escaping from the cluster. We found 15 stars to be RV members that are outside of the proper-motion Gaussian distribution, but are HDBSCAN proper-motion members. Spatially, these stars are distributed in projection from the center of the cluster to its outer radii, which may suggest stars are escaping in all directions.

\begin{figure}
    \centering
    \includegraphics[width=0.45\textwidth]{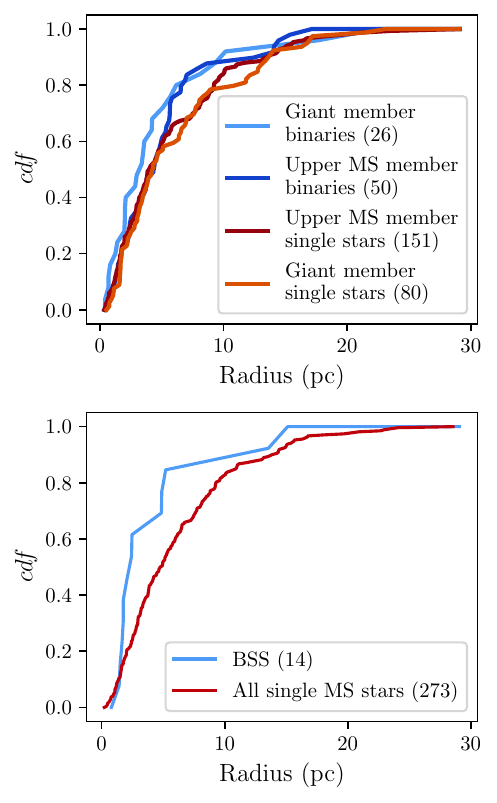}
    \caption{Cumulative distribution functions of different stellar populations in NGC 2506. The upper panel shows the distributions of single and binary giant and main sequence stars. NGC 2506 does not show formally significant mass segregation between binaries and single stars. However, the lower panel shows that the BSS population is centrally segregated from the main-sequence single stars (p = 0.02, bottom panel).}
    \label{fig:mass_seg}
\end{figure}

The extent of mass segregation in the cluster is debated in the literature (Lee et al. 2013; Rangwal et al. 2019). In the upper panel of Figure \ref{fig:mass_seg}, we show the cumulative radial distributions of single and binary giant and main-sequence stars. Although both the giant and main-sequence binary stars are somewhat more centrally concentrated than the single stars of either group, the differences in their spatial distributions are not statistically significant. A Kolmogorov-Smirnov (K-S) test of the cumulative distribution functions of upper main sequence binary and single stars returned a p-value of 0.22. We do, however, find significant mass segregation between the BSS and the main-sequence populations (p = 0.02; shown in the lower panel of Figure \ref{fig:mass_seg}). 

\section{Spectroscopic Binary Orbits}\label{orbits}

Using the RV observations in Table \ref{rv_observations}, we have found orbital solutions for 49 binary member and 11 binary non-member stars using a direct-integration-orbit-fitting code \citep{gellerStellarRadialVelocities2021}. 

\subsection{Single-lined Binaries}

Single-lined binaries (SB1s) are defined by a single peak in the CCF of the spectrum using the solar template. The orbital parameters for our SB1 solutions are given in Table \ref{SB1tab}. For each binary, we give the WOCS ID (ID), period (P), the number of cycles spanned by our observations, center-of-mass RV of the binary ($\gamma$), semi-amplitude RV (K), eccentricity (e), longitude of periastron ($\omega$), the truncated heliocentric Julian date of periastron passage ($\mathrm{T_0}$), the projected semi-major axis ($a \sin i$), the mass function ($\mathrm{f(m)}$), the rms of the RV residuals from the orbital solution ($\sigma$), and the number of RV measurements contributing to the solution. 1-$\sigma$ errors are listed on the second row. We show phase-folded orbital solutions and residuals (see Table \ref{rv_observations} for $O-C$ residual values) for each binary in Figure \ref{fig:sb1fig1}.

\startlongtable
\begin{deluxetable*}{l r c r r r r r r r r c c}
\tabletypesize{\tiny}
\tablewidth{0pt}
\centering
\tablecaption{Orbital Parameters For NGC 2506 Single-Lined Binaries\label{SB1tab}}
\tablehead{\colhead{ID} & \colhead{P} & \colhead{Orbital} & \colhead{$\mathrm{\gamma}$} & \colhead{K} & \colhead{e} & \colhead{$\mathrm{\omega}$} & \colhead{$\mathrm{T_\circ}$} & \colhead{a $\mathrm{\sin}$ i} & \colhead{f(M)} &  \colhead{$\mathrm{\sigma}$} & \colhead{N} \\
\colhead{} &         \colhead{(days)} & \colhead{Cycles} & \colhead{($\mathrm{km\; s^{-1}}$)} & \colhead{($\mathrm{km\; s^{-1}}$)} & \colhead{} & \colhead{(deg)} & \colhead{(HJD-2400000 d)} & \colhead{(10$^6$ km)} & \colhead{($\mathrm{M_\odot}$)} & \colhead{($\mathrm{km\; s^{-1}}$)} & \colhead{}}
\startdata
1017 & 456.6 & 13.5 & 84.27 & 4.63 & 0.34 & 346 & 56541 & 27.3 & $3.9*10^{-3}$ & 0.493 & 31\\
         &     $\pm$0.5  &      &       $\pm$ 0.09 &       $\pm$0.13 &     $\pm$ 0.03 &       $\pm$ 5 &      $\pm$ 5 &       $\pm$ 0.8 &      $\pm$ $3.0*10^{-4}$ &    &      \\
2001 & 103.00 & 38.7 & 83.9 & 15.2 & 0.11 & 320 & 57046 & 21.4 & $3.7*10^{-2}$ & 4.253 & 23\\
         &     $\pm$0.15  &      &       $\pm$ 0.9 &       $\pm$1.4 &     $\pm$ 0.1 &       $\pm$ 40 &      $\pm$ 12 &       $\pm$ 2.0 &      $\pm$ $1.0*10^{-2}$ &    &      \\
2010 & 1323.0 & 10.5 & 82.98 & 9.77 & 0.324 & 181 & 55032 & 168 & $1.08*10^{-1}$ & 0.358 & 26\\
         &     $\pm$1.5  &      &       $\pm$ 0.1 &       $\pm$0.15 &     $\pm$ 0.010 &       $\pm$ 3 &      $\pm$ 10. &       $\pm$ 3 &      $\pm$ $5.0*10^{-3}$ &    &      \\
2011 & 345.5 & 16.8 & 84.6 & 10.97 & 0.06 & 147 & 58251 & 52.0 & $4.7*10^{-2}$ & 0.391 & 17\\
         &     $\pm$0.5  &      &       $\pm$ 0.3 &       $\pm$0.26 &     $\pm$ 0.03 &       $\pm$ 19 &      $\pm$ 17 &       $\pm$ 1.2 &      $\pm$ $3.0*10^{-3}$ &    &      \\
2022 & 2540 & 2.4 & 83.28 & 4.09 & 0.58 & 89 & 58615 & 117 & $9.8*10^{-3}$ & 0.360 & 30\\
         &     $\pm$30  &      &       $\pm$ 0.08 &       $\pm$0.18 &     $\pm$ 0.04 &       $\pm$ 3 &      $\pm$ 18 &       $\pm$ 6 &      $\pm$ $1.4*10^{-3}$ &    &      \\
4001 & 110.39 & 35.5 & 87.2 & 23.3 & 0.20 & 302 & 57527 & 34.6 & $1.36*10^{-1}$ & 2.847 & 19\\
         &     $\pm$0.06  &      &       $\pm$ 0.8 &       $\pm$1.2 &     $\pm$ 0.04 &       $\pm$ 14 &      $\pm$ 4 &       $\pm$ 1.8 &      $\pm$ $2.1*10^{-2}$ &    &      \\
4002 & 816 & 6.2 & 84.77 & 3.2 & 0.37 & 111 & 57270 & 34 & $2.3*10^{-3}$ & 0.839 & 20\\
         &     $\pm$8  &      &       $\pm$ 0.23 &       $\pm$0.5 &     $\pm$ 0.11 &       $\pm$ 16 &      $\pm$ 30 &       $\pm$ 5 &      $\pm$ $1.0*10^{-3}$ &    &      \\
4004 & 472.76 & 29.5 & 84.36 & 5.59 & 0.227 & 16 & 53730. & 35.4 & $7.9*10^{-3}$ & 0.315 & 16\\
         &     $\pm$0.17  &      &       $\pm$ 0.12 &       $\pm$0.16 &     $\pm$ 0.025 &       $\pm$ 6 &      $\pm$ 7 &       $\pm$ 1.0 &      $\pm$ $7.0*10^{-4}$ &    &      \\
4005 & 7580 & 1.9 & 83.3 & 8 & 0.82 & 29 & 53210 & 500 & $6.0*10^{-2}$ & 0.664 & 21\\
         &     $\pm$180  &      &       $\pm$ 0.3 &       $\pm$9 &     $\pm$ 0.21 &       $\pm$ 25 &      $\pm$ 210 &       $\pm$ 600 &      $\pm$ $2.4*10^{-1}$ &    &      \\
5007 & 1988 & 7.3 & 82.6 & 4.8 & 0.42 & 330. & 56210 & 120 & $1.7*10^{-2}$ & 0.388 & 13\\
         &     $\pm$15  &      &       $\pm$ 0.3 &       $\pm$1.5 &     $\pm$ 0.12 &       $\pm$ 13 &      $\pm$ 60 &       $\pm$ 40 &      $\pm$ $1.6*10^{-2}$ &    &      \\
7002 & 9.94962 & 401.2 & 82.6 & 34.0 & 0.047 & 59 & 56348.8 & 4.65 & $4.05*10^{-2}$ & 1.548 & 16\\
         &     $\pm$0.00024  &      &       $\pm$ 0.4 &       $\pm$0.5 &     $\pm$ 0.017 &       $\pm$ 23 &      $\pm$ 0.6 &       $\pm$ 0.07 &      $\pm$ $1.9*10^{-3}$ &    &      \\
7004 & 609 & 10.1 & 85.2 & 3.8 & 0.31 & 30 & 56970 & 30. & $3.0*10^{-3}$ & 1.158 & 21\\
         &     $\pm$4  &      &       $\pm$ 0.3 &       $\pm$0.6 &     $\pm$ 0.1 &       $\pm$ 30 &      $\pm$ 40 &       $\pm$ 5 &      $\pm$ $1.4*10^{-3}$ &    &      \\
7005 & 601.0 & 10.3 & 85.48 & 6.8 & 0.04 & 170 & 57250 & 55.8 & $1.91*10^{-2}$ & 1.035 & 33\\
         &     $\pm$2.1  &      &       $\pm$ 0.20 &       $\pm$0.3 &     $\pm$ 0.04 &       $\pm$ 60 &      $\pm$ 100 &       $\pm$ 2.4 &      $\pm$ $2.5*10^{-3}$ &    &      \\
7008 & 68.428 & 85.9 & 84.86 & 13.5 & 0.447 & 328 & 56872.7 & 11.4 & $1.26*10^{-2}$ & 1.023 & 32\\
         &     $\pm$0.007  &      &       $\pm$ 0.19 &       $\pm$0.3 &     $\pm$ 0.017 &       $\pm$ 3 &      $\pm$ 0.6 &       $\pm$ 0.3 &      $\pm$ $9.0*10^{-4}$ &    &      \\
8009 & 1084 & 12.9 & 84.3 & 6.6 & 0.32 & 210. & 55040 & 93 & $2.7*10^{-2}$ & 1.037 & 21\\
         &     $\pm$3  &      &       $\pm$ 0.3 &       $\pm$0.8 &     $\pm$ 0.08 &       $\pm$ 13 &      $\pm$ 30 &       $\pm$ 11 &      $\pm$ $1.0*10^{-2}$ &    &      \\
8010 & 43.737 & 122.8 & 83.81 & 27.4 & 0.486 & 117.4 & 56314.11 & 14.39 & $6.2*10^{-2}$ & 1.264 & 34\\
         &     $\pm$0.005  &      &       $\pm$ 0.26 &       $\pm$0.4 &     $\pm$ 0.014 &       $\pm$ 1.8 &      $\pm$ 0.17 &       $\pm$ 0.26 &      $\pm$ $3.0*10^{-3}$ &    &      \\
8012 & 3.349210 & 1658.8 & 82.85 & 35.7 & 0.061 & 303 & 56593.18 & 1.641 & $1.57*10^{-2}$ & 0.874 & 22\\
         &     $\pm$0.000008  &      &       $\pm$ 0.22 &       $\pm$0.3 &     $\pm$ 0.007 &       $\pm$ 10. &      $\pm$ 0.1 &       $\pm$ 0.013 &      $\pm$ $4.0*10^{-4}$ &    &      \\
9002 & 827 & 5.7 & 87.0 & 13.3 & 0.59 & 21 & 56945 & 122 & $1.1*10^{-1}$ & 2.416 & 20\\
         &     $\pm$12  &      &       $\pm$ 1 &       $\pm$1.0 &     $\pm$ 0.07 &       $\pm$ 10. &      $\pm$ 15 &       $\pm$ 12 &      $\pm$ $3.0*10^{-2}$ &    &      \\
9005 & 663 & 5.0 & 83.64 & 4.44 & 0.08 & 90 & 57890 & 40.4 & $6.0*10^{-3}$ & 0.568 & 26\\
         &     $\pm$3  &      &       $\pm$ 0.15 &       $\pm$0.15 &     $\pm$ 0.04 &       $\pm$ 30 &      $\pm$ 60 &       $\pm$ 1.3 &      $\pm$ $6.0*10^{-4}$ &    &      \\
9012 & 11.5420 & 376.1 & 79.6 & 23.7 & 0.005 & 300 & 56474 & 3.76 & $1.59*10^{-2}$ & 1.117 & 14\\
         &     $\pm$0.0003  &      &       $\pm$ 0.3 &       $\pm$0.5 &     $\pm$ 0.016 &       $\pm$ 400 &      $\pm$ 13 &       $\pm$ 0.08 &      $\pm$ $1.0*10^{-3}$ &    &      \\
10004 & 16.8991 & 261.8 & 84.1 & 11.1 & 0.13 & 278 & 56703.0 & 2.55 & $2.3*10^{-3}$ & 1.582 & 27\\
         &     $\pm$0.0012  &      &       $\pm$ 0.4 &       $\pm$0.5 &     $\pm$ 0.04 &       $\pm$ 22 &      $\pm$ 1 &       $\pm$ 0.12 &      $\pm$ $3.0*10^{-4}$ &    &      \\
10007 & 385 & 14.4 & 84.0 & 4.7 & 0.56 & 297 & 57477 & 20. & $2.3*10^{-3}$ & 0.815 & 16\\
         &     $\pm$3  &      &       $\pm$ 0.3 &       $\pm$0.5 &     $\pm$ 0.11 &       $\pm$ 13 &      $\pm$ 15 &       $\pm$ 3 &      $\pm$ $9.0*10^{-4}$ &    &      \\
10012 & 1733 & 3.6 & 82.94 & 3.1 & 0.41 & 64 & 58400 & 68 & $4.1*10^{-3}$ & 0.794 & 25\\
         &     $\pm$25  &      &       $\pm$ 0.24 &       $\pm$0.5 &     $\pm$ 0.11 &       $\pm$ 19 &      $\pm$ 60 &       $\pm$ 11 &      $\pm$ $2.0*10^{-3}$ &    &      \\
10016 & 234.62 & 23.3 & 84.21 & 22.8 & 0.171 & 173 & 55442 & 72.5 & $2.75*10^{-1}$ & 0.739 & 18\\
         &     $\pm$0.11  &      &       $\pm$ 0.22 &       $\pm$0.3 &     $\pm$ 0.011 &       $\pm$ 5 &      $\pm$ 3 &       $\pm$ 1.0 &      $\pm$ $1.2*10^{-2}$ &    &      \\
11009 & 216.70 & 15.1 & 84.6 & 16.0 & 0.39 & 124 & 57152.7 & 44 & $7.3*10^{-2}$ & 1.054 & 15\\
         &     $\pm$0.23  &      &       $\pm$ 0.5 &       $\pm$1.2 &     $\pm$ 0.04 &       $\pm$ 5 &      $\pm$ 2.3 &       $\pm$ 3 &      $\pm$ $1.6*10^{-2}$ &    &      \\
11020 & 291.9 & 13.6 & 85.2 & 4.9 & 0.46 & 74 & 56664 & 17 & $2.5*10^{-3}$ & 0.888 & 14\\
         &     $\pm$0.5  &      &       $\pm$ 0.4 &       $\pm$0.8 &     $\pm$ 0.14 &       $\pm$ 17 &      $\pm$ 8 &       $\pm$ 3 &      $\pm$ $1.3*10^{-3}$ &    &      \\
11029 & 3670 & 1.5 & 83.70 & 5.38 & 0.192 & 79 & 59020 & 266 & $5.6*10^{-2}$ & 0.419 & 31\\
         &     $\pm$30  &      &       $\pm$ 0.08 &       $\pm$0.13 &     $\pm$ 0.022 &       $\pm$ 7 &      $\pm$ 70 &       $\pm$ 7 &      $\pm$ $4.0*10^{-3}$ &    &      \\
12014 & 379.5 & 26.0 & 86.4 & 9.3 & 0.38 & 182 & 57094 & 45 & $2.5*10^{-2}$ & 1.666 & 16\\
         &     $\pm$1.3  &      &       $\pm$ 0.5 &       $\pm$0.7 &     $\pm$ 0.07 &       $\pm$ 12 &      $\pm$ 10. &       $\pm$ 4 &      $\pm$ $6.0*10^{-3}$ &    &      \\
13031 & 4070 & 1.1 & 84.4 & 4.5 & 0.45 & 17 & 59570 & 220 & $2.7*10^{-2}$ & 1.478 & 23\\
         &     $\pm$110  &      &       $\pm$ 0.3 &       $\pm$0.5 &     $\pm$ 0.1 &       $\pm$ 17 &      $\pm$ 130 &       $\pm$ 30 &      $\pm$ $9.0*10^{-3}$ &    &      \\
15003 & 13.865 & 347.4 & 83.3 & 12.5 & 0.13 & 300 & 58078 & 2.4 & $2.7*10^{-3}$ & 4.657 & 15\\
         &     $\pm$0.004  &      &       $\pm$ 1.7 &       $\pm$1.8 &     $\pm$ 0.14 &       $\pm$ 70 &      $\pm$ 3 &       $\pm$ 0.3 &      $\pm$ $1.2*10^{-3}$ &    &      \\
16005 & 2.176087 & 1843.2 & 84.4 & 39.3 & 0.027 & 187 & 57044.98 & 1.174 & $1.36*10^{-2}$ & 0.958 & 16\\
         &     $\pm$0.000006  &      &       $\pm$ 0.3 &       $\pm$0.4 &     $\pm$ 0.013 &       $\pm$ 18 &      $\pm$ 0.11 &       $\pm$ 0.011 &      $\pm$ $4.0*10^{-4}$ &    &      \\
16006 & 4.72994 & 1142.5 & 84.05 & 19.94 & 0.115 & 180. & 56696.13 & 1.288 & $3.81*10^{-3}$ & 0.678 & 20\\
         &     $\pm$0.00003  &      &       $\pm$ 0.19 &       $\pm$0.19 &     $\pm$ 0.013 &       $\pm$ 6 &      $\pm$ 0.08 &       $\pm$ 0.013 &      $\pm$ $1.1*10^{-4}$ &    &      \\
16007 & 577.7 & 9.6 & 84.1 & 14.6 & 0.20 & 22 & 56790 & 114 & $1.8*10^{-1}$ & 1.834 & 15\\
         &     $\pm$2.0  &      &       $\pm$ 0.6 &       $\pm$0.9 &     $\pm$ 0.06 &       $\pm$ 17 &      $\pm$ 30 &       $\pm$ 7 &      $\pm$ $3.0*10^{-2}$ &    &      \\
17007 & 20.4951 & 70.7 & 84.01 & 37.79 & 0.395 & 355.1 & 56035.30 & 9.78 & $8.88*10^{-2}$ & 0.380 & 13\\
         &     $\pm$0.0010  &      &       $\pm$ 0.15 &       $\pm$0.16 &     $\pm$ 0.005 &       $\pm$ 0.9 &      $\pm$ 0.03 &       $\pm$ 0.05 &      $\pm$ $1.2*10^{-3}$ &    &      \\
17026 & 16.7476 & 281.5 & 79.7 & 19.1 & 0.22 & 66 & 58498.0 & 4.30 & $1.13*10^{-2}$ & 1.372 & 11\\
         &     $\pm$0.0014  &      &       $\pm$ 0.7 &       $\pm$0.7 &     $\pm$ 0.03 &       $\pm$ 16 &      $\pm$ 0.7 &       $\pm$ 0.16 &      $\pm$ $1.3*10^{-3}$ &    &      \\
20002 & 3.12146 & 1977.3 & 84.79 & 3.7 & 0.36 & 165 & 57437.93 & 0.146 & $1.3*10^{-5}$ & 0.551 & 14\\
         &     $\pm$0.00005  &      &       $\pm$ 0.17 &       $\pm$0.4 &     $\pm$ 0.06 &       $\pm$ 15 &      $\pm$ 0.13 &       $\pm$ 0.016 &      $\pm$ $4.0*10^{-6}$ &    &      \\
20003 & 26.5696 & 153.1 & 85.5 & 11.8 & 0.20 & 313 & 56874.3 & 4.23 & $4.3*10^{-3}$ & 1.020 & 17\\
         &     $\pm$0.0020  &      &       $\pm$ 0.3 &       $\pm$0.6 &     $\pm$ 0.04 &       $\pm$ 10. &      $\pm$ 0.7 &       $\pm$ 0.21 &      $\pm$ $6.0*10^{-4}$ &    &      \\
20032 & 8.5070 & 166.6 & 84.30 & 32.7 & 0.015 & 170 & 56423.0 & 3.82 & $3.07*10^{-2}$ & 0.761 & 14\\
         &     $\pm$0.0003  &      &       $\pm$ 0.22 &       $\pm$0.3 &     $\pm$ 0.008 &       $\pm$ 50 &      $\pm$ 1.1 &       $\pm$ 0.04 &      $\pm$ $9.0*10^{-4}$ &    &      \\
24003 & 16.30503 & 287.4 & 83.30 & 21.9 & 0.398 & 270. & 57688.49 & 4.50 & $1.37*10^{-2}$ & 0.717 & 16\\
         &     $\pm$0.00022  &      &       $\pm$ 0.20 &       $\pm$0.5 &     $\pm$ 0.018 &       $\pm$ 3 &      $\pm$ 0.09 &       $\pm$ 0.10 &      $\pm$ $9.0*10^{-4}$ &    &      \\
24012 & 126.79 & 31.6 & 84.9 & 8.5 & 0.61 & 341 & 57186.6 & 11.8 & $4.0*10^{-3}$ & 1.354 & 19\\
         &     $\pm$0.11  &      &       $\pm$ 0.4 &       $\pm$1.3 &     $\pm$ 0.06 &       $\pm$ 7 &      $\pm$ 2.1 &       $\pm$ 2.0 &      $\pm$ $1.9*10^{-3}$ &    &      \\
26003 & 111.33 & 41.3 & 82.6 & 16.3 & 0.38 & 275 & 57490. & 23.0 & $3.9*10^{-2}$ & 1.933 & 14\\
         &     $\pm$0.05  &      &       $\pm$ 0.7 &       $\pm$1.6 &     $\pm$ 0.08 &       $\pm$ 16 &      $\pm$ 4 &       $\pm$ 2.3 &      $\pm$ $1.2*10^{-2}$ &    &      \\
26004 & 2000 & 2.3 & 85.4 & 10.6 & 0.46 & 176 & 56960 & 260 & $1.7*10^{-1}$ & 2.201 & 16\\
         &     $\pm$40  &      &       $\pm$ 0.6 &       $\pm$1.8 &     $\pm$ 0.11 &       $\pm$ 12 &      $\pm$ 60 &       $\pm$ 50 &      $\pm$ $9.0*10^{-2}$ &    &      \\
27007 & 1.934675 & 2870.1 & 84.98 & 9.4 & 0.06 & 00 & 56144.79 & 0.251 & $1.68*10^{-4}$ & 0.928 & 19\\
         &     $\pm$0.000015  &      &       $\pm$ 0.22 &       $\pm$0.4 &     $\pm$ 0.05 &       $\pm$ 30 &      $\pm$ 0.14 &       $\pm$ 0.010 &      $\pm$ $2.0*10^{-5}$ &    &      \\
31040 & 1.210046 & 3268.5 & 86.5 & 17.0 & 0.03 & 330 & 58323.93 & 0.282 & $6.1*10^{-4}$ & 1.152 & 21\\
         &     $\pm$0.000003  &      &       $\pm$ 0.3 &       $\pm$0.4 &     $\pm$ 0.03 &       $\pm$ 40 &      $\pm$ 0.15 &       $\pm$ 0.007 &      $\pm$ $5.0*10^{-5}$ &    &      \\
\enddata
\end{deluxetable*}

\begin{figure*}
    \centering
    \includegraphics{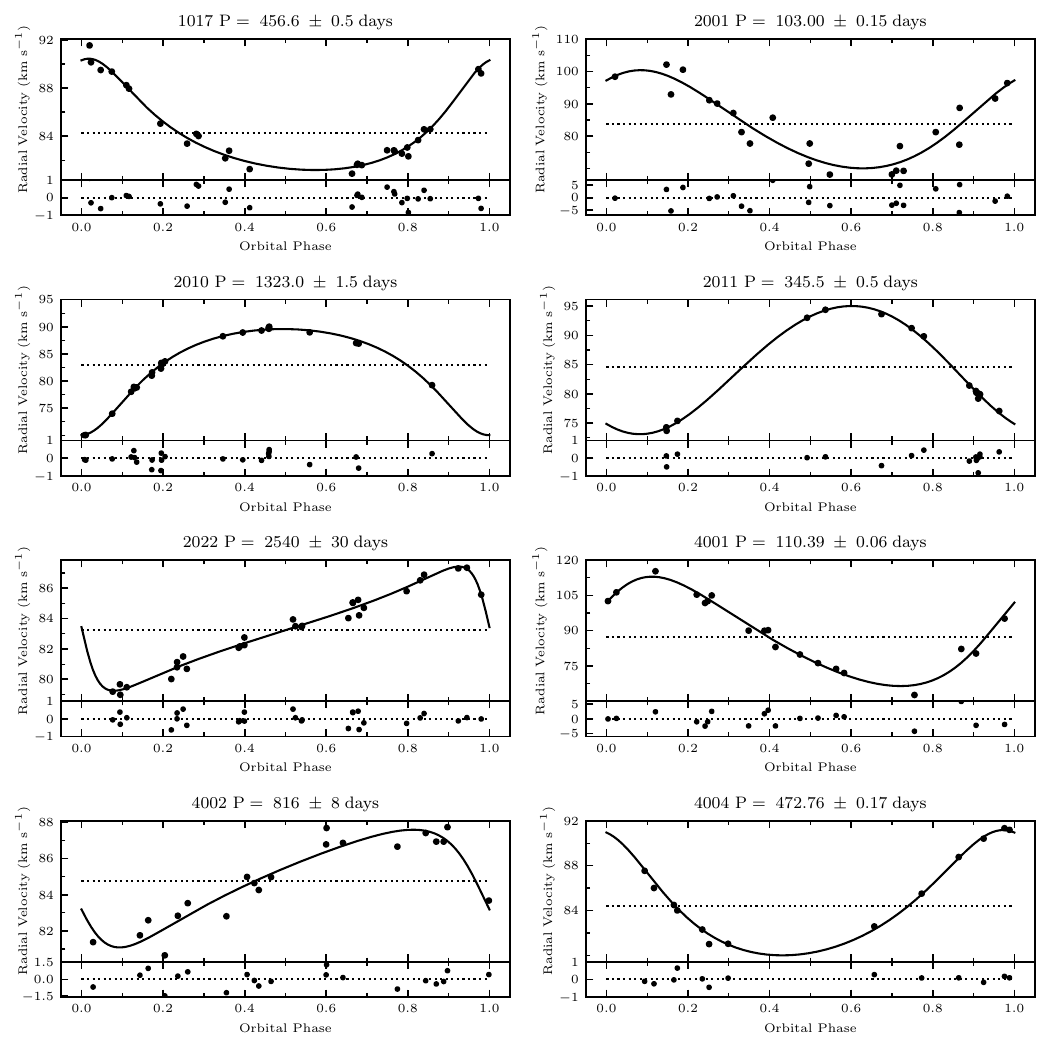}
    \caption{NGC 2506 SB1 orbit plots. For each binary listed in Table \ref{SB1tab}, we plot RV measurements against orbital phase. In the top panel, our best-fit orbit is the solid line and $\gamma$-velocity is marked as the dotted line. In the bottom panel, we plot phase-folded residuals and a dotted line on residuals of 0 $\mathrm{km\;s^{-1}}$ for reference. Above each plot we give the star's WOCS ID and orbital period.}
    \label{fig:sb1fig1}
\end{figure*}
\begin{figure*}
    \addtocounter{figure}{-1}
    \centering
    \includegraphics{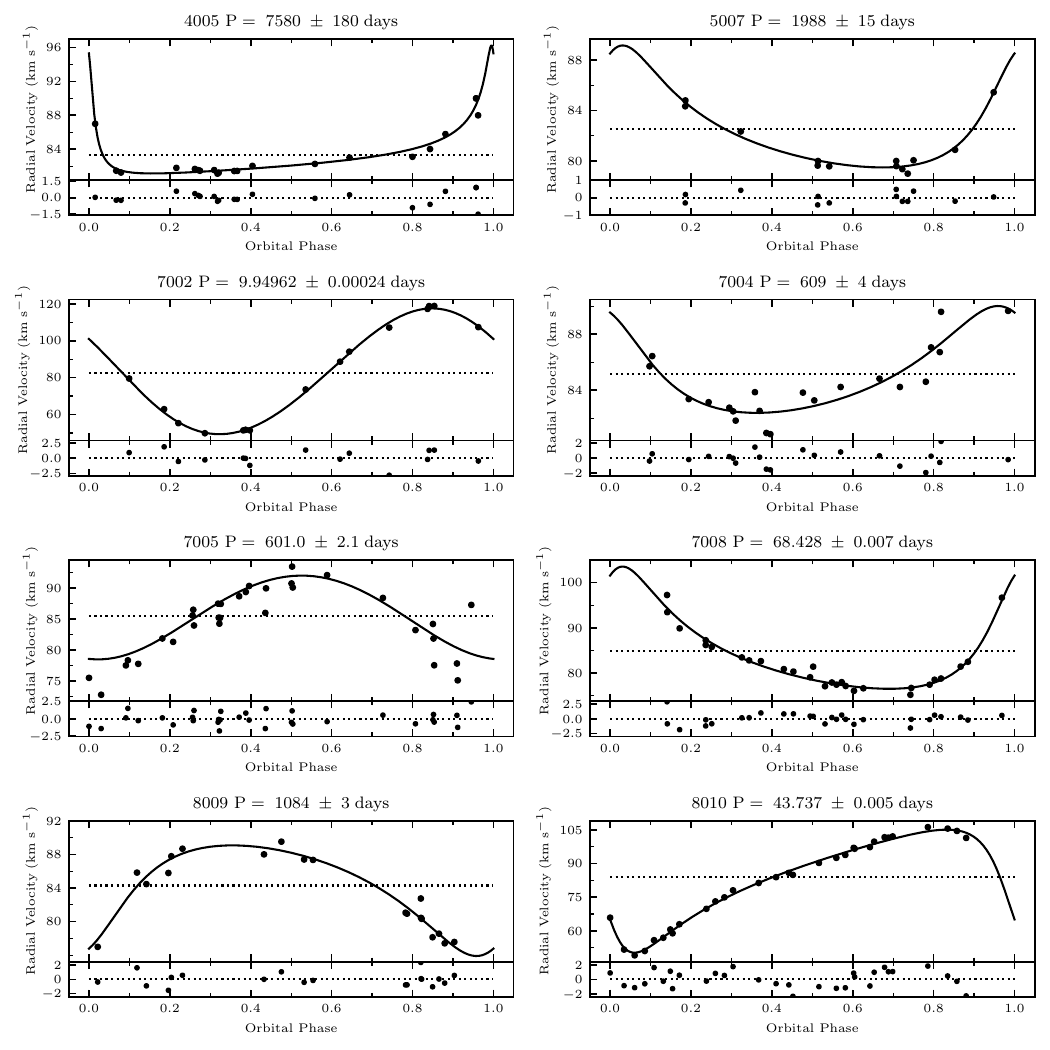}
    \caption{(Continued)}
    \label{fig:sb1fig2}
\end{figure*}
\begin{figure*}
    \addtocounter{figure}{-1}
    \centering
    \includegraphics{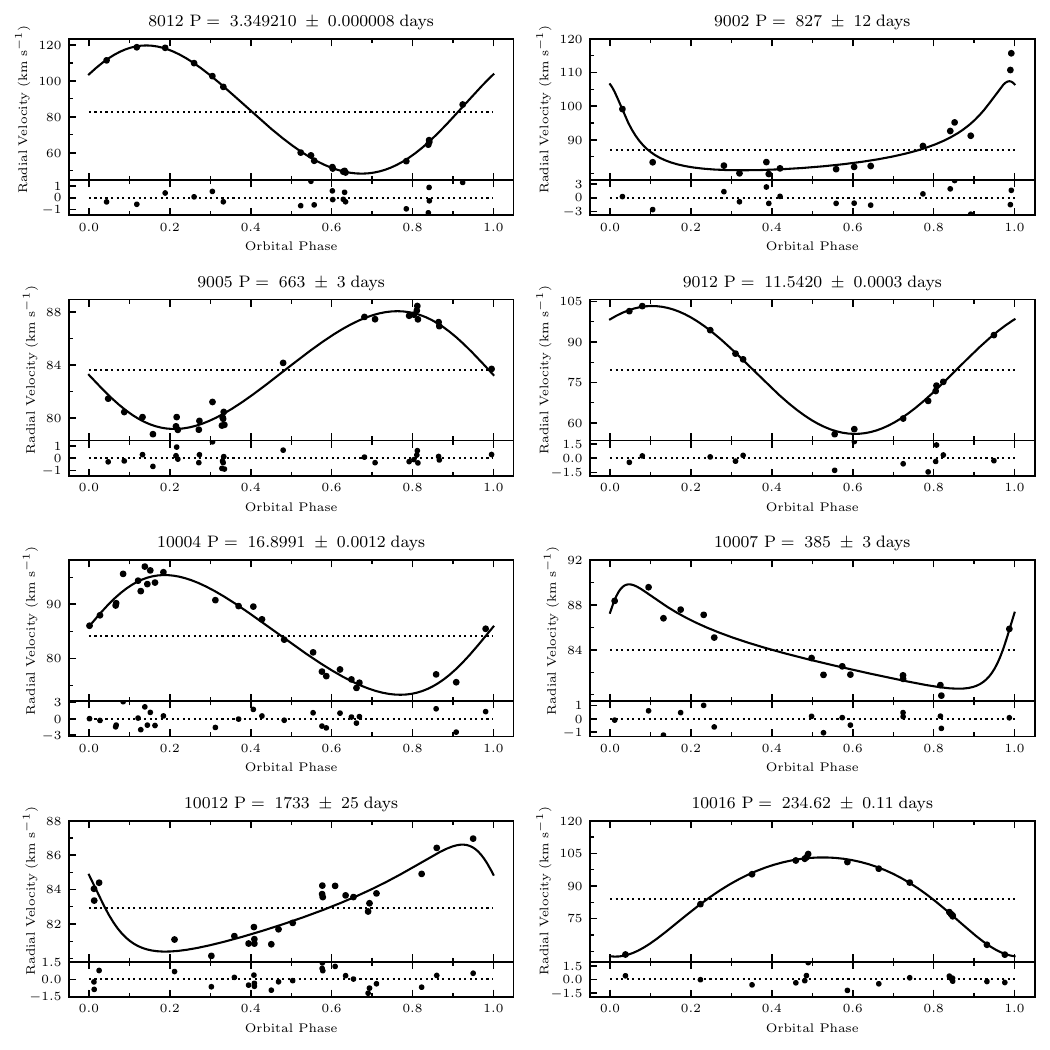}
    \caption{(Continued)}
    \label{fig:sb1fig3}
\end{figure*}
\begin{figure*}
    \addtocounter{figure}{-1}
    \centering
    \includegraphics{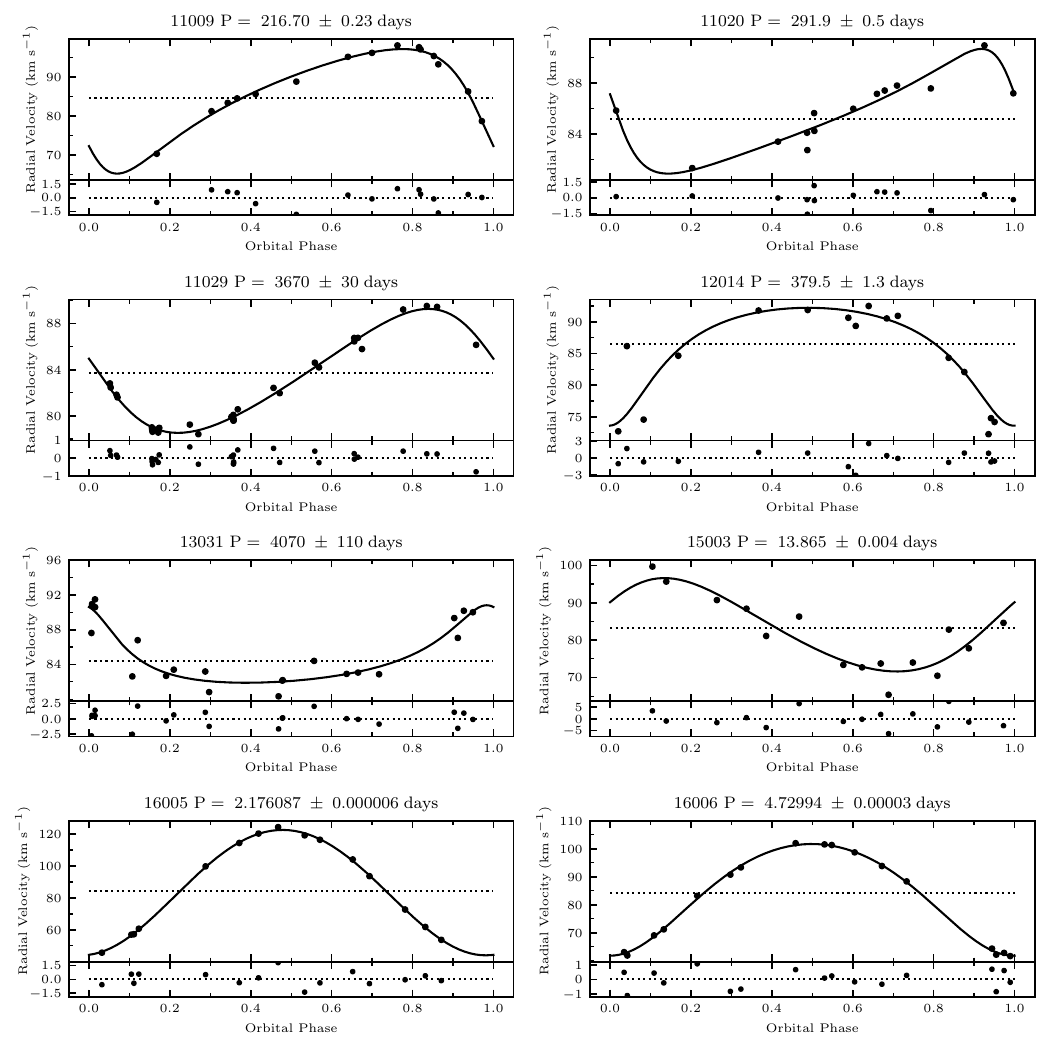}
    \caption{(Continued)}
    \label{fig:sb1fig4}
\end{figure*}
\begin{figure*}
    \addtocounter{figure}{-1}
    \centering
    \includegraphics{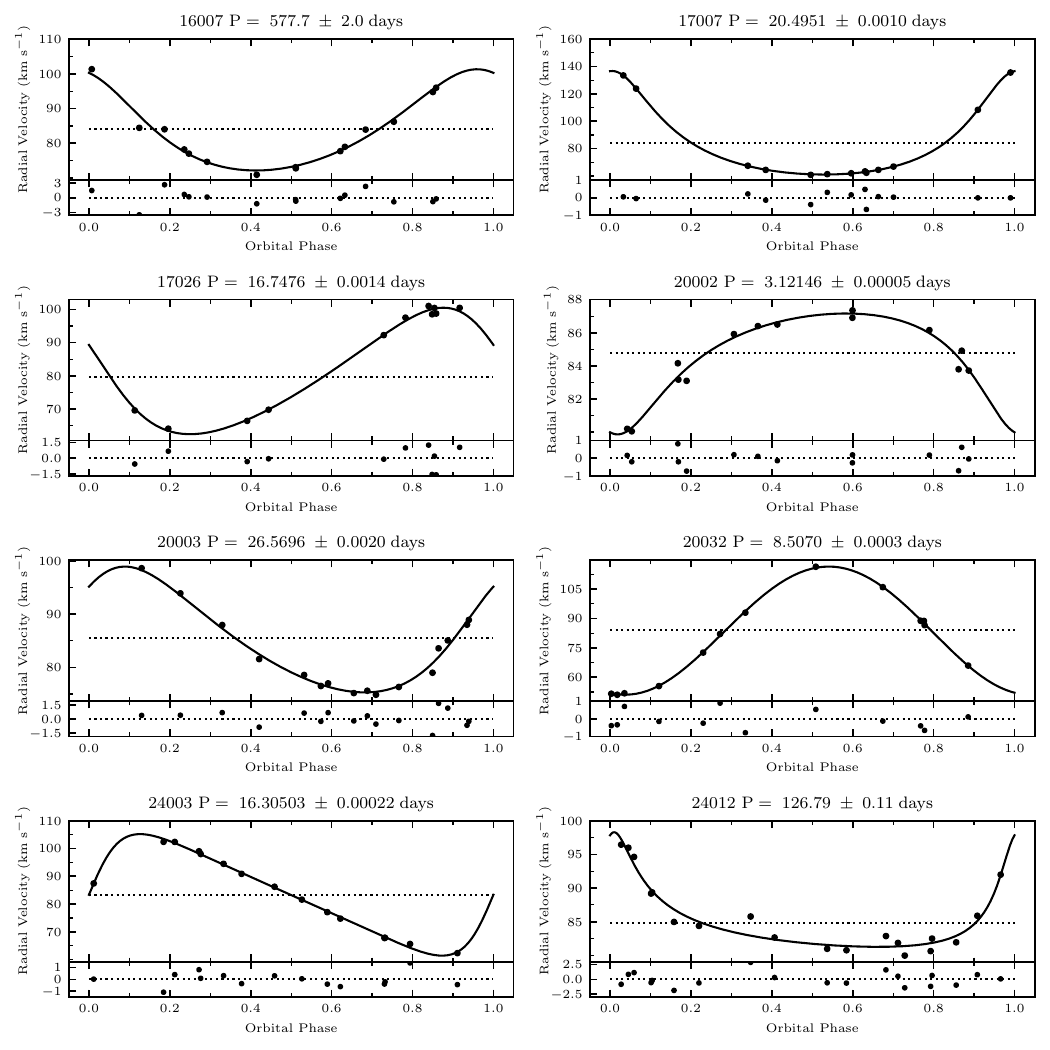}
    \caption{(Continued)}
    \label{fig:sb1fig5}
\end{figure*}

\subsection{Double-lined Binaries}\label{sb2_orbits}

Double-lined binaries (SB2s) are manually detected by identifying two distinct peaks in a CCF. Section \ref{RV_measurements} discusses our procedure for identifying and measuring two peaks. The orbital parameters for our SB2 solutions are in Table \ref{SB2tab}. For each binary, we present the primary orbital parameters and 1-$\sigma$ errors on the first two lines and the secondary orbital parameters and 1-$\sigma$ errors on the second set of two lines. We list the WOCS ID (ID), period (P), the number of cycles spanned by our observations, center-of-mass RV of the binary ($\gamma$), semi-amplitude RV of each component (K), eccentricity (e), longitude of periastron ($\omega$), the Julian date of periastron passage ($\mathrm{T_0}$), the projected semi-major axis of each component ($a \sin i$), the $m \sin^3 i$ of each component, the mass ratio ($q$), the rms of the RV residuals for both the primary and secondary ($\sigma$), and the number of RV measurements contributing to the solution. We show phase-folded orbital solutions and residuals for each binary in Figure \ref{fig:sb2fig}. 

\begin{figure*}
    \centering
    \includegraphics{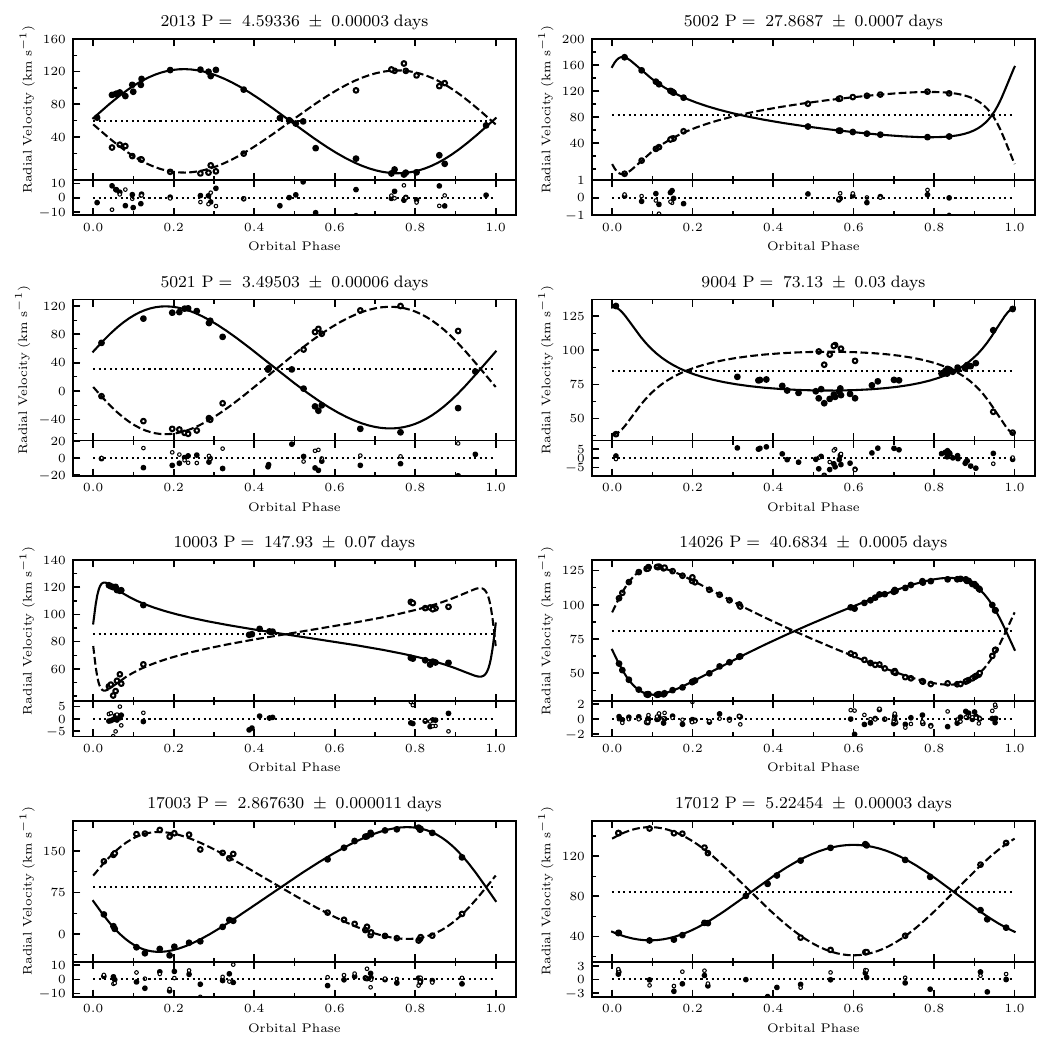}
    \caption{NGC 2506 SB2 orbit plots. For each binary listed in Table \ref{SB2tab}, we plot RV measurements against orbital phase. Primary data points are plotted with filled circles; secondary data points are plotted as open circles. The primary best fit orbit is plotted as a solid line, the secondary's orbit is plotted with a dashed line, and $\gamma$-velocity is marked as the dotted line. Phase-folded residuals are plotted below each orbit plot. Above each plot we give the star ID and orbital period.}
    \label{fig:sb2fig}
\end{figure*}
\begin{figure*}
    \addtocounter{figure}{-1}
    \centering
    \includegraphics{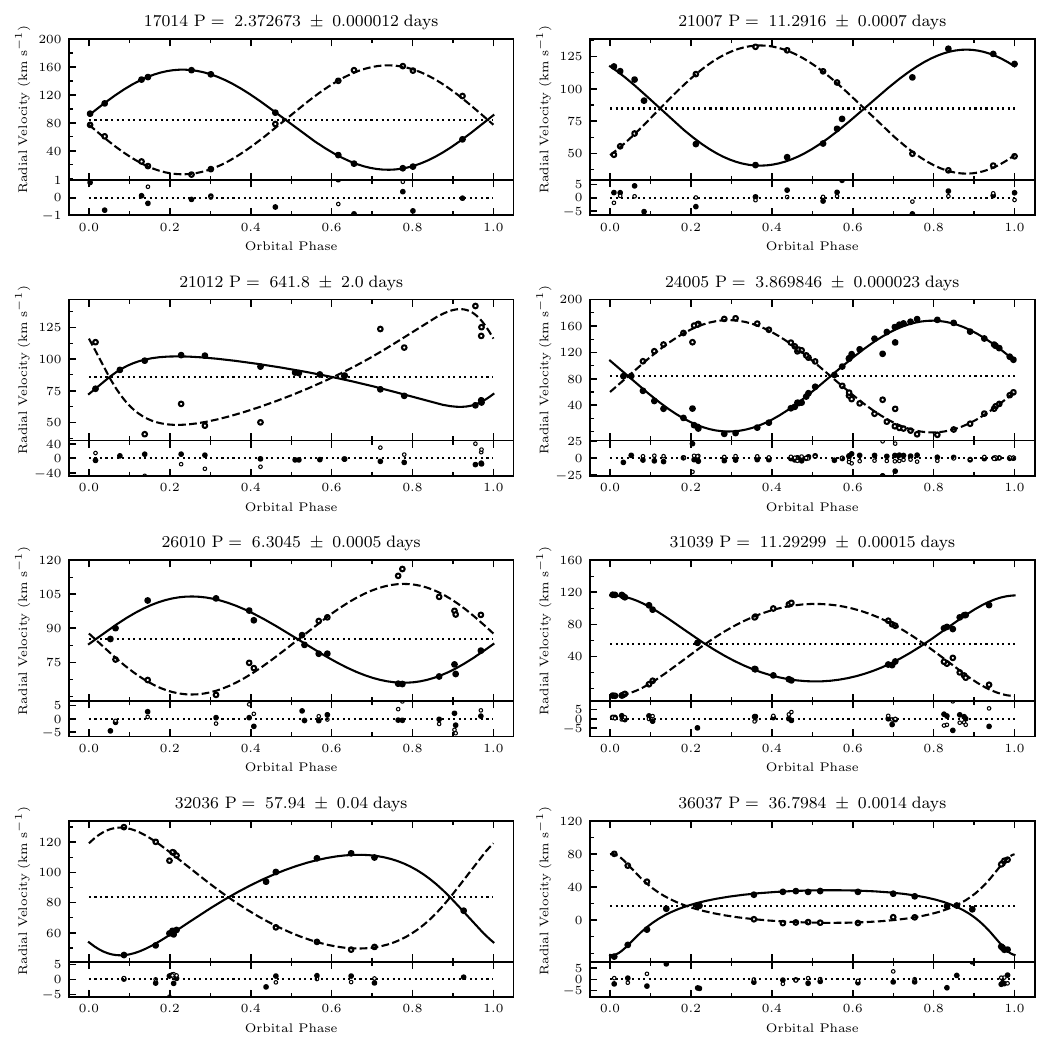}
    \caption{(Continued)}
    \label{fig:sb2fig2}
\end{figure*}

\begin{deluxetable*}{l r c r r r r r r r r c c}
\tabletypesize{\tiny}
\tablewidth{0pt}
\centering
\tablecaption{Orbital Parameters For NGC 2506 Double-Lined Binaries\label{SB2tab}}
\tablehead{\colhead{ID} & \colhead{$P$} & \colhead{Orbital} & \colhead{$\gamma$} & \colhead{$K$} & \colhead{$e$} & \colhead{$\omega$} & \colhead{$T_\circ$} & \colhead{a$\,\sin\,$i} & \colhead{m$\,\sin^3\,$i} & \colhead{q} & \colhead{$\sigma$} & \colhead{$N$} \\
\colhead{} & \colhead{(days)} & \colhead{Cycles} & \colhead{($\mathrm{km\; s^{-1}}$)} & \colhead{($\mathrm{km\; s^{-1}}$)} & \colhead{} & \colhead{(deg)} & \colhead{(HJD-2400000 d)} & \colhead{(10$^6$ km)} & \colhead{($M_\odot$)} & \colhead{} & \colhead{($\mathrm{km\; s^{-1}}$)} & \colhead{}}
\startdata
2013 & 4.59336 & 1371.1 & 59.1 & 63.8 & 0.049 & 273 & 56492.9 & 4.03 & 0.48 & 1.019 & 5.49 & 30\\
         &     $\pm$0.00003  &      &       $\pm$ 0.8 &       $\pm$1.5 &     $\pm$ 0.020 &       $\pm$ 22 &      $\pm$ 0.3 &       $\pm$ 0.10 &      $\pm$ 0.03 &     $\pm$ 0.035 &     &      \\
         &         &     &          &            62.6 &         &          &         &            3.95 &            0.48 &         &        5.28 &          20 \\
         &          &      &          &       $\pm$ 1.4&          &         &         &       $\pm$ 0.1 &      $\pm$ 0.03 &          &      &     \\
5002 & 27.8687 & 207.5 & 83.34 & 61.7 & 0.586 & 319.0 & 55915.210 & 19.17 & 1.52 & 0.9754 & 0.25 & 16\\
         &     $\pm$0.0007  &      &       $\pm$ 0.12 &       $\pm$0.4 &     $\pm$ 0.003 &       $\pm$ 0.3 &      $\pm$ 0.025 &       $\pm$ 0.1 &      $\pm$ 0.03 &     $\pm$ 0.0065 &     &      \\
         &         &     &          &            63.3 &         &          &         &            19.65 &            1.48 &         &        0.95 &          16 \\
         &          &      &          &       $\pm$ 0.6&          &         &         &       $\pm$ 0.17 &      $\pm$ 0.03 &          &      &     \\
5021 & 3.49503 & 1739.1 & 31.2 & 86 & 0.1 & 285 & 57598.62 & 4.12 & 0.99 & 0.959 & 10.19 & 21\\
         &     $\pm$0.00006  &      &       $\pm$ 1.6 &       $\pm$4 &     $\pm$ 0.03 &       $\pm$ 17 &      $\pm$ 0.16 &       $\pm$ 0.19 &      $\pm$ 0.1 &     $\pm$ 0.052 &     &      \\
         &         &     &          &            90. &         &          &         &            4.30 &            0.95 &         &        8.18 &          17 \\
         &          &      &          &       $\pm$ 3&          &         &         &       $\pm$ 0.16 &      $\pm$ 0.10 &          &      &     \\
9004 & 73.13 & 55.7 & 84.8 & 30.6 & 0.53 & 355 & 57479.2 & 26.0 & 0.51 & 1.013 & 4.09 & 41\\
         &     $\pm$0.03  &      &       $\pm$ 0.8 &       $\pm$1.5 &     $\pm$ 0.03 &       $\pm$ 5 &      $\pm$ 0.8 &       $\pm$ 1.3 &      $\pm$ 0.08 &     $\pm$ 0.081 &     &      \\
         &         &     &          &            30.2 &         &          &         &            25.7 &            0.52 &         &        4.71 &          10 \\
         &          &      &          &       $\pm$ 1.8&          &         &         &       $\pm$ 1.8 &      $\pm$ 0.07 &          &      &     \\
10003 & 147.93 & 27.1 & 85.4 & 35 & 0.76 & 277 & 57290.1 & 45.4 & 0.81 & 0.918 & 2.24 & 26\\
         &     $\pm$0.07  &      &       $\pm$ 0.6 &       $\pm$3 &     $\pm$ 0.04 &       $\pm$ 6 &      $\pm$ 0.5 &       $\pm$ 1.7 &      $\pm$ 0.09 &     $\pm$ 0.044 &     &      \\
         &         &     &          &            38 &         &          &         &            49.5 &            0.74 &         &        4.30 &          16 \\
         &          &      &          &       $\pm$ 4&          &         &         &       $\pm$ 2.3 &      $\pm$ 0.08 &          &      &     \\
14026 & 40.6834 & 100.1 & 80.76 & 42.73 & 0.3830 & 103.4 & 57117.49 & 22.08 & 1.048 & 0.9946 & 0.58 & 53\\
         &     $\pm$0.0005  &      &       $\pm$ 0.06 &       $\pm$0.10 &     $\pm$ 0.0020 &       $\pm$ 0.4 &      $\pm$ 0.03 &       $\pm$ 0.06 &      $\pm$ 0.007 &     $\pm$ 0.0039 &     &      \\
         &         &     &          &            42.96 &         &          &         &            22.20 &            1.042 &         &        0.73 &          53 \\
         &          &      &          &       $\pm$ 0.13&          &         &         &       $\pm$ 0.07 &      $\pm$ 0.006 &          &      &     \\
17003 & 2.867630 & 1285.4 & 84.2 & 112.8 & 0.188 & 100.6 & 55792.290 & 4.37 & 1.19 & 1.166 & 3.49 & 27\\
         &     $\pm$0.000011  &      &       $\pm$ 0.6 &       $\pm$0.9 &     $\pm$ 0.007 &       $\pm$ 2.5 &      $\pm$ 0.018 &       $\pm$ 0.04 &      $\pm$ 0.03 &     $\pm$ 0.017 &     &      \\
         &         &     &          &            96.7 &         &          &         &            3.74 &            1.39 &         &        4.86 &          27 \\
         &          &      &          &       $\pm$ 1.2&          &         &         &       $\pm$ 0.05 &      $\pm$ 0.03 &          &      &     \\
17012 & 5.22454 & 762.0 & 84.2 & 47.6 & 0.015 & 150 & 56638.6 & 3.42 & 0.430 & 0.745 & 1.87 & 19\\
         &     $\pm$0.00003  &      &       $\pm$ 0.3 &       $\pm$0.6 &     $\pm$ 0.009 &       $\pm$ 30 &      $\pm$ 0.5 &       $\pm$ 0.05 &      $\pm$ 0.011 &     $\pm$ 0.013 &     &      \\
         &         &     &          &            63.9 &         &          &         &            4.59 &            0.320 &         &        1.82 &          14 \\
         &          &      &          &       $\pm$ 0.6&          &         &         &       $\pm$ 0.05 &      $\pm$ 0.009 &          &      &     \\
17014 & 2.372673 & 573.2 & 84.43 & 71.7 & 0.013 & 275 & 56180.69 & 2.338 & 0.429 & 0.919 & 0.70 & 12\\
         &     $\pm$0.000012  &      &       $\pm$ 0.23 &       $\pm$0.3 &     $\pm$ 0.005 &       $\pm$ 19 &      $\pm$ 0.12 &       $\pm$ 0.011 &      $\pm$ 0.016 &     $\pm$ 0.017 &     &      \\
         &         &     &          &            78.0 &         &          &         &            2.54 &            0.394 &         &        3.02 &          12 \\
         &          &      &          &       $\pm$ 1.2&          &         &         &       $\pm$ 0.05 &      $\pm$ 0.008 &          &      &     \\
21007 & 11.2916 & 123.2 & 84.8 & 44.9 & 0.022 & 40 & 56097.3 & 7.0 & 0.52 & 0.904 & 4.12 & 14\\
         &     $\pm$0.0007  &      &       $\pm$ 0.4 &       $\pm$1.6 &     $\pm$ 0.012 &       $\pm$ 30 &      $\pm$ 0.8 &       $\pm$ 0.3 &      $\pm$ 0.03 &     $\pm$ 0.040 &     &      \\
         &         &     &          &            49.6 &         &          &         &            7.70 &            0.47 &         &        1.17 &          12 \\
         &          &      &          &       $\pm$ 0.6&          &         &         &       $\pm$ 0.10 &      $\pm$ 0.04 &          &      &     \\
21012 & 641.8 & 6.3 & 85.7 & 19.9 & 0.36 & 241 & 56943 & 164 & 11 & 0.436 & 1.81 & 16\\
         &     $\pm$2.0  &      &       $\pm$ 0.5 &       $\pm$0.9 &     $\pm$ 0.03 &       $\pm$ 7 &      $\pm$ 10. &       $\pm$ 9 &      $\pm$ 4 &     $\pm$ 0.063 &     &      \\
         &         &     &          &            46 &         &          &         &            380 &            4.7 &         &        13.22 &          10 \\
         &          &      &          &       $\pm$ 6&          &         &         &       $\pm$ 50 &      $\pm$ 1.1 &          &      &     \\
24005 & 3.869846 & 1435.4 & 83.9 & 84.0 & 0.006 & 70 & 57434.2 & 4.47 & 0.98 & 0.987 & 6.85 & 46\\
         &     $\pm$0.000023  &      &       $\pm$ 0.7 &       $\pm$1.5 &     $\pm$ 0.013 &       $\pm$ 130 &      $\pm$ 1.4 &       $\pm$ 0.09 &      $\pm$ 0.04 &     $\pm$ 0.027 &     &      \\
         &         &     &          &            85.2 &         &          &         &            4.53 &            0.96 &         &        6.95 &          43 \\
         &          &      &          &       $\pm$ 1.6&          &         &         &       $\pm$ 0.09 &      $\pm$ 0.04 &          &      &     \\
26010 & 6.3045 & 232.7 & 85.1 & 19.0 & 0.04 & 260 & 56178.6 & 1.64 & 0.030 & 0.780 & 2.26 & 16\\
         &     $\pm$0.0005  &      &       $\pm$ 0.5 &       $\pm$0.9 &     $\pm$ 0.04 &       $\pm$ 60 &      $\pm$ 1.1 &       $\pm$ 0.09 &      $\pm$ 0.005 &     $\pm$ 0.073 &     &      \\
         &         &     &          &            24.3 &         &          &         &            2.11 &            0.023 &         &        3.98 &          13 \\
         &          &      &          &       $\pm$ 1.6&          &         &         &       $\pm$ 0.16 &      $\pm$ 0.003 &          &      &     \\
31039 & 11.29299 & 491.5 & 55.6 & 53.6 & 0.130 & 358 & 57311.12 & 8.25 & 0.81 & 0.933 & 2.47 & 23\\
         &     $\pm$0.00015  &      &       $\pm$ 0.5 &       $\pm$0.7 &     $\pm$ 0.014 &       $\pm$ 5 &      $\pm$ 0.15 &       $\pm$ 0.13 &      $\pm$ 0.03 &     $\pm$ 0.022 &     &      \\
         &         &     &          &            57.4 &         &          &         &            8.84 &            0.75 &         &        3.25 &          22 \\
         &          &      &          &       $\pm$ 0.9&          &         &         &       $\pm$ 0.16 &      $\pm$ 0.03 &          &      &     \\
32036 & 57.94 & 14.5 & 83.5 & 33.2 & 0.207 & 138 & 59647.7 & 25.9 & 1.20 & 0.832 & 1.53 & 12\\
         &     $\pm$0.04  &      &       $\pm$ 0.4 &       $\pm$0.8 &     $\pm$ 0.024 &       $\pm$ 6 &      $\pm$ 0.8 &       $\pm$ 0.7 &      $\pm$ 0.10 &     $\pm$ 0.032 &     &      \\
         &         &     &          &            39.9 &         &          &         &            31.1 &            1.0 &         &        2.47 &          10 \\
         &          &      &          &       $\pm$ 1.2&          &         &         &       $\pm$ 1.1 &      $\pm$ 0.07 &          &      &     \\
36037 & 36.7984 & 109.1 & 17.0 & 39.2 & 0.515 & 173.6 & 57717.19 & 17.0 & 0.65 & 0.943 & 3.41 & 21\\
         &     $\pm$0.0014  &      &       $\pm$ 0.4 &       $\pm$1.0 &     $\pm$ 0.013 &       $\pm$ 2.1 &      $\pm$ 0.10 &       $\pm$ 0.5 &      $\pm$ 0.03 &     $\pm$ 0.030 &     &      \\
         &         &     &          &            41.6 &         &          &         &            18.03 &            0.61 &         &        1.60 &          15 \\
         &          &      &          &       $\pm$ 0.6&          &         &         &       $\pm$ 0.25 &      $\pm$ 0.04 &          &      &     \\
\enddata
\end{deluxetable*}

\section{Stars of Note}\label{starsofnote}

In the several subsections below, we discuss notable stars. We consider their binarity and orbits based on our findings in Sections \ref{Membership} and \ref{orbits}, MIST evolutionary track masses and characteristics from Section \ref{cluster_isochrone}, best-fit spun-up Synspec spectral templates \citep{hubenySynspecGeneralSpectrum2011} from Section \ref{RV_measurements}, TESS CDIPS Sector 7 light curves \citep{boumaClusterDifferenceImaging2019}, and other findings from the literature. After we completed our light curve analyses, the CDIPS team released light curves for Sector 34, which includes NGC 2506; these are not included in the analyses below. 

\subsection{Blue Stragglers}\label{bss_discussion}

Typically, BSSs are defined by their location on the CMD, blueward and often more luminous than the single-star and binary-star MSTOs. Their CMD locations suggest that they are stars more massive than the MSTO, and too massive to not yet have evolved at the cluster age. Three mechanisms for BSS formation are currently under discussion: collisions between stars during dynamical encounters of binary stars or in dense environments \citep{leonardStellarCollisionsGlobular1989,sillsDistributionCollisionallyInduced1999}; mergers of binary stars from angular momentum loss \citep{andronovMergersClosePrimordial2006, peretsTRIPLEORIGINBLUE2009}; and mass transfer between companions in multiple-star systems \citep{mccreaExtendedMainSequenceStellar1964,chenBlueStragglersPrimordial2008,boffinRochelobeFillingFactor2014}.

\begin{figure*}
    \centering
    \includegraphics{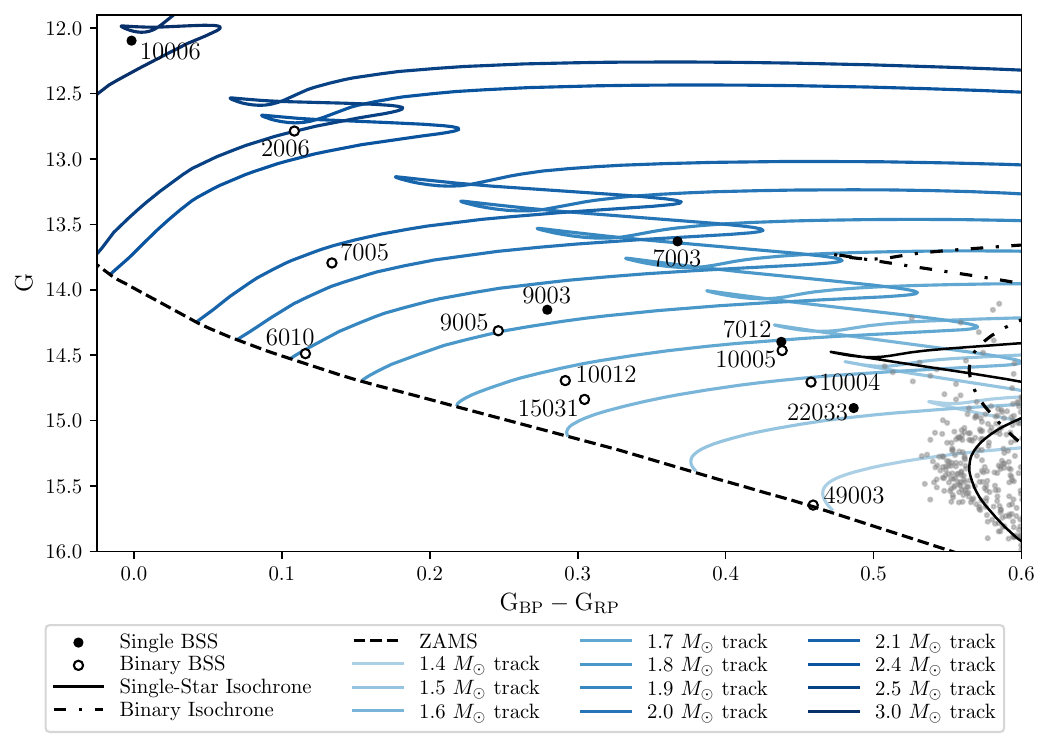}
    \caption{The blue stragglers of NGC 2506. Single BSSs are plotted as black circles; binary BSSs are plotted as black rings; other member stars as plotted as grey dots. MIST evolutionary tracks of relevant masses are plotted in varying shades of blue. We plot the single-star
isochrone (solid line), equal-mass binary isochrone (dash-dot
line), and zero-age main-sequence MIST isochrone (dashed line)
for reference.
    }
    \label{fig:BSS_cmd}
\end{figure*}

Based on the proper-motion and RV membership probabilities and the single-star isochrone found in this work, NGC 2506 contains 14 BSSs. Figure \ref{fig:BSS_cmd} shows their location in the CMD, along with MIST evolutionary tracks. These stars all lie blueward of the theoretical location of the blue hook of the isochrone and, perhaps as importantly, are bluer than the main sequence stars on the red side of the blue hook by at least 0.07 magnitude.  

In comparison to the 14 BSSs identified in this study, \cite{ahumadaNewCatalogueBlue2007} found 15 BSSs in NGC 2506, of which we confirm 4 to be cluster members. Using Gaia DR2 and PARSEC isochrones, \cite{rainNewGaiaBased2021} found 14 BSSs in NGC 2506. The reddest two BSSs in this paper (WOCS 22033 and 10004) are not included in their list of BSSs, although they are considered cluster members based on their source list from \cite{cantat-gaudinClustersMiragesCataloguing2020}. On the other hand, we find one of their BSSs to be a RV non-member (WOCS 5022) and another BSS (WOCS 27005) to have Gaia BP measurements that differ by 0.12 mag between Gaia DR2 and Gaia DR3, making this star a main-sequence star in our analysis. 

\cite{panthiUOCSVIIIUV2022} identified nine BSS after fitting a PARSEC isochrone to 2,175 Gaia EDR3 member stars identified using ML-MOC \citep{Agarwal+2021}---an algorithm that rejects probable field stars with a k-nearest neighbors algorithm before fitting remaining stars with a GMM to  identify members---on position, proper motion, and parallax. We also find all nine of these stars to be BSS.   

\cite{vaidyaBlueStragglerPopulations2020} found 28 BSSs candidates using PARSEC isochrones. The authors do not provide a list of their BSSs; comparing by eye their published CMD to ours, all of our BSSs are included in the 28 of \cite{vaidyaBlueStragglerPopulations2020}. The difference appears to be primarily due to choice of isochrone, as the PARSEC isochrone has a blue hook that is significantly smaller (and thus redder) than the MIST isochrone. Finally, one BSS in \cite{vaidyaBlueStragglerPopulations2020} appears to be WOCS 27005, which we do not find to be a BSS in this work. 

Similarly, several member stars previously identified as BSSs in the literature on NGC 2506 \citep[e.g.,][] {mcclureOldOpenCluster1981, ahumadaCatalogueBlueStragglers1995, arentoftOscillatingBlueStragglers2007} fall within the single-star blue hook of this and other recent works \citep[e.g.,][] {knudstrupExtremelyPreciseAge2020}. These stars will be addressed in subsection \ref{otherstars} if they are RV members of the cluster. 

Based on the MIST evolutionary tracks, the BSSs of NGC 2506 have masses between 1.4 $M_\odot$ and $\sim3.0 M_\odot$. As the MSTO is $1.5 M_\odot$, the upper range in masses would suggest that these BSSs formed through different mechanisms. 

At least 9 of these 14 BSSs (64 $\pm$ 21\%) are in binary systems with periods less than $10^4$ days, with a tenth star in a possible very-long-period binary. This high BSS binary frequency is consistent with frequencies of 80\% found in NGC 188 and M67 \citep{mathieuBlueStragglersOld2015,gellerLARRADIALVELOCITIES2015}.

Below, we discuss what is known about each BSS based on our orbital solutions and other results presented earlier and in the literature.  

{In order of brightness:}

\textit{WOCS 10006}---This star is the brightest and bluest BSS of NGC 2506. Due to the star's effective temperature ($T_{\mathrm{eff}}$), our solar template did not produce good CCF peaks from our spectra. We instead used a spun-up template of $T_{\mathrm{eff}}$ = 9750 K, log g = 3.0, and $v\sin i = 20\mathrm{\;km\;s^{-1}}$ for cross correlation, as it produced the highest CCF peak height. The star has a RV of 84.87 $\mathrm{\pm\; 0.34 \;km\;s^{-1}}$, peak-to-peak RV difference of 4.26 $\mathrm{km\;s^{-1}}$ and e/i = 2.0, which is below our threshold for velocity variability. The evolutionary track mass of the star is 3.0-3.2 $M_\odot$. Notably, this mass is at least twice the mass of NGC 2506's MSTO mass of 1.5 $M_\odot$.

\textit{WOCS 2006}---This SB1 is the second brightest BSS of NGC 2506. The spectra have very broad lines, indicative of rapid rotation. A deep 6563 \r{A} $\mathrm{H\alpha}$ absorption feature seen in spectra available on the European Southern Observatory archive (taken by F. Grundahl for \cite{knudstrupExtremelyPreciseAge2020}) is suggestive of an early A-type star. The evolutionary track mass is between 2.4 and 2.55 $M_\odot$ with $T_{\mathrm{eff}}$ of approximately 9,150 K, in good agreement with this spectral feature. 

Our spectral analyses are somewhat of a mystery. Surprisingly, a spun-up $T_{\mathrm{eff}}$ = 5,500 K, log g = 5.0 template produced significantly higher cross-correlation peaks than a template for $T_{\mathrm{eff}}$ = 9,250 K, log g = 4.0, as appropriate for an A-type star. A $v\sin i = 140 \mathrm{\;km\;s^{-1}}$ template produced the highest correlation peaks. We fit every CCF manually, but as we expect our errors to be at least 10 $\mathrm{km\;s^{-1}}$, we do not report our measured RVs. Even so, we do find this star to be velocity variable, with a peak-to-peak RV difference of 44.5 $\mathrm{km\;s^{-1}}$. The measured RVs overlap the cluster mean RV. Hence we classify this star as a BLM. Further observations are required to confirm RV variability, when possible derive an orbital solution, and determine this star's membership probability. 

\textit{WOCS 7003}---This narrow-lined star is the oscillating BSS V1 from \cite{kimThreeDeltaScuti2000}, the brightest and bluest $\delta$ Scuti of NGC 2506. \cite{panthiUOCSVIIIUV2022} found this star, which they label BSS1, to have a $T_{\mathrm{eff}}$ = 7,750 K, log g = 3.5 primary with an extremely low-mass (ELM) WD companion of 0.17 $M_\odot$. Our evolutionary tracks indicate the BSS is between 1.85 and 2.0 $M_\odot$, and has a $T_{\mathrm{eff}}$ of 7,750 K and log g = 3.75, in good agreement with their findings. 
\cite{panthiUOCSVIIIUV2022} state that the ELM companion is indicative of Case A/Case B mass transfer for formation of the BSS.

The compilation in Figure 1 of \cite{khuranaDynamicallyFormingExtremely2023} shows the vast majority of known ELM WDs to be in binaries with periods of less than approximately 1 day and in double-degenerate systems, toward which search criteria may have been biased. ELM candidates with main sequence or red giant companions have been found up to orbital periods of 30 days. One field ELM WD with a main sequence companion also has been found in a 455-day orbit \citep{masudaSelflensingDiscoveryWhite2019}; \cite{khuranaDynamicallyFormingExtremely2023} suggest such longer-period systems may originate from dynamical exchange in clusters (and subsequent ejection) rather than from mass transfer.

We have collected 31 RV measurements of WOCS 7003 over 14 years and found them to have a 2.7 $\mathrm{km\;s^{-1}}$ peak-to-peak range in RV and a standard deviation of 0.71 $\mathrm{km\;s^{-1}}$. These small RV variations are below our threshold for binary velocity variation (e/i = 1.4). This system would need to be close to face-on ($i$ $<$ $3.6^{\circ}$ for a primary mass of 2.0 $M_\odot$) for the peak-to-peak RV variation range to permit a 0.17 $M_\odot$ companion with a period below 1 day. More broadly, the inclination limit becomes $11.2^{\circ}$ at a period of 30 days. Alternatively, a dynamically undetected ELM companion could have a very long orbital period, albeit without a proposed formation mechanism that also would produce a BSS as the primary star.

Further, this star exhibits $\delta$ Scuti photometric pulsations. The dispersion of our RV measurements are characteristic of such stars. If much of the RV variation is due to radial pulsations, this would further limit the inclination angle of any possible binary orbit.

\textit{WOCS 7005}---This SB1 is among the bluest BSSs in the cluster. Using SED fitting, \cite{panthiUOCSVIIIUV2022} found this star (their BSS3) to have a $T_{\mathrm{eff}}$ = 9,250 K, log g = 4.0 BSS. Our evolutionary tracks support this finding, with the closest model having $T\mathrm{_{eff}\;=\; 9,230\; K,\; log\; g \;=\; 4.1,\;}$ and $\mathrm{M\; =\; 2.05\; M_\odot}$. \cite{panthiUOCSVIIIUV2022} find no UV excess indicative of a hot companion. 

The template that produced the highest peak CCF had $T_{\mathrm{eff}}$ = 7,000 K, log g = 4.0 star with a $v\sin i = 45\mathrm{\;km\;s^{-1}}$. Using this template, our best-fit orbital solution is a $P = 596.7 \pm 2.1$ days, circular ($e = 0.04 \pm 0.04$) orbit. Using a primary mass of 2.05 $M_\odot$, the orbital mass function yields a minimum mass of the companion of $0.50 M_\odot$. 

\textit{WOCS 9003}---This BSS is a narrow-lined star with e/i = 1.4, well below our criterion for velocity variability. This star has an evolutionary track mass of 1.85 $M_\odot$, $T_{\mathrm{eff}}$ of 8,225 K, and log g of 4.0, which aligns well with our best-fit spectral template of $T_{\mathrm{eff}}$ = 8,000 K, log g = 4.0 and low $v \sin i$. \cite{panthiUOCSVIIIUV2022} found this star (their BSS2) to be an $T_{\mathrm{eff}}$ = 8,000 K, log g = 3.0 primary star with an ELM WD (0.18 $M_\odot$) companion from SED fitting and WD cooling curves. Like with WOCS 7003, for the star's RV peak-to-peak range of 1.98 $\mathrm{km\;s^{-1}}$ to be consistent with a 0.18 $M_\odot$ companion with a period below 1 day, the binary star's orbit again would need to be nearly face-on ($i$ $<$ $2.37^{\circ}$). More generally, the inclination limit becomes $7.4^{\circ}$ at a period of 30 days. 

\textit{WOCS 9005}---This SB1 has a circular ($e = 0.08 \pm 0.04$), 663-day orbit and $v\sin i= 12.8 \mathrm{\;km\;s^{-1}}$. With an evolutionary track mass for the primary of 1.80 $M_\odot$, the mass function gives a minimum secondary mass of 0.29 $M_\odot$, which would permit a WD companion. GALEX \citep{morrisseyCalibrationDataProducts2007} did not detect this star as UV bright, so if the companion is a WD, it would have cooled below detectable levels. 

\textit{WOCS 7012}---This narrow-lined BSS is 0.03 mag blueward of the theoretical blue hook of the isochrone. We have 3 RV measurements from 2009-2010 that are 2-3 $\mathrm{km\;s^{-1}}$ higher than our 8 measurements since 2020, possibly indicating that the star is a very long-period binary. The star has an evolutionary track mass between 1.56 and 1.69 $M_\odot$. 

\textit{WOCS 10005}---This narrow-lined SB1 is the oscillating BSS V2 from \cite{kimThreeDeltaScuti2000}, the $\delta$ Scuti in NGC 2506 with the highest amplitude photometric oscillations (0.12 in B mag \citep{arentoftOscillatingBlueStragglers2007}). The star is just blueward of the theoretical blue hook of the isochrone (0.03 mag) and has an evolutionary track mass between 1.55 and 1.67 $M_\odot$. Most of our RV measurements are consistent with $\delta$ Scuti RV variations of several $\mathrm{km\;s^{-1}}$ and show no obvious periodicity or a Keplerian orbit. However, there are 6 measurements from 3 observing runs that are much higher (10-40 $\mathrm{km\;s^{-1}}$) than the others and suggest orbital motion with high eccentricity. Our best orbital solution has a 61-day orbit with e = 0.8. Periastron is unconstrained, so there are high errors on the semi-major amplitude, and as such, we do not consider this system solved. Additionally, we found possible solutions at periods of 19.9 days and 31.8 days with somewhat higher residuals; while less likely to be the correct orbital solution, they cannot be ruled out. Further observations at periastron are required to confirm a tentative 61-day orbit.

We note that the star's proper-motion membership from GMM is low (${<10}$\%), but its proper-motion membership from HDBSCAN (100\%), RV membership (89\%), low-precision parallax, and central location in the cluster suggest that it is a member. Additionally, that the star is a $\delta$ Scuti and is in the instability strip at the distance of NGC 2506 also suggest it to be a cluster member.

\textit{WOCS 6010}---This SB1 is one of the two BSSs closest to the zero-age main sequence (ZAMS). This star is rapidly rotating with a $v\sin i= 56 \mathrm{\;km\;s^{-1}}$ and has e/i = 6.29. Evolutionary tracks suggest this star has a mass of 1.90 $M_\odot$, $T_{\mathrm{eff}}$ of 9,500 K, and log g of 4.4. This star is also BSS7 ($T_{\mathrm{eff}}$ = 9,750 K, log g = 4.5) of \cite{panthiUOCSVIIIUV2022}, who did not detect a hot companion. Their temperature and log g match our evolutionary track results well. 

We have 10 RV measurements of this star and see weak lines indicative of an A-type star. The Synspec template that fit best was $T_{\mathrm{eff}}$ = 8,000 K, log g = 1.5, and  $v\sin i= 45 \mathrm{\;km\;s^{-1}}$, which we use for the RV measurements published here. We note that the temperature and especially log g do not agree well with the stellar characteristics derived from photometry, similarly to WOCS 2006.  

\textit{WOCS 10012}---This narrow-lined SB1 has an evolutionary track mass of 1.65 $M_\odot$. We find it to have a 1733-day, e = 0.41 orbit. Additionally, the solution has cyclic residuals indicative of being a triple system. Given the evolutionary track primary mass, the orbital solution gives a minimum secondary mass of 0.24 $M_\odot$, which, if edge on, marginally permits the WD mass of 0.20 $M_\odot$ put forward by \cite{panthiUOCSVIIIUV2022} for this star (BSS8; $T_{\mathrm{eff}}$ = 8,250 K, log g = 3.0). 

\textit{WOCS 10004}---This BSS is a narrow-lined SB1 with a 16.9-day period and eccentricity of 0.13. WOCS 10004 is variable star V3, discovered to be both a $\delta$ Scuti star by \cite{kimThreeDeltaScuti2000} and a $\gamma$ Doradus candidate by \cite{arentoftOscillatingBlueStragglers2007} due to low- and high-frequency oscillation modes, respectively. Both papers found that the star pulsates approximately 12.265 times per day (once every 1.95 hours). Unfortunately, our usual integration time of 1 hour is almost half of one oscillation period, so our RV measurements will have increased scatter depending in which half of the oscillation period is observed. Also, the true peak-to-peak RV range of the radial pulsations is likely slightly greater, as our measurements integrate over such a large fraction of the period. Our orbital solution has the largest residuals occurring near periastron, which we suggest is due to radial pulsations. Our evolutionary track model suggests this star has a mass of 1.58 $M_\odot$, $T_{\mathrm{eff}}$ of 7,400 K and log g of 4.0. The orbital mass function thus gives a minimum secondary mass of 0.19 $M_\odot$. If this star is confirmed to be a $\gamma$ Doradus, asteroseismology techniques \citep{sanchezariasAsteroseismologyHybridScuti2017} may enable directly studying stellar structure of a post-interaction star.

\textit{WOCS 15031}---This BSS is an SB1 rotating with a $v\sin i= 15 \mathrm{\;km\;s^{-1}}$. With only 10 observations, we were not able to find an orbital solution, although inspection of the RV data suggests it may be long period (at least several thousand days) and possibly eccentric. The star has an evolutionary track mass of 1.63 $M_\odot$, $T_{\mathrm{eff}}$ = 8,175 K and log g = 4.2. \cite{panthiUOCSVIIIUV2022} found this star (BSS9) to have $T_{\mathrm{eff}}$ = 8,000 K and log g = 3.5. They did not find any UV excess indicative of a hot companion. 
 
\textit{WOCS 22033}---This BSS is the reddest BSS in the cluster, tucked below the location of the theoretical blue hook with an evolutionary track mass of 1.52 $M_\odot$. It has e/i = 1.96, so is below our velocity-variable threshold. The spectrum shows some rotation with $v\sin i = 21.4 \mathrm{\;km\;s^{-1}}$.  

\textit{WOCS 49003}---This SB1 is the faintest BSS of NGC 2506. It is on the ZAMS and has an evolutionary track mass of 1.41 $M_\odot$, about 0.1 $M_\odot$ below the turn-off mass. The star is rotating with $v\sin i= 51 \mathrm{\;km\;s^{-1}}$. The maximum CCF heights were obtained with a template of $T_{\mathrm{eff}}$ = 6,750 K, log g = 4.0, and $v\sin i= 40 \mathrm{\;km\;s^{-1}}$. Our 8 observations have a peak-to-peak difference of 13.5 $\mathrm{km\;s^{-1}}$ and so confirm velocity variability. Additional observations are needed to determine the orbital solution. Further, although not explicitly analyzed in \cite{panthiUOCSVIIIUV2022}, this star appears to be more UV-bright than nearby main-sequence stars in Figure 2 of \cite{panthiUOCSVIIIUV2022}. This may warrant further investigation to determine if it has a hot WD companion.  

\hfill \break
NGC 2506 has a BSS binary frequency of 64 $\pm$ 21\% (9/14); at least one additional BSS (7012) may be in a very long-period binary. This BSS binary frequency is similar to those found in NGC 188 and M67 of $\sim$ 80\% \citep{mathieuBlueStragglersOld2015,gellerLARRADIALVELOCITIES2015}.
Further, the BSS period distribution is similar to that of NGC 188, with most periods ranging from 600 days and longer, but a few—such as that of WOCS 10004—at much shorter periods. 

Among the nine velocity-variable BSSs in NGC 2506, four have measured orbital eccentricities. Both WOCS 7005 and 9005 have circular orbits at $\sim600$ days, suggestive of tidal effects from a prior evolved companion, even though neither shows UV excess from a current hot WD companion. The minimum secondary mass of 0.5 $M_\odot$ for WOCS 7005 strongly suggests an AGB progenitor if the companion is a cool WD. The rapid rotation of the BSS could also indicate prior mass transfer from an evolved companion. Two other stars, WOCS 10004 and 10012, show moderate eccentricities of 0.1 and 0.4, respectively, but at very different periods (17 days and 1,733 days, respectively), suggestive of different formation mechanisms. However, the connection of orbital eccentricity with formation mechanism remains uncertain, especially for mass transfer origins.

Both WOCS 6010 and 49003 lie on the ZAMS, indicative of having hydrogen cores without helium enrichment. This may suggest that both formed through mass transfer onto progenitor secondaries that had not yet evolved off the ZAMS (M $<$ 1.25 $M_\odot$). These two BSSs have rather different masses of 1.9 and 1.4 $M_\odot$, perhaps indicative of different progenitor secondary masses, different mass transfer efficiencies, or both. Their rapid rotations are suggestive of recent formation \citep{leinerObservationsSpindownPostmasstransfer2018} and thus insufficient time to evolve off the ZAMS, although neither show ultraviolet excess from a hot WD.

WOCS 49003 is also notable for having a mass 0.1 $M_\odot$ below the MSTO mass, which would have placed it on the main-sequence as a blue lurker were it not so blue. WOCS 22033 has a mass of 1.5 $M_\odot$, much the same as the MSTO turnoff mass, but the different locations of these two stars on the CMD indicate different core helium abundances. In a mass transfer scenario, the progenitor of WOCS 22033 may have had more helium in its core at formation than the progenitor of WOCS 49003, or WOCS 22033 may have spent a longer amount of time as a BSS. Both are rapidly rotating, but only WOCS 49033 is velocity variable. 

The two most luminous BSSs in the cluster, WOCS 10006, and WOCS 2006, are notable for their masses from evolutionary tracks. WOCS 10006 is twice the MSTO mass and a non-velocity variable star, which would require a multi-star dynamical encounter or perhaps a merger to form. WOCS 2006 is $\sim1\;M_\odot$ more massive than the MSTO mass. Notably, it is spinning with a $v \sin i > 120\mathrm{\;km\;s^{-1}}$, and likely a binary. This would require very high mass-transfer efficiency or again, a merger or collision.  

Altogether, the BSSs of NGC 2506 have physical characteristics that suggest they represent many different evolutionary histories and likely formation mechanisms, from collisions or mergers to mass-transfer at many stages of stellar evolution.

\subsection{Yellow Stragglers and Candidates}\label{yss}

YSSs are stars brighter than the MSTO and sub-giant branch, redder than the BSSs, and bluer than the RGB. Like with BSSs, these stars appear to be too massive to exist at the age of the cluster, and have been suggested to be evolved BSSs \citep{renziniTestsEvolutionarySequences1988, mathieuOrbits22Spectroscopic1990, leinerK2M67STUDY2016}. Relatedly, asteroseismic studies have found over-massive stars in the red clumps of some clusters \citep[e.g., in NGC 6819][]{corsaroASTEROSEISMOLOGYOPENCLUSTERS2012,handbergNGC6819Testing2017}, suggesting these to also be evolved BSSs.

\begin{figure*}
    \centering
    \includegraphics{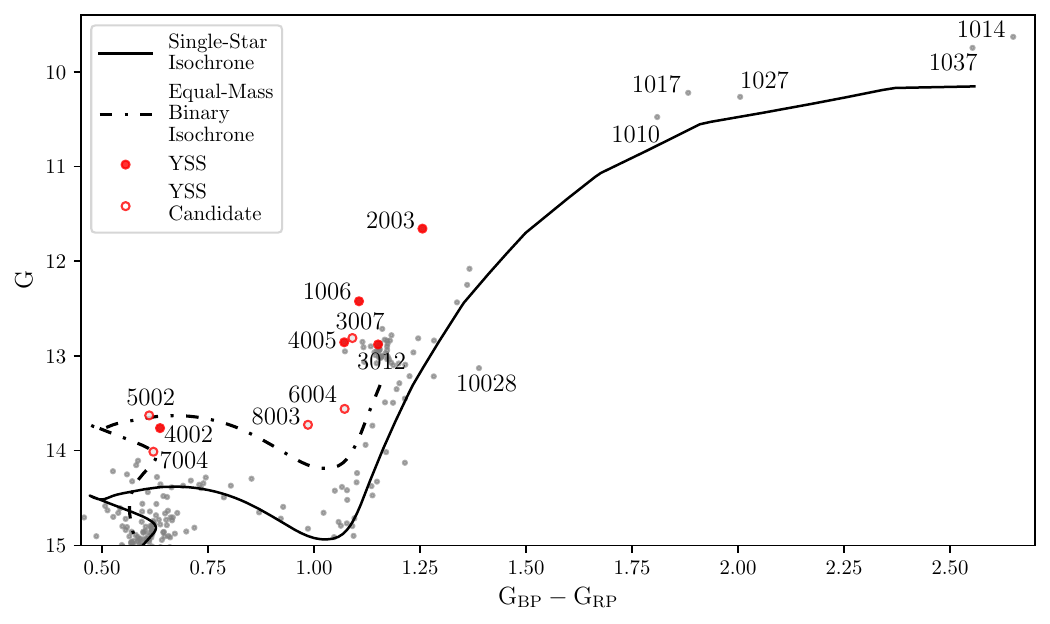}
         \caption{The YSSs and candidates of NGC 2506. YSSs are plotted as red dots. Stars that do not lie upon the isochrone but have possible alternative explanations to being YSSs are considered YSS candidates and plotted as red rings. All other stars of the cluster are plotted in grey, and names are listed if referenced elsewhere in the text. We also plot the single star isochrone (solid line) through the He flash and an equal-mass binary isochrone (dash-dot line) through the base of the RGB.
    }
    \label{fig:yss_cmd}
\end{figure*}

Figure \ref{fig:yss_cmd} shows the CMD location of YSSs and YSS candidates in NGC 2506. As discussed in Section \ref{cluster_isochrone}, MIST isochrones may not fit the brightest giants well \citep[e.g., in M67; ][]{choiStarClusterAges2018}. Thus, we do not take the location of the brightest giants off of the best-fit isochrone to indicate their status as YSS candidates. We note that WOCS 1014, 1037, 1027, and 1010 have low e/i and no indication of being binaries. WOCS 1017 is a binary star but shows no evidence of being post-mass transfer (see Section \ref{otherstars}). None of these stars have measurable projected rotation velocities.

At lower luminosities where the MIST isochrone fits most giants well, there are also stars that lie off the isochrone. Here we discuss these in the context of being possible YSSs. We first present the stars that we do consider to be YSSs in order of brightness, and then examine the remaining YSS candidates. 

\textit{WOCS 2003}---This star is perhaps the most striking YSS photometrically, being far removed from both the giants and the red clump. It is a narrow-lined giant star with e/i = 2.0, below our threshold of velocity variability. Even so, RV measurements over the last 6 years show a continuous downward trend in RV of $\mathrm{2\;km\;s^{-1}}$, suggesting a long-period binary. Strictly, the photometry of the star can be reproduced with a pairing of the brightest red clump star and a giant around G = 12, which perhaps could be the result of a near-twin binary. This pairing would have a flux ratio of roughly 50\%. If this is a long-period binary, both RVs may be close enough to preclude detection as an SB2.

\textit{WOCS 1006}---This narrow-lined giant star has an e/i = 6.0, which is driven by a single RV measurement of 77 $\mathrm{km\;s^{-1}}$; the other 8 measurements give an e/i of 1.3. We inspected each spectrum and CCF for this star and the one outlier RV measurement appears of similar quality to the other measurements. This star may be a highly eccentric, long-period binary. Strictly, the photometry of the star can be reproduced with a pairing of the bluest red clump star and a giant around G = 13.6, which again would require a near-twin binary. Such a combination of stars could show signs of being an SB2 at periapsis as the fainter star would have roughly 50\% of the flux of the primary. However, the CCF does not show any evidence of this star being an SB2. 

\textit{WOCS 4005}---This SB1 is RC1 of \cite{panthiUOCSVIIIUV2022}, a red clump star with UV excess. We find this star to have a highly eccentric (e = 0.82), 7580-day orbit and to have a $v \sin i < 10 \mathrm{\;km\;s^{-1}}$. Adopting a red clump mass of 1.6 $M_\odot$ from evolutionary tracks, the minimum mass of the secondary star is 0.7 $M_\odot$ from the mass function, which marginally permits the WD mass of 0.6 $M_\odot$ from SED fitting by \cite{panthiUOCSVIIIUV2022}. 

This star is also listed as V10 in \cite{arentoftOscillatingBlueStragglers2007}, a star the authors note as having $\delta$ Scuti-like mmag-level variability at $\mathrm{10\;day^{-1}}$. Our RV measurements do not show RV pulsations found in $\delta$ Scuti stars. The \cite{arentoftOscillatingBlueStragglers2007} photometric measurements were obtained between January and April of 2005, which was 6-months after the prior periastron passage on July 20, 2004. The TESS Sector 7 CDIPS \citep{boumaClusterDifferenceImaging2019} light curve of this star does not show periodic mmag-level variability at cycles near $\mathrm{10\;day^{-1}}$. Sector 7 was taken during January 2019, which is about 6 years before the next periastron passage. We suggest that the pulsations observed by \cite{arentoftOscillatingBlueStragglers2007} could be the result of the close encounter during periastron.

\textit{WOCS 3012}---This SB1 (e/i = 6.0) is RC2 of \cite{panthiUOCSVIIIUV2022}, another red clump star with UV excess. We have 19 RV measurements over almost 5,000 days. Over the first 2,500 days, the RVs rose 10.3 $\mathrm{km\;s^{-1}}$. They then slowly decreased over the last 2,250 days and had not cycled back to the lowest recorded RV, suggesting that the binary has not yet completed a full orbital cycle. We have not yet found an orbital solution for this star. However for scale, the minimum mass of a secondary for an orbit with P = 5,000 days, K = 5.15 $\mathrm{km\;s^{-1}}$, and $\mathrm{M_1}\; =\; 1.6\; M_\odot$ is 0.7 $M_\odot$. The real orbit likely has a longer period (and possibly a larger K), which would yield a larger secondary mass. \cite{panthiUOCSVIIIUV2022} found this system to contain an approximately 0.6 $M_\odot$ WD, which the RV data presented here could support.

\textit{WOCS 4002} is a rapidly rotating ($v\sin i= 55.2 \mathrm{\;km\;s^{-1}}$) star. We find this star to have a 816-day eccentric (e=0.37) orbit. Photometrically, this star lies almost upon the equal-mass binary isochrone but is not observed to be an SB2, in contrast to its photometric neighbor, the SB2 WOCS 5002. \cite{panthiUOCSVIIIUV2022} did not find this system to have UV excess. If this star is an evolved BSS, it formed as a BSS long enough ago for the WD to cool below detection limits.

The remaining stars are not secure YSSs due to either possibly being photometric combinations of giant and main-sequence stars, a marginal probability of being proper-motion members, or being identified as SB2s. In the last case, because the spectrum of a WD companion would not be detected in the CCF with a solar template within this study's wavelength regime and signal-to-noise, spectroscopic detection of a companion indicates that these binaries at least comprise two main-sequence/subgiant stars near the MSTO. 

\textit{WOCS 3007}---This star is RC3 of \cite{panthiUOCSVIIIUV2022}, a third red clump star with UV excess. Considering the proper-motion (19\%) and RV (66\%) Gaussian memberships and a Gaia parallax at the edge of the distribution of parallax for cluster member stars, we do not consider this star to be a secure member of NGC 2506. However, the proper-motion membership from HDBSCAN is 100\%, suggesting the star may be associated with the cluster. The star has e/i = 1.4, below our threshold of velocity variability. For the star's measured RVs to be consistent with the 0.8 $M_\odot$ WD mass estimate of \cite{panthiUOCSVIIIUV2022}, the binary star's orbit would again need to be nearly face-on. Assuming a primary mass of $1.6\;M_{\odot}$ based on the red clump mass, the inclination limit becomes $12.9^{\circ}$ at a period of $10^4$ days, our upper limit on period for RV variability detection. 

\textit{WOCS 6004 and 8003} are both narrow-lined stars that do not show significant velocity variability (e/i = 2.0 and 2.6, respectively) nor UV excess. Both stars have photometry that could be explained as the blended light of a long-period giant-main sequence binary pair, but would require very specific combinations of such stars to recreate their blueness. Even so, we consider them to be YSS candidates.

\textit{WOCS 5002 and 7004} lie above the MSTO, and are YSS1 and YSS3 in \cite{panthiUOCSVIIIUV2022}. These authors find both stars to have UV excesses. As shown in Figure \ref{fig:yss_cmd}, both stars lie on the equal-mass binary isochrone. Both are binary stars with orbital solutions.\textit{WOCS 5002} is the equal-mass, double-lined binary V2032 from \cite{knudstrupExtremelyPreciseAge2020}. Our orbital solution (P = 27.869 days, e = 0.586) is in excellent agreement with \cite{knudstrupExtremelyPreciseAge2020}. We find this star system to have a mass-ratio of 0.975, which could suggest it is a photometric combination of a sub-giant and main-sequence star on the blue hook. \textit {WOCS 7004} has a 609-day eccentric (e=0.31) orbit. The star has several CCFs that show signs of being an SB2, but which could not be deconvolved to identify a secondary component. If this star is an SB2, that could explain the position above the MSTO. 

To determine whether these stars have UV excesses, \cite{panthiUOCSVIIIUV2022} fit single-star SEDs, and if their UV data showed excesses above the best-fit stellar SED, they added WD SEDs to their fit. However, if both binaries are SB2s, the interpretation of the combined SED is more complex. Thus we do not yet consider these stars to be secure YSSs. If re-analysis confirms the UV excesses, the presence of a WD tertiary (especially for 5002) may point to a triple mass-transfer scenario \citep{zwartTripleOriginTwin2019}.

\hfill \break
Of the five YSSs, four are confirmed or suspected multi-thousand-day period binaries in the region of the red clump and giant branch and one is an eccentric binary above the MSTO. 

In the vicinity of the red clump, WOCS 3012 is embedded within the red clump. In contrast, WOCS 4005 is bluer and WOCS 1006 is bluer and more luminous, possibly indicating higher masses then the red clump population \citep{rosvickBVPhotometryGyr1998}. For WOCS 3012 and 4005, and for the YSS candidate WOCS 3007 also blueward of the red clump, \cite{panthiUOCSVIIIUV2022} gave WD ages of 3 to 30 Myr (log ages of 6.5 to 7.5 years). If these YSSs are interpreted as evolved BSSs, such low WD ages present challenges for evolutionary timescales. Even considering the more massive BSSs, from MIST evolutionary tracks a 1.9 $M_{\odot}$ takes 110 Myr to evolve from the blue hook to the red clump and then spends 130 Myr in the red clump. The \cite{panthiUOCSVIIIUV2022} ages would require that these red clump YSSs were already giant stars when their companions evolved to WDs. While somewhat challenging for 2 or 3 cases, perhaps their progenitor binaries were of very nearly equal mass and went through a wind mass-transfer phase \citep[e.g.,][]{stassunEclipsingBinaryComprising2023}.

The YSS period distribution may be different than that of the BSSs. 4 of 5 YSSs appear to have at least multi-thousand day orbits, whereas at most 6 (1 binary and 5 apparent single stars) of the 14 BSSs could have periods that long. Perhaps this too is indicative of more complex formation and evolutionary paths of the YSS near the red clump.

\subsection{Other Stars of Note}\label{otherstars}

This section discusses our observations of other stars of note that have appeared in the literature on NGC 2506 or have interesting features. 

\textit{WOCS 17003}---This star is V4 \citep{kimSearchVariableStars2001}, an eclipsing binary system that \cite{arentoftOscillatingBlueStragglers2007} found to have a 2.868-day photometric period. We clearly identify this star as an SB2 and recover this period spectroscopically. \cite{knudstrupExtremelyPreciseAge2020} suggest this system contains a third body due to shifts in eclipse timings, and found the CCF to have a small third peak that may be indicative of light from a third star near the systemic velocity of the system. From modeling, they suggest the tertiary is orbiting the inner binary with a 443-day eccentric orbit. In one of our CCFs, we do see signs of a third star at 30 $\mathrm{km\;s^{-1}}$ above the systemic velocity.

\textit{WOCS 1017, 2011, and 4004}---WOCS 1017 is an SB1 giant with a 456.6-day, e = 0.34 orbit. Adopting a mass of 1.6 $M_\odot$ from our isochrone fit and a minimum secondary mass of 0.23 $M_\odot$ from our mass function, we calculate the Roche lobe radius using \cite{leahyCalculatorRocheLobe2015}. With a current radius of 71 $R_\odot$, the primary star approximately half fills its Roche lobe. The star will overflow its Roche lobe prior to reaching the tip of the RGB. WOCS 2011 and 4004 are both red clump binaries with similar periods to WOCS 1017 (345 and 472 days, respectively), suggesting WOCS 1017 may be an analog system to these two red clump stars when they were on the RGB. Together, these stars may illustrate the limits of mass transfer interactions at the tip of the RGB. Notably, WOCS 4004 has a slightly eccentric orbit (e=0.227), which may have been expected to circularize due to tidal interactions. 

\textit{WOCS 13017} (TIC 169716430) is a previously unidentified $\delta$ Scuti star. Using TESS CDIPS light curves \citep{boumaClusterDifferenceImaging2019} and the software Lightkurve \citep{2018lightkurve}, we find that the star undergoes pulsations at 11.394 cycles per day. Additionally, we have 40 RV measurements of the star that show RV variations of 14.3 $\mathrm{km\;s^{-1}}$ but do not yield a Keplerian orbital solution, further suggesting this star is a  $\delta$ Scuti star.

\textit{WOCS 10008 and 9006} are the oscillating BSS V6 and V7 from \cite{arentoftOscillatingBlueStragglers2007}, respectively. These stars are among the bluest main-sequence stars that match our isochrone fit, but still are redder than the theoretical end of the blue hook, so we do not classify either as a BSS. We find RV variations for both stars that are consistent with $\delta$ Scuti stars. These variations do not satisfy our binary criterion due to the rapid rotation of the stars; in addition neither star yields a Keplerian orbital solution. 

\textit{WOCS 10028} is redder than the RGB, and thus is a red straggler candidate. It has a low e/i of 0.89 $\mathrm{\;km\;s^{-1}}$ and low rotation ($v\sin i < 10 \mathrm{\;km\;s^{-1}}$).

\textit{WOCS 23006} is the eclipsing binary V9 of \cite{arentoftOscillatingBlueStragglers2007}, who found a period of 0.93 days from light curves and suggested it may be an Algol or $\beta$ Lyrae binary. We have obtained 5 spectra of this main-sequence star; the spectra are flat with very weak, but broadened, Mg b absorption features. Among the highest signal spectra obtained, the low-quality RVs obtained with a solar template are near the cluster's RV, so this may be a member star. Due to the line broadening, we classify this star as a VRR, M.

\section{Summary}\label{summary}
In this paper we discuss the results of the WIYN Open Cluster Survey RV study of the open cluster NGC 2506. We present 12,157 spectroscopic RV measurements taken over 41 years with the WIYN 3.5m telescope and the CfA Digital Speedometers on the MMT and 1.5-m Tillinghast Reflector of 2,442 stars in the NGC 2506 field. Our observations are complete (3 or more RV observations) for all proper-motion members brighter than G = 15.5, yielding 320 proper-motion and RV member stars. Further, we identify 469 three-dimensional cluster members to G = 16.5. 

We find NGC 2506 to have a mean RV of $\mathrm{84.29 \pm 0.06\;km\;s^{-1}}$, a true RV dispersion of $\mathrm{0.54\pm 0.06\;km\;s^{-1}}$, a core radius of 
$2.46 \pm 0.13$ pc, a tidal radius of $44.58 \pm 2.92$ pc, and a virial mass of 3,400 $\pm\;900 M_\odot$. 

We present orbital solutions for 60 binary stars, with periods between 1 and 7,580 days. We find NGC 2506 to have an incompleteness-corrected binary fraction for binaries with periods less than $10^4$ days of $35 \pm 5\%$ among the upper main sequence stars. Additionally, we find at least 64\% of the 14 BSSs to be in binary systems. We do not find statistically significant mass segregation among binaries and single stars on the upper main sequence and giant branch. We do, however, find significant central concentration of the BSSs relative to single stars. 

We also discuss stars of note among this population, including BSSs and YSSs. Although the BSSs and YSSs all have photometry that separates them from the single-star evolution of NGC 2506, these populations contain stars with a wide diversity of stellar and orbital characteristics. The BSS population in particular comprises stellar systems with evolutionary track masses from 0.1 $M_\odot$ below to double the MSTO mass, projected rotation rates from below our sensitivity limits ($v\sin i < 10 \mathrm{\;km\;s^{-1}}$) to $v\sin i > 120 \mathrm{\;km\;s^{-1}}$, and CMD locations from the ZAMS to the terminal-age main-sequence. Likewise, the binary properties of these stars are varied: at least two binaries have periods $<100$ days, at least three binaries (up to six) have periods clustered around $10^3$ days, potentially two binaries have multi-thousand day periods, and four to five stars show no velocity variation to our detection limits. Of the four known BSS orbits, two are circular and two are eccentric. On the other hand, the YSSs have primarily multi-thousand day periods. The diversity of observable characteristics of both BSSs and YSSs in NGC 2506 suggests these stars span a variety of evolutionary histories, and possibly formed through a variety of mechanisms. Continued RV observations of these stars are necessary to determine more orbital solutions, which can provide insight on their progenitors and probe the frequencies of formation mechanisms of populations on alternative stellar evolution pathways.

This study of NGC 2506 completes the time-series RV surveys of the initial set of WOCS clusters \citep{mathieuWIYNOpenCluster2000}. NGC 2506 is important as a rich, metal-poor open cluster situated at an age between well-studied young clusters like M35 and NGC 7789 and intermediate-age clusters such as NGC 6819 and M67. Together, these clusters mapped the many facets of stellar evolution at different ages, identified new stellar populations such as sub-subgiants and blue lurkers, and elucidated the fundamental role binary stars play in creating alternative stellar evolution pathways. These RV studies and their identified binary populations lay the foundation for further discoveries in these clusters through investigations of chemical abundances, stellar rotation, asteroseismology, and more. Coupled with detailed binary evolution simulations, these clusters will continue to strengthen our understanding of stellar evolution in binary stars.

\begin{acknowledgments}
The authors would like to express their gratitude to the staff of the WIYN Observatory without whom we would not have been able to acquire these thousands of stellar spectra. We want to thank the many undergraduate and graduate students who have obtained spectra over the years at WIYN for this project. The reported observations were founded on the superb astrometric work of Imants Platais. Observations reported here were obtained at the MMT Observatory, a joint facility of the Smithsonian Institution and the University of Arizona. Finally, we acknowledge the support of the National Science Foundation through award NSF AST-1714506, the Wisconsin Space Grant Consortium through award RFP22\_10-0, and the Wisconsin Alumni Research Fund.

This work has made use of data from the European Space Agency (ESA) mission Gaia (https://www.cosmos.esa.int/gaia), processed by the Gaia Data Processing and Analysis Consortium (DPAC, https://www.cosmos.esa.int/web/gaia/dpac/ consortium). Funding for the DPAC has been provided by national institutions, in particular the institutions participating in the Gaia Multilateral Agreement. This paper includes data collected by the TESS mission. Funding for the TESS mission is provided by the NASA's Science Mission Directorate. 

This work was conducted at the University of Wisconsin-Madison, which is located on occupied ancestral land of the Ho-Chunk people, a place their nation has called Teejop since time immemorial. In an 1832 treaty, the Ho-Chunk were forced to cede this territory. The university was founded on and funded through this seized land; this legacy enabled the science presented here. Observations for this work were conducted at the WIYN telescope on Kitt Peak, which is part of the lands of the Tohono O’odham Nation.
\end{acknowledgments}

\facilities{WIYN (Hydra MOS), FLWO:1.5m (Digital Speedometer), MMT (Digital Speedometer) Gaia, TESS, GALEX}
\software{Astropy \citep{astropycollaborationAstropyCommunityPython2013,astropycollaborationAstropyProjectBuilding2018, astropycollaborationAstropyProjectSustaining2022}, 
HDBSCAN Clustering Library \citep{mcinnesHdbscanHierarchicalDensity2017},
Lightkurve \citep{2018lightkurve},
LIMEPY \citep{gielesFamilyLoweredIsothermal2015},
MIST \citep{dotterMESAISOCHRONESLAR2016, choiMESAISOCHRONESSTELLAR2016, paxtonMODULESEXPERIMENTSSTELLAR2011, paxtonMODULESEXPERIMENTSSTELLAR2013,paxtonMODULESEXPERIMENTSSTELLAR2015}, 
NumPy \citep{harrisArrayProgrammingNumPy2020},
pandas \citep[doi:10.5281/zenodo.5774815]{mckinney-proc-scipy-2010},
Scikit-learn \citep{scikit-learn},
SciPy \citep{virtanenSciPyFundamentalAlgorithms2020},
VizieR catalogue access tool \citep[doi: 10.26093/cds/vizier]{ochsenbeinVizieRDatabaseAstronomical2000}
}

\bibliographystyle{aasjournal.bst}
\bibliography{all_references}

\begin{thebibliography}{}
\expandafter\ifx\csname natexlab\endcsname\relax\def\natexlab#1{#1}\fi
\providecommand{\url}[1]{\href{#1}{#1}}
\providecommand{\dodoi}[1]{doi:~\href{http://doi.org/#1}{\nolinkurl{#1}}}
\providecommand{\doeprint}[1]{\href{http://ascl.net/#1}{\nolinkurl{http://ascl.net/#1}}}
\providecommand{\doarXiv}[1]{\href{https://arxiv.org/abs/#1}{\nolinkurl{https://arxiv.org/abs/#1}}}

\bibitem[{Agarwal {et~al.}(2021)Agarwal, Rao, Vaidya, \&
  Bhattacharya}]{Agarwal+2021}
Agarwal, M., Rao, K.~K., Vaidya, K., \& Bhattacharya, S. 2021, MNRAS, 502,
  2582, \dodoi{10.1093/mnras/stab118}

\bibitem[{Ahumada \& Lapasset(1995)}]{ahumadaCatalogueBlueStragglers1995}
Ahumada, J., \& Lapasset, E. 1995, A\&AS, 109, 375

\bibitem[{Ahumada \& Lapasset(2007)}]{ahumadaNewCatalogueBlue2007}
Ahumada, J.~A., \& Lapasset, E. 2007, A\&A, 463, 789,
  \dodoi{10.1051/0004-6361:20054590}

\bibitem[{Andronov {et~al.}(2006)Andronov, Pinsonneault, \&
  Terndrup}]{andronovMergersClosePrimordial2006}
Andronov, N., Pinsonneault, M.~H., \& Terndrup, D.~M. 2006, ApJ, 646, 1160,
  \dodoi{10.1086/505127}

\bibitem[{{Anthony-Twarog} {et~al.}(2016){Anthony-Twarog}, Deliyannis, \&
  Twarog}]{anthony-twarogWIYNOPENCLUSTER2016}
{Anthony-Twarog}, B.~J., Deliyannis, C.~P., \& Twarog, B.~A. 2016, AJ, 152,
  192, \dodoi{10.3847/0004-6256/152/6/192}

\bibitem[{{Anthony-Twarog} {et~al.}(2018){Anthony-Twarog}, {Lee-Brown},
  Deliyannis, \& Twarog}]{anthony-twarogWIYNOpenCluster2018}
{Anthony-Twarog}, B.~J., {Lee-Brown}, D.~B., Deliyannis, C.~P., \& Twarog,
  B.~A. 2018, AJ, 155, 138, \dodoi{10.3847/1538-3881/aaad66}

\bibitem[{Arentoft {et~al.}(2007)Arentoft, De~Ridder, Grundahl, Glowienka,
  Waelkens, Dupret, Grigahc{\`e}ne, Lefever, Jensen, Reyniers, Frandsen, \&
  Kjeldsen}]{arentoftOscillatingBlueStragglers2007}
Arentoft, T., De~Ridder, J., Grundahl, F., {et~al.} 2007, A\&A, 465, 965,
  \dodoi{10.1051/0004-6361:20066931}

\bibitem[{{Astropy Collaboration} {et~al.}(2013){Astropy Collaboration},
  Robitaille, Tollerud, Greenfield, Droettboom, Bray, Aldcroft, Davis,
  Ginsburg, {Price-Whelan}, Kerzendorf, Conley, Crighton, Barbary, Muna,
  Ferguson, Grollier, Parikh, Nair, Unther, Deil, Woillez, Conseil, Kramer,
  Turner, Singer, Fox, Weaver, Zabalza, Edwards, Azalee~Bostroem, Burke, Casey,
  Crawford, Dencheva, Ely, Jenness, Labrie, Lim, Pierfederici, Pontzen, Ptak,
  Refsdal, Servillat, \&
  Streicher}]{astropycollaborationAstropyCommunityPython2013}
{Astropy Collaboration}, Robitaille, T.~P., Tollerud, E.~J., {et~al.} 2013,
  A\&A, 558, A33, \dodoi{10.1051/0004-6361/201322068}

\bibitem[{{Astropy Collaboration} {et~al.}(2018){Astropy Collaboration},
  {Price-Whelan}, Sip{\H o}cz, G{\"u}nther, Lim, Crawford, Conseil, Shupe,
  Craig, Dencheva, Ginsburg, VanderPlas, Bradley, {P{\'e}rez-Su{\'a}rez}, {de
  Val-Borro}, Aldcroft, Cruz, Robitaille, Tollerud, Ardelean, Babej, Bach,
  Bachetti, Bakanov, Bamford, Barentsen, Barmby, Baumbach, Berry, Biscani,
  Boquien, Bostroem, Bouma, Brammer, Bray, Breytenbach, Buddelmeijer, Burke,
  Calderone, Cano~Rodr{\'i}guez, Cara, Cardoso, Cheedella, Copin, Corrales,
  Crichton, D'Avella, Deil, Depagne, Dietrich, Donath, Droettboom, Earl, Erben,
  Fabbro, Ferreira, Finethy, Fox, Garrison, Gibbons, Goldstein, Gommers, Greco,
  Greenfield, Groener, Grollier, Hagen, Hirst, Homeier, Horton, Hosseinzadeh,
  Hu, Hunkeler, Ivezi{\'c}, Jain, Jenness, Kanarek, Kendrew, Kern, Kerzendorf,
  Khvalko, King, Kirkby, Kulkarni, Kumar, Lee, Lenz, Littlefair, Ma, Macleod,
  Mastropietro, McCully, Montagnac, Morris, Mueller, Mumford, Muna, Murphy,
  Nelson, Nguyen, Ninan, N{\"o}the, Ogaz, Oh, Parejko, Parley, Pascual, Patil,
  Patil, Plunkett, Prochaska, Rastogi, Reddy~Janga, Sabater, Sakurikar,
  Seifert, Sherbert, {Sherwood-Taylor}, Shih, Sick, Silbiger, Singanamalla,
  Singer, Sladen, Sooley, Sornarajah, Streicher, Teuben, Thomas, Tremblay,
  Turner, Terr{\'o}n, {van Kerkwijk}, {de la Vega}, Watkins, Weaver, Whitmore,
  Woillez, Zabalza, \& {Astropy
  Contributors}}]{astropycollaborationAstropyProjectBuilding2018}
{Astropy Collaboration}, {Price-Whelan}, A.~M., Sip{\H o}cz, B.~M., {et~al.}
  2018, AJ, 156, 123, \dodoi{10.3847/1538-3881/aabc4f}

\bibitem[{{Astropy Collaboration} {et~al.}(2022){Astropy Collaboration},
  {Price-Whelan}, Lim, Earl, Starkman, Bradley, Shupe, Patil, Corrales,
  Brasseur, N{\"o}the, Donath, Tollerud, Morris, Ginsburg, Vaher, Weaver,
  Tocknell, Jamieson, {van Kerkwijk}, Robitaille, Merry, Bachetti, G{\"u}nther,
  Aldcroft, {Alvarado-Montes}, Archibald, B{\'o}di, Bapat, Barentsen,
  Baz{\'a}n, Biswas, Boquien, Burke, Cara, Cara, Conroy, Conseil, Craig, Cross,
  Cruz, D'Eugenio, Dencheva, Devillepoix, Dietrich, Eigenbrot, Erben, Ferreira,
  {Foreman-Mackey}, Fox, Freij, Garg, Geda, Glattly, Gondhalekar, Gordon,
  Grant, Greenfield, Groener, Guest, Gurovich, Handberg, Hart,
  {Hatfield-Dodds}, Homeier, Hosseinzadeh, Jenness, Jones, Joseph, Kalmbach,
  Karamehmetoglu, Ka{\l}uszy{\'n}ski, Kelley, Kern, Kerzendorf, Koch, Kulumani,
  Lee, Ly, Ma, MacBride, Maljaars, Muna, Murphy, Norman, O'Steen, Oman,
  Pacifici, Pascual, {Pascual-Granado}, Patil, Perren, Pickering, Rastogi,
  Roulston, Ryan, Rykoff, Sabater, Sakurikar, Salgado, Sanghi, Saunders,
  Savchenko, Schwardt, {Seifert-Eckert}, Shih, Jain, Shukla, Sick, Simpson,
  Singanamalla, Singer, Singhal, Sinha, Sip{\H o}cz, Spitler, Stansby,
  Streicher, {\v S}umak, Swinbank, Taranu, Tewary, Tremblay, {de Val-Borro},
  Van~Kooten, Vasovi{\'c}, Verma, {de Miranda Cardoso}, Williams, Wilson,
  Winkel, {Wood-Vasey}, Xue, Yoachim, Zhang, Zonca, \& {Astropy Project
  Contributors}}]{astropycollaborationAstropyProjectSustaining2022}
{Astropy Collaboration}, {Price-Whelan}, A.~M., Lim, P.~L., {et~al.} 2022, AJ,
  935, 167, \dodoi{10.3847/1538-4357/ac7c74}

\bibitem[{Boffin {et~al.}(2014)Boffin, Hillen, Berger, Jorissen, Blind,
  Bouquin, Miko{\l}ajewska, \& Lazareff}]{boffinRochelobeFillingFactor2014}
Boffin, H. M.~J., Hillen, M., Berger, J.~P., {et~al.} 2014, A\&A, 564, A1,
  \dodoi{10.1051/0004-6361/201323194}

\bibitem[{Bouma {et~al.}(2019)Bouma, Hartman, Bhatti, Winn, \&
  Bakos}]{boumaClusterDifferenceImaging2019}
Bouma, L.~G., Hartman, J.~D., Bhatti, W., Winn, J.~N., \& Bakos, G.~A. 2019,
  ApJS, 245, 13, \dodoi{10.3847/1538-4365/ab4a7e}

\bibitem[{Bovy(2015)}]{bovyGalpyPythonLIBRARY2015}
Bovy, J. 2015, ApJS, 216, 29, \dodoi{10.1088/0067-0049/216/2/29}

\bibitem[{Campello {et~al.}(2013)Campello, Moulavi, \&
  Sander}]{campelloDensityBasedClusteringBased2013}
Campello, R. J. G.~B., Moulavi, D., \& Sander, J. 2013, in Adv. {{Knowl}}.
  {{Discov}}. {{Data Min}}., ed. J.~Pei, V.~S. Tseng, L.~Cao, H.~Motoda, \&
  G.~Xu, Lecture {{Notes}} in {{Computer Science}} (Berlin, Heidelberg:
  Springer), 160--172, \dodoi{10.1007/978-3-642-37456-2_14}

\bibitem[{{Cantat-Gaudin} \&
  Anders(2020)}]{cantat-gaudinClustersMiragesCataloguing2020}
{Cantat-Gaudin}, T., \& Anders, F. 2020, A\&A, 633, A99,
  \dodoi{10.1051/0004-6361/201936691}

\bibitem[{{Cantat-Gaudin} {et~al.}(2018){Cantat-Gaudin}, Jordi, Vallenari,
  Bragaglia, {Balaguer-N{\'u}{\~n}ez}, Soubiran, Bossini, Moitinho,
  {Castro-Ginard}, {Krone-Martins}, Casamiquela, Sordo, \&
  Carrera}]{cantat-gaudinGaiaDR2View2018}
{Cantat-Gaudin}, T., Jordi, C., Vallenari, A., {et~al.} 2018, A\&A, 618, A93,
  \dodoi{10.1051/0004-6361/201833476}

\bibitem[{Carlberg(2014)}]{carlbergROTATIONALRADIALVELOCITIES2014}
Carlberg, J.~K. 2014, AJ, 147, 138, \dodoi{10.1088/0004-6256/147/6/138}

\bibitem[{Chen \& Han(2008)}]{chenBlueStragglersPrimordial2008}
Chen, X., \& Han, Z. 2008, Proc. IAU, 4, 417, \dodoi{10.1017/S1743921308023351}

\bibitem[{Chiu \& {van Altena}(1981)}]{chiuMembershipOldOpen1981}
Chiu, L.-T.~G., \& {van Altena}, W.~F. 1981, ApJ, 243, 827,
  \dodoi{10.1086/158649}

\bibitem[{Choi {et~al.}(2018)Choi, Conroy, Ting, Cargile, Dotter, \&
  Johnson}]{choiStarClusterAges2018}
Choi, J., Conroy, C., Ting, Y.-S., {et~al.} 2018, ApJ, 863, 65,
  \dodoi{10.3847/1538-4357/aad18c}

\bibitem[{Choi {et~al.}(2016)Choi, Dotter, Conroy, Cantiello, Paxton, \&
  Johnson}]{choiMESAISOCHRONESSTELLAR2016}
Choi, J., Dotter, A., Conroy, C., {et~al.} 2016, ApJ, 823, 102,
  \dodoi{10.3847/0004-637X/823/2/102}

\bibitem[{Corsaro {et~al.}(2012)Corsaro, Stello, Huber, Bedding, Bonanno,
  Brogaard, Kallinger, Benomar, White, Mosser, Basu, Chaplin,
  {Christensen-Dalsgaard}, Elsworth, Garc{\'i}a, Hekker, Kjeldsen, Mathur,
  Meibom, Hall, Ibrahim, \& Klaus}]{corsaroASTEROSEISMOLOGYOPENCLUSTERS2012}
Corsaro, E., Stello, D., Huber, D., {et~al.} 2012, ApJ, 757, 190,
  \dodoi{10.1088/0004-637X/757/2/190}

\bibitem[{Cutri {et~al.}(2003)Cutri, Skrutskie, {van Dyk}, Beichman, Carpenter,
  Chester, Cambresy, Evans, Fowler, Gizis, Howard, Huchra, Jarrett, Kopan,
  Kirkpatrick, Light, Marsh, McCallon, Schneider, Stiening, Sykes, Weinberg,
  Wheaton, Wheelock, \& Zacarias}]{2003_2MASS_pointsourcecat}
Cutri, R.~M., Skrutskie, M.~F., {van Dyk}, S., {et~al.} 2003, {{2MASS}} All Sky
  Catalog of Point Sources.,  NASA/IPAC Infrared Science Archive

\bibitem[{Dotter(2016)}]{dotterMESAISOCHRONESLAR2016}
Dotter, A. 2016, ApJS, 222, 8, \dodoi{10.3847/0067-0049/222/1/8}

\bibitem[{{Gaia Collaboration} {et~al.}(2018){Gaia Collaboration}, Brown,
  Vallenari, Prusti, {de Bruijne}, Babusiaux, {Bailer-Jones}, Biermann, Evans,
  Eyer, Jansen, Jordi, Klioner, Lammers, Lindegren, Luri, Mignard, Panem,
  Pourbaix, Randich, Sartoretti, Siddiqui, Soubiran, {van Leeuwen}, Walton,
  Arenou, Bastian, Cropper, Drimmel, Katz, Lattanzi, Bakker, Cacciari,
  Casta{\~n}eda, Chaoul, Cheek, De~Angeli, Fabricius, Guerra, Holl, Masana,
  Messineo, Mowlavi, Nienartowicz, Panuzzo, Portell, Riello, Seabroke, Tanga,
  Th{\'e}venin, {Gracia-Abril}, Comoretto, {Garcia-Reinaldos}, Teyssier,
  Altmann, Andrae, Audard, {Bellas-Velidis}, Benson, Berthier, Blomme, Burgess,
  Busso, Carry, Cellino, Clementini, Clotet, Creevey, Davidson, De~Ridder,
  Delchambre, Dell'Oro, Ducourant, {Fern{\'a}ndez-Hern{\'a}ndez}, Fouesneau,
  Fr{\'e}mat, Galluccio, {Garc{\'i}a-Torres}, {Gonz{\'a}lez-N{\'u}{\~n}ez},
  {Gonz{\'a}lez-Vidal}, Gosset, Guy, Halbwachs, Hambly, Harrison,
  Hern{\'a}ndez, Hestroffer, Hodgkin, Hutton, Jasniewicz,
  {Jean-Antoine-Piccolo}, Jordan, Korn, {Krone-Martins}, Lanzafame, Lebzelter,
  L{\"o}ffler, Manteiga, Marrese, {Mart{\'i}n-Fleitas}, Moitinho, Mora,
  Muinonen, Osinde, Pancino, Pauwels, Petit, {Recio-Blanco}, Richards,
  Rimoldini, Robin, Sarro, Siopis, Smith, Sozzetti, S{\"u}veges, Torra, {van
  Reeven}, Abbas, Abreu~Aramburu, Accart, Aerts, Altavilla, {\'A}lvarez,
  Alvarez, Alves, Anderson, Andrei, Anglada~Varela, Antiche, Antoja, Arcay,
  Astraatmadja, Bach, Baker, {Balaguer-N{\'u}{\~n}ez}, Balm, Barache, Barata,
  Barbato, Barblan, Barklem, Barrado, Barros, Barstow,
  Bartholom{\'e}~Mu{\~n}oz, Bassilana, Becciani, Bellazzini, Berihuete,
  Bertone, Bianchi, Bienaym{\'e}, {Blanco-Cuaresma}, Boch, Boeche, Bombrun,
  Borrachero, Bossini, Bouquillon, Bourda, Bragaglia, Bramante, Breddels,
  Bressan, Brouillet, Br{\"u}semeister, Brugaletta, Bucciarelli, Burlacu,
  Busonero, Butkevich, Buzzi, Caffau, Cancelliere, Cannizzaro, {Cantat-Gaudin},
  Carballo, Carlucci, Carrasco, Casamiquela, Castellani, {Castro-Ginard},
  Charlot, Chemin, Chiavassa, Cocozza, Costigan, Cowell, Crifo, Crosta,
  Crowley, Cuypers{\dag}, Dafonte, Damerdji, Dapergolas, David, David, {de
  Laverny}, De~Luise, De~March, {de Martino}, {de Souza}, {de Torres},
  Debosscher, {del Pozo}, Delbo, Delgado, Delgado, Di~Matteo, Diakite, Diener,
  Distefano, Dolding, Drazinos, Dur{\'a}n, Edvardsson, Enke, Eriksson, Esquej,
  Eynard~Bontemps, Fabre, Fabrizio, Faigler, Falc{\~a}o, Farr{\`a}s~Casas,
  Federici, Fedorets, Fernique, Figueras, Filippi, Findeisen, Fonti, Fraile,
  Fraser, Fr{\'e}zouls, Gai, Galleti, Garabato, {Garc{\'i}a-Sedano}, Garofalo,
  Garralda, Gavel, Gavras, Gerssen, Geyer, Giacobbe, Gilmore, Girona,
  Giuffrida, Glass, Gomes, Granvik, Gueguen, Guerrier, Guiraud,
  {Guti{\'e}rrez-S{\'a}nchez}, Haigron, Hatzidimitriou, Hauser, Haywood,
  Heiter, Helmi, Heu, Hilger, Hobbs, Hofmann, Holland, Huckle, Hypki, Icardi,
  Jan{\ss}en, {Jevardat de Fombelle}, Jonker, Juh{\'a}sz, Julbe, Karampelas,
  Kewley, Klar, Kochoska, Kohley, Kolenberg, Kontizas, Kontizas, Koposov,
  Kordopatis, {Kostrzewa-Rutkowska}, Koubsky, Lambert, Lanza, Lasne, Lavigne,
  Le~Fustec, {Le Poncin-Lafitte}, Lebreton, Leccia, Leclerc, {Lecoeur-Taibi},
  Lenhardt, Leroux, Liao, Licata, Lindstr{\o}m, Lister, Livanou, Lobel,
  L{\'o}pez, Managau, Mann, Mantelet, Marchal, Marchant, Marconi, Marinoni,
  Marschalk{\'o}, Marshall, Martino, Marton, Mary, Massari, Matijevi{\v c},
  Mazeh, McMillan, Messina, Michalik, Millar, Molina, Molinaro, Moln{\'a}r,
  Montegriffo, Mor, Morbidelli, Morel, Morris, Mulone, Muraveva, Musella,
  Nelemans, Nicastro, Noval, O'Mullane, Ord{\'e}novic,
  {Ord{\'o}{\~n}ez-Blanco}, Osborne, Pagani, Pagano, Pailler, Palacin,
  Palaversa, Panahi, Pawlak, Piersimoni, Pineau, Plachy, Plum, Poggio,
  Poujoulet, Pr{\v s}a, Pulone, Racero, Ragaini, Rambaux, {Ramos-Lerate},
  Regibo, Reyl{\'e}, Riclet, Ripepi, Riva, Rivard, Rixon, Roegiers, Roelens,
  {Romero-G{\'o}mez}, Rowell, Royer, {Ruiz-Dern}, Sadowski,
  Sagrist{\`a}~Sell{\'e}s, Sahlmann, Salgado, Salguero, Sanna, {Santana-Ros},
  Sarasso, Savietto, Schultheis, Sciacca, Segol, Segovia, S{\'e}gransan, Shih,
  Siltala, Silva, Smart, Smith, Solano, Solitro, Sordo, Soria~Nieto, Souchay,
  Spagna, Spoto, Stampa, Steele, Steidelm{\"u}ller, Stephenson, Stoev, Suess,
  Surdej, Szabados, {Szegedi-Elek}, Tapiador, Taris, Tauran, Taylor, Teixeira,
  Terrett, Teyssandier, Thuillot, Titarenko, Torra~Clotet, Turon, Ulla,
  Utrilla, Uzzi, Vaillant, Valentini, Valette, {van Elteren}, Van~Hemelryck,
  {van Leeuwen}, Vaschetto, Vecchiato, Veljanoski, Viala, Vicente, Vogt, {von
  Essen}, Voss, Votruba, Voutsinas, Walmsley, Weiler, Wertz, Wevers,
  Wyrzykowski, Yoldas, {\v Z}erjal, Ziaeepour, Zorec, Zschocke, Zucker,
  Zurbach, \& Zwitter}]{gaiacollaborationGaiaDataRelease2018}
{Gaia Collaboration}, Brown, A. G.~A., Vallenari, A., {et~al.} 2018, A\&A, 616,
  A1, \dodoi{10.1051/0004-6361/201833051}

\bibitem[{{Gaia Collaboration} {et~al.}(2023){Gaia Collaboration}, Vallenari,
  Brown, Prusti, de~Bruijne, Arenou, Babusiaux, Biermann, Creevey, Ducourant,
  Evans, Eyer, Guerra, Hutton, Jordi, Klioner, Lammers, Lindegren, Luri,
  Mignard, Panem, Pourbaix, Randich, Sartoretti, Soubiran, Tanga, Walton,
  {Bailer-Jones}, Bastian, Drimmel, Jansen, Katz, Lattanzi, van Leeuwen,
  Bakker, Cacciari, Casta{\~n}eda, Angeli, Fabricius, Fouesneau, Fr{\'e}mat,
  Galluccio, Guerrier, Heiter, Masana, Messineo, Mowlavi, Nicolas,
  Nienartowicz, Pailler, Panuzzo, Riclet, Roux, Seabroke, Sordo, Th{\'e}venin,
  {Gracia-Abril}, Portell, Teyssier, Altmann, Andrae, Audard, {Bellas-Velidis},
  Benson, Berthier, Blomme, Burgess, Busonero, Busso, C{\'a}novas, Carry,
  Cellino, Cheek, Clementini, Damerdji, Davidson, de~Teodoro, Campos,
  Delchambre, Dell'Oro, Esquej, {Fern{\'a}ndez-Hern{\'a}ndez}, Fraile,
  Garabato, {Garc{\'i}a-Lario}, Gosset, Haigron, Halbwachs, Hambly, Harrison,
  Hern{\'a}ndez, Hestroffer, Hodgkin, Holl, Jan{\ss}en, de~Fombelle, Jordan,
  {Krone-Martins}, Lanzafame, L{\"o}ffler, Marchal, Marrese, Moitinho,
  Muinonen, Osborne, Pancino, Pauwels, {Recio-Blanco}, Reyl{\'e}, Riello,
  Rimoldini, Roegiers, Rybizki, Sarro, Siopis, Smith, Sozzetti, Utrilla, van
  Leeuwen, Abbas, {\'A}brah{\'a}m, Aramburu, Aerts, Aguado, Ajaj,
  {Aldea-Montero}, Altavilla, {\'A}lvarez, Alves, Anders, Anderson, Varela,
  Antoja, Baines, Baker, {Balaguer-N{\'u}{\~n}ez}, Balbinot, Balog, Barache,
  Barbato, Barros, Barstow, Bartolom{\'e}, Bassilana, Bauchet, Becciani,
  Bellazzini, Berihuete, Bernet, Bertone, Bianchi, Binnenfeld,
  {Blanco-Cuaresma}, Blazere, Boch, Bombrun, Bossini, Bouquillon, Bragaglia,
  Bramante, Breedt, Bressan, Brouillet, Brugaletta, Bucciarelli, Burlacu,
  Butkevich, Buzzi, Caffau, Cancelliere, {Cantat-Gaudin}, Carballo, Carlucci,
  Carnerero, Carrasco, Casamiquela, Castellani, {Castro-Ginard}, Chaoul,
  Charlot, Chemin, Chiaramida, Chiavassa, Chornay, Comoretto, Contursi, Cooper,
  Cornez, Cowell, Crifo, Cropper, Crosta, Crowley, Dafonte, Dapergolas, David,
  David, de~Laverny, Luise, March, Ridder, de~Souza, de~Torres, del Peloso, del
  Pozo, Delbo, Delgado, Delisle, Demouchy, Dharmawardena, Matteo, Diakite,
  Diener, Distefano, Dolding, Edvardsson, Enke, Fabre, Fabrizio, Faigler,
  Fedorets, Fernique, Fienga, Figueras, Fournier, Fouron, Fragkoudi, Gai,
  {Garcia-Gutierrez}, {Garcia-Reinaldos}, {Garc{\'i}a-Torres}, Garofalo, Gavel,
  Gavras, Gerlach, Geyer, Giacobbe, Gilmore, Girona, Giuffrida, Gomel, Gomez,
  {Gonz{\'a}lez-N{\'u}{\~n}ez}, {Gonz{\'a}lez-Santamar{\'i}a},
  {Gonz{\'a}lez-Vidal}, Granvik, Guillout, Guiraud,
  {Guti{\'e}rrez-S{\'a}nchez}, Guy, Hatzidimitriou, Hauser, Haywood, Helmer,
  Helmi, Sarmiento, Hidalgo, Hilger, H{\l}adczuk, Hobbs, Holland, Huckle,
  Jardine, Jasniewicz, Piccolo, {Jim{\'e}nez-Arranz}, Jorissen, Campillo,
  Julbe, Karbevska, Kervella, Khanna, Kontizas, Kordopatis, Korn,
  K{\'o}sp{\'a}l, {Kostrzewa-Rutkowska}, Kruszy{\'n}ska, Kun, Laizeau, Lambert,
  Lanza, Lasne, Campion, Lebreton, Lebzelter, Leccia, Leclerc, {Lecoeur-Taibi},
  Liao, Licata, Lindstr{\o}m, Lister, Livanou, Lobel, Lorca, Loup, Pardo,
  Romeo, Managau, Mann, Manteiga, Marchant, Marconi, Marcos, Santos, Pina,
  Marinoni, Marocco, Marshall, Polo, {Mart{\'i}n-Fleitas}, Marton, Mary, Masip,
  Massari, {Mastrobuono-Battisti}, Mazeh, McMillan, Messina, Michalik, Millar,
  Mints, Molina, Molinaro, Moln{\'a}r, Monari, Mongui{\'o}, Montegriffo,
  Montero, Mor, Mora, Morbidelli, Morel, Morris, Muraveva, Murphy, Musella,
  Nagy, Noval, Oca{\~n}a, Ogden, Ordenovic, Osinde, Pagani, Pagano, Palaversa,
  Palicio, {Pallas-Quintela}, Panahi, {Payne-Wardenaar}, Esteller,
  Penttil{\"a}, Pichon, Piersimoni, Pineau, Plachy, Plum, Poggio, Pr{\v s}a,
  Pulone, Racero, Ragaini, Rainer, Raiteri, Rambaux, Ramos, {Ramos-Lerate},
  Fiorentin, Regibo, Richards, Diaz, Ripepi, Riva, Rix, Rixon, Robichon, Robin,
  Robin, Roelens, Rogues, Rohrbasser, {Romero-G{\'o}mez}, Rowell, Royer,
  Mieres, Rybicki, Sadowski, N{\'u}{\~n}ez, Sell{\'e}s, Sahlmann, Salguero,
  Samaras, Gimenez, Sanna, Santove{\~n}a, Sarasso, Schultheis, Sciacca, Segol,
  Segovia, S{\'e}gransan, Semeux, Shahaf, Siddiqui, Siebert, Siltala, Silvelo,
  Slezak, Slezak, Smart, Snaith, Solano, Solitro, Souami, Souchay, Spagna,
  Spina, Spoto, Steele, Steidelm{\"u}ller, Stephenson, S{\"u}veges, Surdej,
  Szabados, {Szegedi-Elek}, Taris, Taylor, Teixeira, Tolomei, Tonello, Torra,
  Torra, Elipe, Trabucchi, Tsounis, Turon, Ulla, Unger, Vaillant, van Dillen,
  van Reeven, Vanel, Vecchiato, Viala, Vicente, Voutsinas, Weiler, Wevers,
  Wyrzykowski, Yoldas, Yvard, Zhao, Zorec, Zucker, \&
  Zwitter}]{GaiaDataRelease2023}
{Gaia Collaboration}, Vallenari, A., Brown, A. G.~A., {et~al.} 2023, A\&A, 674,
  A1, \dodoi{10.1051/0004-6361/202243940}

\bibitem[{Gao(2020)}]{gaoDiscoveryTidalTails2020}
Gao, X. 2020, ApJ, 894, 48, \dodoi{10.3847/1538-4357/ab8560}

\bibitem[{Geller {et~al.}(2015)Geller, Latham, \&
  Mathieu}]{gellerLARRADIALVELOCITIES2015}
Geller, A.~M., Latham, D.~W., \& Mathieu, R.~D. 2015, AJ, 150, 97,
  \dodoi{10.1088/0004-6256/150/3/97}

\bibitem[{Geller \& Mathieu(2012)}]{gellerWIYNOPENCLUSTER2012}
Geller, A.~M., \& Mathieu, R.~D. 2012, AJ, 22

\bibitem[{Geller {et~al.}(2010{\natexlab{a}})Geller, Mathieu, Braden, Meibom,
  Platais, \& Dolan}]{gellerWIYNOPENCLUSTER2010}
Geller, A.~M., Mathieu, R.~D., Braden, E.~K., {et~al.} 2010{\natexlab{a}}, The
  Astronomical Journal, 139, 1383, \dodoi{10.1088/0004-6256/139/4/1383}

\bibitem[{Geller {et~al.}(2010{\natexlab{b}})Geller, Mathieu, Braden, Meibom,
  Platais, \& Dolan}]{gellerWIYNOPENCLUSTER2010a}
---. 2010{\natexlab{b}}, AJ, 139, 1383, \dodoi{10.1088/0004-6256/139/4/1383}

\bibitem[{Geller {et~al.}(2008)Geller, Mathieu, Harris, \&
  McClure}]{gellerWIYNOPENCLUSTER2008}
Geller, A.~M., Mathieu, R.~D., Harris, H.~C., \& McClure, R.~D. 2008, AJ, 135,
  2264, \dodoi{10.1088/0004-6256/135/6/2264}

\bibitem[{Geller {et~al.}(2009)Geller, Mathieu, Harris, \&
  McClure}]{gellerWIYNOPENCLUSTER2009}
---. 2009, AJ, 137, 3743, \dodoi{10.1088/0004-6256/137/4/3743}

\bibitem[{Geller {et~al.}(2021)Geller, Mathieu, Latham, Pollack, Torres, \&
  Leiner}]{gellerStellarRadialVelocities2021}
Geller, A.~M., Mathieu, R.~D., Latham, D.~W., {et~al.} 2021, AJ, 161, 190,
  \dodoi{10.3847/1538-3881/abdd23}

\bibitem[{Gieles \& Zocchi(2015)}]{gielesFamilyLoweredIsothermal2015}
Gieles, M., \& Zocchi, A. 2015, MNRAS, 454, 576, \dodoi{10.1093/mnras/stv1848}

\bibitem[{Handberg {et~al.}(2017)Handberg, Brogaard, Miglio, Bossini, Elsworth,
  Slumstrup, Davies, \& Chaplin}]{handbergNGC6819Testing2017}
Handberg, R., Brogaard, K., Miglio, A., {et~al.} 2017, MNRAS, 472, 979,
  \dodoi{10.1093/mnras/stx1929}

\bibitem[{Harris {et~al.}(2020)Harris, Millman, {van der Walt}, Gommers,
  Virtanen, Cournapeau, Wieser, Taylor, Berg, Smith, Kern, Picus, Hoyer, {van
  Kerkwijk}, Brett, Haldane, {del R{\'i}o}, Wiebe, Peterson,
  {G{\'e}rard-Marchant}, Sheppard, Reddy, Weckesser, Abbasi, Gohlke, \&
  Oliphant}]{harrisArrayProgrammingNumPy2020}
Harris, C.~R., Millman, K.~J., {van der Walt}, S.~J., {et~al.} 2020, Natur,
  585, 357, \dodoi{10.1038/s41586-020-2649-2}

\bibitem[{Hubeny \& Lanz(2011)}]{hubenySynspecGeneralSpectrum2011}
Hubeny, I., \& Lanz, T. 2011, Astrophys. Source Code Libr., ascl:1109.022

\bibitem[{Hunt \& Reffert(2021)}]{huntImprovingOpenCluster2021}
Hunt, E.~L., \& Reffert, S. 2021, A\&A, 646, A104,
  \dodoi{10.1051/0004-6361/202039341}

\bibitem[{Khurana {et~al.}(2023)Khurana, Chawla, \&
  Chatterjee}]{khuranaDynamicallyFormingExtremely2023}
Khurana, A., Chawla, C., \& Chatterjee, S. 2023, ApJ, 949, 102,
  \dodoi{10.3847/1538-4357/acc8d6}

\bibitem[{Kim \& Chun(2000)}]{kimThreeDeltaScuti2000}
Kim, S.-L., \& Chun, M.-Y. 2000, IBVS, 4964, 1

\bibitem[{Kim {et~al.}(2001)Kim, Chun, Park, Kim, Lee, Lee, Ann, Sung, Jeon, \&
  Yuk}]{kimSearchVariableStars2001}
Kim, S.~L., Chun, M.~Y., Park, B.~G., {et~al.} 2001, AcA, 51, 49

\bibitem[{King(1962)}]{kingStructureStarClusters1962}
King, I. 1962, AJ, 67, 471, \dodoi{10.1086/108756}

\bibitem[{King(1966)}]{kingStructureStarClusters1966}
King, I.~R. 1966, AJ, 71, 276, \dodoi{10.1086/109918}

\bibitem[{Knudstrup {et~al.}(2020)Knudstrup, Grundahl, Brogaard, Slumstrup,
  Orosz, Sandquist, {Jessen-Hansen}, Lund, Arentoft, Tronsgaard, Yong,
  Frandsen, \& Bruntt}]{knudstrupExtremelyPreciseAge2020}
Knudstrup, E., Grundahl, F., Brogaard, K., {et~al.} 2020, MNRAS, 499, 1312,
  \dodoi{10.1093/mnras/staa2855}

\bibitem[{Kroupa(2002)}]{kroupaInitialMassFunction2002}
Kroupa, P. 2002, Science, 295, 82, \dodoi{10.1126/science.1067524}

\bibitem[{Leahy \& Leahy(2015)}]{leahyCalculatorRocheLobe2015}
Leahy, D.~A., \& Leahy, J.~C. 2015, ComAC, 2, 4,
  \dodoi{10.1186/s40668-015-0008-8}

\bibitem[{Lee {et~al.}(2013)Lee, Kang, \& Ann}]{leeDeepWidePhotometry2013}
Lee, S.~H., Kang, Y.-W., \& Ann, H.~B. 2013, MNRAS, 432, 1672,
  \dodoi{10.1093/mnras/stt588}

\bibitem[{Leiner {et~al.}(2018)Leiner, Mathieu, Gosnell, \&
  Sills}]{leinerObservationsSpindownPostmasstransfer2018}
Leiner, E., Mathieu, R.~D., Gosnell, N.~M., \& Sills, A. 2018, ApJ, 869, L29,
  \dodoi{10.3847/2041-8213/aaf4ed}

\bibitem[{Leiner {et~al.}(2016)Leiner, Mathieu, Stello, Vanderburg, \&
  Sandquist}]{leinerK2M67STUDY2016}
Leiner, E., Mathieu, R.~D., Stello, D., Vanderburg, A., \& Sandquist, E. 2016,
  ApJ, 832, L13, \dodoi{10.3847/2041-8205/832/1/L13}

\bibitem[{Leiner {et~al.}(2015)Leiner, Mathieu, Gosnell, \&
  Geller}]{leinerWIYNOPENCLUSTER2015}
Leiner, E.~M., Mathieu, R.~D., Gosnell, N.~M., \& Geller, A.~M. 2015, AJ, 150,
  10, \dodoi{10.1088/0004-6256/150/1/10}

\bibitem[{Leonard(1989)}]{leonardStellarCollisionsGlobular1989}
Leonard, P. J.~T. 1989, AJ, 98, 217, \dodoi{10.1086/115138}

\bibitem[{{Lightkurve Collaboration} {et~al.}(2018){Lightkurve Collaboration},
  Cardoso, Hedges, {Gully-Santiago}, Saunders, Cody, Barclay, Hall, Sagear,
  Turtelboom, Zhang, Tzanidakis, Mighell, Coughlin, Bell, {Berta-Thompson},
  Williams, Dotson, \& Barentsen}]{2018lightkurve}
{Lightkurve Collaboration}, Cardoso, J. V. d.~M., Hedges, C., {et~al.} 2018,
  Lightkurve: {{Kepler}} and {{TESS}} Time Series Analysis in {{Python}},
  Astrophysics Source Code Library.
\newblock \doeprint{1812.013}

\bibitem[{Marconi {et~al.}(1997)Marconi, Hamilton, Tosi, \&
  Bragaglia}]{marconiOldOpenClusters1997}
Marconi, G., Hamilton, D., Tosi, M., \& Bragaglia, A. 1997, MNRAS, 291, 763,
  \dodoi{10.1093/mnras/291.4.763}

\bibitem[{Masuda {et~al.}(2019)Masuda, Kawahara, Latham, Bieryla, Kunitomo,
  MacLeod, \& Aoki}]{masudaSelflensingDiscoveryWhite2019}
Masuda, K., Kawahara, H., Latham, D.~W., {et~al.} 2019, ApJL, 881, L3,
  \dodoi{10.3847/2041-8213/ab321b}

\bibitem[{Mathieu(1985)}]{mathieuStructureInternalKinematics1985}
Mathieu, R.~D. 1985, IAU Symp., 113, 427

\bibitem[{Mathieu(2000)}]{mathieuWIYNOpenCluster2000}
---. 2000, Stellar Clust. Assoc. Convect. Rotat. Dynamos Proc. ASP Conf., 198,
  517

\bibitem[{Mathieu \& Geller(2015)}]{mathieuBlueStragglersOld2015}
Mathieu, R.~D., \& Geller, A.~M. 2015, in Astrophysics and {{Space Science
  Library}}, Vol. 413, Ecology of {{Blue Straggler Stars}}, ed. {Boffin, H.},
  {Carraro, G.}, \& {Beccari, G.} (Berlin, Heidelberg: Springer), 29--63,
  \dodoi{10.1007/978-3-662-44434-4_3}

\bibitem[{Mathieu {et~al.}(1990)Mathieu, Latham, \&
  Griffin}]{mathieuOrbits22Spectroscopic1990}
Mathieu, R.~D., Latham, D.~W., \& Griffin, R.~F. 1990, AJ, 100, 1859,
  \dodoi{10.1086/115643}

\bibitem[{McClure {et~al.}(1981)McClure, Twarog, \&
  Forrester}]{mcclureOldOpenCluster1981}
McClure, R.~D., Twarog, B.~A., \& Forrester, W.~T. 1981, ApJ, 243, 841,
  \dodoi{10.1086/158650}

\bibitem[{McCrea(1964)}]{mccreaExtendedMainSequenceStellar1964}
McCrea, W.~H. 1964, MNRAS, 128, 147, \dodoi{10.1093/mnras/128.2.147}

\bibitem[{McInnes {et~al.}(2017)McInnes, Healy, \&
  Astels}]{mcinnesHdbscanHierarchicalDensity2017}
McInnes, L., Healy, J., \& Astels, S. 2017, JOSS, 2, 205,
  \dodoi{10.21105/joss.00205}

\bibitem[{Mikolaitis {et~al.}(2012)Mikolaitis, Tautvai{\v s}ien{\.e}, Gratton,
  Bragaglia, \& Carretta}]{mikolaitisAbundancesCarbonIsotope2012}
Mikolaitis, S., Tautvai{\v s}ien{\.e}, G., Gratton, R., Bragaglia, A., \&
  Carretta, E. 2012, A\&A, 541, A137, \dodoi{10.1051/0004-6361/201218831}

\bibitem[{Milliman {et~al.}(2016)Milliman, Leiner, Mathieu, Tofflemire, \&
  Platais}]{millimanWIYNOPENCLUSTER2016}
Milliman, K.~E., Leiner, E., Mathieu, R.~D., Tofflemire, B.~M., \& Platais, I.
  2016, AJ, 151, 152, \dodoi{10.3847/0004-6256/151/6/152}

\bibitem[{Milliman {et~al.}(2014)Milliman, Mathieu, Geller, Gosnell, Meibom, \&
  Platais}]{millimanWIYNOPENCLUSTER2014}
Milliman, K.~E., Mathieu, R.~D., Geller, A.~M., {et~al.} 2014, AJ, 148, 38,
  \dodoi{10.1088/0004-6256/148/2/38}

\bibitem[{Mishenina {et~al.}(2015)Mishenina, Pignatari, Carraro, Kovtyukh,
  Monaco, Korotin, Shereta, Yegorova, \& Herwig}]{misheninaNewInsightsBa2015}
Mishenina, T., Pignatari, M., Carraro, G., {et~al.} 2015, MNRAS, 446, 3651,
  \dodoi{10.1093/mnras/stu2337}

\bibitem[{Momany {et~al.}(2001)Momany, Vandame, Zaggia, Mignani, {da Costa},
  Arnouts, Groenewegen, Hatziminaoglou, Madejsky, Rit{\'e}, Schirmer, \&
  Slijkhuis}]{momanyESOImagingSurvey2001}
Momany, Y., Vandame, B., Zaggia, S., {et~al.} 2001, A\&A, 379, 436,
  \dodoi{10.1051/0004-6361:20011325}

\bibitem[{Morrissey {et~al.}(2007)Morrissey, Conrow, Barlow, Small, Seibert,
  Wyder, Budavari, Arnouts, Friedman, Forster, Martin, Neff, Schiminovich,
  Bianchi, Donas, Heckman, Lee, Madore, Milliard, Rich, Szalay, Welsh, \&
  Yi}]{morrisseyCalibrationDataProducts2007}
Morrissey, P., Conrow, T., Barlow, T.~A., {et~al.} 2007, ApJS, 173, 682,
  \dodoi{10.1086/520512}

\bibitem[{Morton(2015)}]{mortonIsochronesStellarModel2015}
Morton, T.~D. 2015, Astrophys. Source Code Libr., ascl:1503.010

\bibitem[{Netopil {et~al.}(2016)Netopil, Paunzen, Heiter, \&
  Soubiran}]{netopilMetallicityOpenClusters2016}
Netopil, M., Paunzen, E., Heiter, U., \& Soubiran, C. 2016, A\&A, 585, A150,
  \dodoi{10.1051/0004-6361/201526370}

\bibitem[{Nine {et~al.}(2020)Nine, Milliman, Mathieu, Geller, Leiner, Platais,
  \& Tofflemire}]{nineWIYNOpenCluster2020}
Nine, A.~C., Milliman, K.~E., Mathieu, R.~D., {et~al.} 2020, AJ, 160, 169,
  \dodoi{10.3847/1538-3881/abad3b}

\bibitem[{Ochsenbein {et~al.}(2000)Ochsenbein, Bauer, \&
  Marcout}]{ochsenbeinVizieRDatabaseAstronomical2000}
Ochsenbein, F., Bauer, P., \& Marcout, J. 2000, A\&AS, 143, 23,
  \dodoi{10.1051/aas:2000169}

\bibitem[{Panthi {et~al.}(2022)Panthi, Vaidya, Jadhav, Rao, Subramaniam,
  Agarwal, \& Pandey}]{panthiUOCSVIIIUV2022}
Panthi, A., Vaidya, K., Jadhav, V., {et~al.} 2022, MNRAS, 516, 5318,
  \dodoi{10.1093/mnras/stac2421}

\bibitem[{Paxton {et~al.}(2011)Paxton, Bildsten, Dotter, Herwig, Lesaffre, \&
  Timmes}]{paxtonMODULESEXPERIMENTSSTELLAR2011}
Paxton, B., Bildsten, L., Dotter, A., {et~al.} 2011, ApJS, 192, 3,
  \dodoi{10.1088/0067-0049/192/1/3}

\bibitem[{Paxton {et~al.}(2013)Paxton, Cantiello, Arras, Bildsten, Brown,
  Dotter, Mankovich, Montgomery, Stello, Timmes, \&
  Townsend}]{paxtonMODULESEXPERIMENTSSTELLAR2013}
Paxton, B., Cantiello, M., Arras, P., {et~al.} 2013, ApJS, 208, 4,
  \dodoi{10.1088/0067-0049/208/1/4}

\bibitem[{Paxton {et~al.}(2015)Paxton, Marchant, Schwab, Bauer, Bildsten,
  Cantiello, Dessart, Farmer, Hu, Langer, Townsend, Townsley, \&
  Timmes}]{paxtonMODULESEXPERIMENTSSTELLAR2015}
Paxton, B., Marchant, P., Schwab, J., {et~al.} 2015, ApJS, 220, 15,
  \dodoi{10.1088/0067-0049/220/1/15}

\bibitem[{Pedregosa {et~al.}(2011)Pedregosa, Varoquaux, Gramfort, Michel,
  Thirion, Grisel, Blondel, Prettenhofer, Weiss, Dubourg, Vanderplas, Passos,
  Cournapeau, Brucher, Perrot, \& Duchesnay}]{scikit-learn}
Pedregosa, F., Varoquaux, G., Gramfort, A., {et~al.} 2011, J. Mach. Learn.
  Res., 12, 2825

\bibitem[{Perets \& Fabrycky(2009)}]{peretsTRIPLEORIGINBLUE2009}
Perets, H.~B., \& Fabrycky, D.~C. 2009, ApJ, 697, 1048,
  \dodoi{10.1088/0004-637X/697/2/1048}

\bibitem[{Portegies~Zwart \& Leigh(2019)}]{zwartTripleOriginTwin2019}
Portegies~Zwart, S., \& Leigh, N. W.~C. 2019, ApJL, 876, L33,
  \dodoi{10.3847/2041-8213/ab1b75}

\bibitem[{Rain {et~al.}(2021)Rain, Ahumada, \& Carraro}]{rainNewGaiaBased2021}
Rain, M.~J., Ahumada, J.~A., \& Carraro, G. 2021, A\&A, 650, A67,
  \dodoi{10.1051/0004-6361/202040072}

\bibitem[{Rangwal {et~al.}(2019)Rangwal, Yadav, Durgapal, Bisht, \&
  Nardiello}]{rangwalAstrometricPhotometricStudy2019}
Rangwal, G., Yadav, R. K.~S., Durgapal, A., Bisht, D., \& Nardiello, D. 2019,
  MNRAS, 490, 1383, \dodoi{10.1093/mnras/stz2642}

\bibitem[{Reddy {et~al.}(2012)Reddy, Giridhar, \&
  Lambert}]{reddyComprehensiveAbundanceAnalysis2012}
Reddy, A. B.~S., Giridhar, S., \& Lambert, D.~L. 2012, MNRAS, 419, 1350,
  \dodoi{10.1111/j.1365-2966.2011.19791.x}

\bibitem[{Renzini \& Fusi~Pecci(1988)}]{renziniTestsEvolutionarySequences1988}
Renzini, A., \& Fusi~Pecci, F. 1988, ARA\&A, 26, 199,
  \dodoi{10.1146/annurev.aa.26.090188.001215}

\bibitem[{Rosvick \& VandenBerg(1998)}]{rosvickBVPhotometryGyr1998}
Rosvick, J.~M., \& VandenBerg, D.~A. 1998, AJ, 115, 1516,
  \dodoi{10.1086/300304}

\bibitem[{S{\'a}nchez~Arias {et~al.}(2017)S{\'a}nchez~Arias, C{\'o}rsico, \&
  Althaus}]{sanchezariasAsteroseismologyHybridScuti2017}
S{\'a}nchez~Arias, J.~P., C{\'o}rsico, A.~H., \& Althaus, L.~G. 2017, A\&A,
  597, A29, \dodoi{10.1051/0004-6361/201629126}

\bibitem[{Sills \& Bailyn(1999)}]{sillsDistributionCollisionallyInduced1999}
Sills, A., \& Bailyn, C.~D. 1999, ApJ, 513, 428, \dodoi{10.1086/306840}

\bibitem[{Spitzer(1987)}]{spitzerDynamicalEvolutionGlobular1987}
Spitzer, L. 1987, Dynamical Evolution of Globular Clusters, Princeton
  {{Series}} in {{Astrophysics}} (Princeton University Press)

\bibitem[{Stassun {et~al.}(2023)Stassun, Torres, Kounkel, Tofflemire, Leiner,
  Feliz, Dixon, Mathieu, Gosnell, \&
  {Gully-Santiago}}]{stassunEclipsingBinaryComprising2023}
Stassun, K.~G., Torres, G., Kounkel, M., {et~al.} 2023, ApJ, 950, 99,
  \dodoi{10.3847/1538-4357/acd17c}

\bibitem[{Tabetha~Hole {et~al.}(2009)Tabetha~Hole, Geller, Mathieu, Platais,
  Meibom, \& Latham}]{tabethaholeWIYNOPENCLUSTER2009}
Tabetha~Hole, K., Geller, A.~M., Mathieu, R.~D., {et~al.} 2009, AJ, 138, 159,
  \dodoi{10.1088/0004-6256/138/1/159}

\bibitem[{Tofflemire {et~al.}(2014)Tofflemire, Gosnell, Mathieu, \&
  Platais}]{tofflemireWIYNOPENCLUSTER2014}
Tofflemire, B.~M., Gosnell, N.~M., Mathieu, R.~D., \& Platais, I. 2014, AJ,
  148, 61, \dodoi{10.1088/0004-6256/148/4/61}

\bibitem[{Vaidya {et~al.}(2020)Vaidya, Rao, Agarwal, \&
  Bhattacharya}]{vaidyaBlueStragglerPopulations2020}
Vaidya, K., Rao, K.~K., Agarwal, M., \& Bhattacharya, S. 2020, MNRAS, 496,
  2402, \dodoi{10.1093/mnras/staa1667}

\bibitem[{van Dokkum(2001)}]{dokkumCosmicRayRejection2001}
van Dokkum, P.~G. 2001, PASP, 113, 1420, \dodoi{10.1086/323894}

\bibitem[{Virtanen {et~al.}(2020)Virtanen, Gommers, Oliphant, Haberland, Reddy,
  Cournapeau, Burovski, Peterson, Weckesser, Bright, {van der Walt}, Brett,
  Wilson, Millman, Mayorov, Nelson, Jones, Kern, Larson, Carey, Polat, Feng,
  Moore, VanderPlas, Laxalde, Perktold, Cimrman, Henriksen, Quintero, Harris,
  Archibald, Ribeiro, Pedregosa, \& {van
  Mulbregt}}]{virtanenSciPyFundamentalAlgorithms2020}
Virtanen, P., Gommers, R., Oliphant, T.~E., {et~al.} 2020, Nat Methods, 17,
  261, \dodoi{10.1038/s41592-019-0686-2}

\bibitem[{{Wes McKinney}(2010)}]{mckinney-proc-scipy-2010}
{Wes McKinney}. 2010, in Proc. 9th {{Python Sci}}. {{Conf}}., ed. S.~{van der
  Walt} \& {Jarrod Millman}, 56--61, \dodoi{10.25080/Majora-92bf1922-00a}

\bibitem[{Zucker \& Mazeh(1994)}]{zuckerStudySpectroscopicBinaries1994}
Zucker, S., \& Mazeh, T. 1994, ApJ, 420, 806, \dodoi{10.1086/173605}

\end{thebibliography}

\end{document}